\documentclass{jfm}
\usepackage{graphicx}
\usepackage{epstopdf, epsfig}

\usepackage{setspace} 
\usepackage{graphicx}
\usepackage{amssymb}
\usepackage{epstopdf}
\usepackage{color,graphicx}
\usepackage{color, xcolor}
\usepackage{tikz}
\usetikzlibrary{shapes}
\usepackage{psfrag}
\usepackage{subfigure}
\usepackage{amsmath}
\usepackage{adjustbox}
\usepackage{fancyhdr}
\usepackage{fancybox}
\usepackage{wrapfig}
\usepackage[version=4]{mhchem}
\usepackage{tabularx}
\usepackage{array} 
\usepackage{multirow}
\usepackage{soul}

\usepackage{bm}

\def\be{\begin{equation}}
\def\ee{\end{equation}}
\def\beno{\begin{eqnarray}}
\def\eeno{\end{eqnarray}}
\def\beqno{\begin{eqnarray*}}
\def\eeqno{\end{eqnarray*}}

\def\del{\partial}

\newcommand{\closure}[2][3]{%
{}\mkern#1mu\overline{\mkern-#1mu#2}}

\newcommand{\blkcircle}{\raisebox{0.5pt}{\protect\tikz{\protect\node[draw,scale=0.7,circle,fill=none](){};}}}
\newcommand{\blkfillcircle}{\raisebox{0.5pt}{\protect\tikz{\protect\node[draw,scale=0.7,circle,fill=black](){};}}}

\newcommand{\redcircle}{\raisebox{0.5pt}{\protect\tikz{\protect\node[draw=red,scale=0.7,circle,fill=none](){};}}}

\newcommand{\graysquare}{\raisebox{0.5pt}{\protect\tikz{\protect\node[draw=gray,scale=0.7,regular polygon, regular polygon sides=4,fill=none,rotate=0](){};}}}

\newcommand{\blksquare}{\raisebox{0.5pt}{\protect\tikz{\protect\node[draw=black,scale=0.7,regular polygon, regular polygon sides=4,fill=none,rotate=0](){};}}}

\newcommand{\bluediamond}{\raisebox{0pt}{\protect\tikz{\protect\node[draw=blue,scale=0.6,diamond,fill=none](){};}}}
\newcommand{\bluefilldiamond}{\raisebox{0pt}{\protect\tikz{\protect\node[draw=blue,scale=0.6,diamond,fill=blue](){};}}}

\newcommand{\blkline}{\raisebox{2pt}{\protect\tikz{\protect\draw[-,black!40!black,solid,line width = 1.0pt](0,0) -- (3.0mm,0);}}}
\newcommand{\greenline}{\raisebox{2pt}{\protect\tikz{\protect\draw[-,green!40!green,solid,line width = 1.0pt](0,0) -- (3.0mm,0);}}}

\newcommand{\cyanline}{\raisebox{2pt}{\protect\tikz{\protect\draw[-,blue!0.5!cyan,fill=blue!0.5!cyan,solid,line width = 1.0pt](0,0) -- (3.0mm,0);}}}

\newcommand{\blkdashdotline}{\raisebox{2pt}{\protect\tikz{\protect\draw[-,black!40!black, dashdotted,line width = 1.1pt](0,0) -- (4.0mm,0);}}}
\newcommand{\reddashdotline}{\raisebox{2pt}{\protect\tikz{\protect\draw[-,red!40!red, dashdotted,line width = 1.1pt](0,0) -- (4.0mm,0);}}}
\newcommand{\blkdashline}{\raisebox{2pt}{\protect\tikz{\protect\draw[-,black!40!black, dashed,line width = 1.1pt](0,0) -- (3.0mm,0);}}}

\newcommand{\bluedottedline}{\raisebox{2pt}{\protect\tikz{\protect\draw[-,blue!40!blue,dotted,line width = 1.2pt](0,0) -- (3.0mm,0);}}}
\newcommand{\grayline}{\raisebox{2pt}{\protect\tikz{\protect\draw[-,gray!40!gray,solid,line width = 1.0pt](0,0) -- (3.0mm,0);}}}
\newcommand{\graydottedline}{\raisebox{2pt}{\protect\tikz{\protect\draw[-,gray!40!gray,dotted,line width = 1.2pt](0,0) -- (3.0mm,0);}}}
\newcommand{\graydashdotline}{\raisebox{2pt}{\protect\tikz{\protect\draw[-,gray!40!gray, dashdotted,line width = 1.1pt](0,0) -- (4.0mm,0);}}}
\newcommand{\graydashline}{\raisebox{2pt}{\protect\tikz{\protect\draw[-,gray!40!gray, dashed,line width = 1.1pt](0,0) -- (3.0mm,0);}}}

\newcommand{\greencirclesolidline}{\raisebox{0pt}{\protect\tikz{\protect\node[draw=green,scale=0.7,circle,fill=none](){};\protect\draw[-,green!40!green,solid,line width = 1.0pt](-0.2,0.1mm) -- (2.0mm,0.1mm)}}}
\newcommand{\bluerectangledottedline}{\raisebox{0pt}{\protect\tikz{\protect\node[draw=blue,scale=0.6,regular polygon, regular polygon sides=4,fill=none,rotate=0](){};\protect\draw[-,blue!40!blue,dotted,line width = 1.2pt](-0.2,0.1mm) -- (2.0mm,0.1mm)}}}
\newcommand{\blacktriangledashdotline}{\raisebox{0pt}{\protect\tikz{\protect\node[draw,scale=0.5,regular polygon, regular polygon sides=3,fill=none,rotate=0](){};\protect\draw[-,black!40!black, dashdotted,line width = 1.1pt](-0.2,0.1mm) -- (2.0mm,0.1mm)}}}
\newcommand{\reddiamonddashdotline}{\raisebox{0pt}{\protect\tikz{\protect\node[draw=red,scale=0.5,diamond,fill=none](){};\protect\draw[-,red!40!red, dashdotted,line width = 1.1pt](-0.2,0.1mm) -- (2.0mm,0.1mm)}}}

\shorttitle{Pore-resolved open channel flow over a permeable bed}
\shortauthor{S. K. Karra, S. V. Apte, X. He and T. D. Scheibe}

\title{
Pore-resolved investigation of turbulent open channel flow over a randomly packed permeable sediment bed}

\author{Shashank K. Karra\aff{1},
 Sourabh V. Apte\aff{1} \corresp{\email{sourabh.apte@oregonstate.edu}}, \\ 
 Xiaoliang He\aff{2}, \and Timothy D. Scheibe\aff{2}\\}

\affiliation{\aff{1}School of Mechanical, Industrial and Manufacturing Engineering, Oregon State University, Corvallis, OR, 97331
\aff{2}Pacific Northwest National Laboratory, Richland, WA, 99354}

\begin{document}

\maketitle

\begin{abstract}
Pore-resolved direct numerical simulations (DNS) are performed to investigate the interactions between streamflow turbulence and groundwater flow through a randomly packed porous sediment bed for three permeability Reynolds numbers, $Re_K$, of 2.56, 5.17, and 8.94, representative of natural stream or river systems. Time-space averaging is used to quantify the Reynolds stress, form-induced stress, mean flow and shear penetration depths, and mixing length at the sediment-water interface (SWI). The mean flow and shear penetration depths increase with $Re_K$ and are found to be nonlinear functions of non-dimensional permeability. The peaks and significant values of the Reynolds stresses, form-induced stresses, and pressure variations are shown to occur in the top layer of the bed, which is also confirmed by conducting simulations of just the top layer as roughness elements over an impermeable wall. 
The probability distribution functions (PDFs) of normalized local bed stress are found to collapse for all Reynolds numbers and their root mean-squared fluctuations are assumed to follow logarithmic correlations. The fluctuations in local bed stress and resultant drag and lift forces on sediment grains are mainly a result of the top layer, their PDFs are symmetric with heavy tails, and can be well represented by a non-Gaussian model fit.  
The bed stress statistics and the pressure data at the SWI can potentially be used in providing better boundary conditions in modeling of incipient motion and reach-scale transport in the hyporheic zone.
\end{abstract}

\section{Introduction}\label{sec:intro}
The interchange of mass and momentum between streamflow and ground water occurs across the sediment-water interface (SWI) and into the porous bed underneath, termed as the hyporheic zone.
Hyporheic transient storage or retention and transport of solutes such as chemicals and pollutants, dissolved oxygen, nutrients, and heat across the SWI is one of the most important concepts for stream ecology, and has enormous societal value in predicting source of fresh drinking water, transport, biogeochemical processing of 
nutrients, and sustaining diverse aquatic ecosystems~\citep{bencala1983rhodamine,d1993transient,valett1996parent,harvey1996evaluating,anderson2008groundwater,briggs2009method,grant2018hypor}.

A broad range of spatio-temporal scales corresponding to disparate physical and chemical processes contribute to the mixing within the hyporheic zone~\citep{hester2017importance}. Turbulent transport across the SWI, coherent flow structures, and non-Darcy flow within the sediment bed have been hypothesized as critical mechanisms impacting transient storage. The importance of penetration of turbulence within the bed and near bed pressure fluctuations are crucial and their impact on the hyporheic transient storage is poorly understood~\citep{hester2017importance}. 
Moreover, turbulence characteristics over a permeable bed are different compared to a rough, impermeable wall~\citep{jimenez2004turbulent}; impacting long time-scales of retention within the bed. 

\begin{figure}
   \centering
   \includegraphics[width=9cm,height=6cm,keepaspectratio]{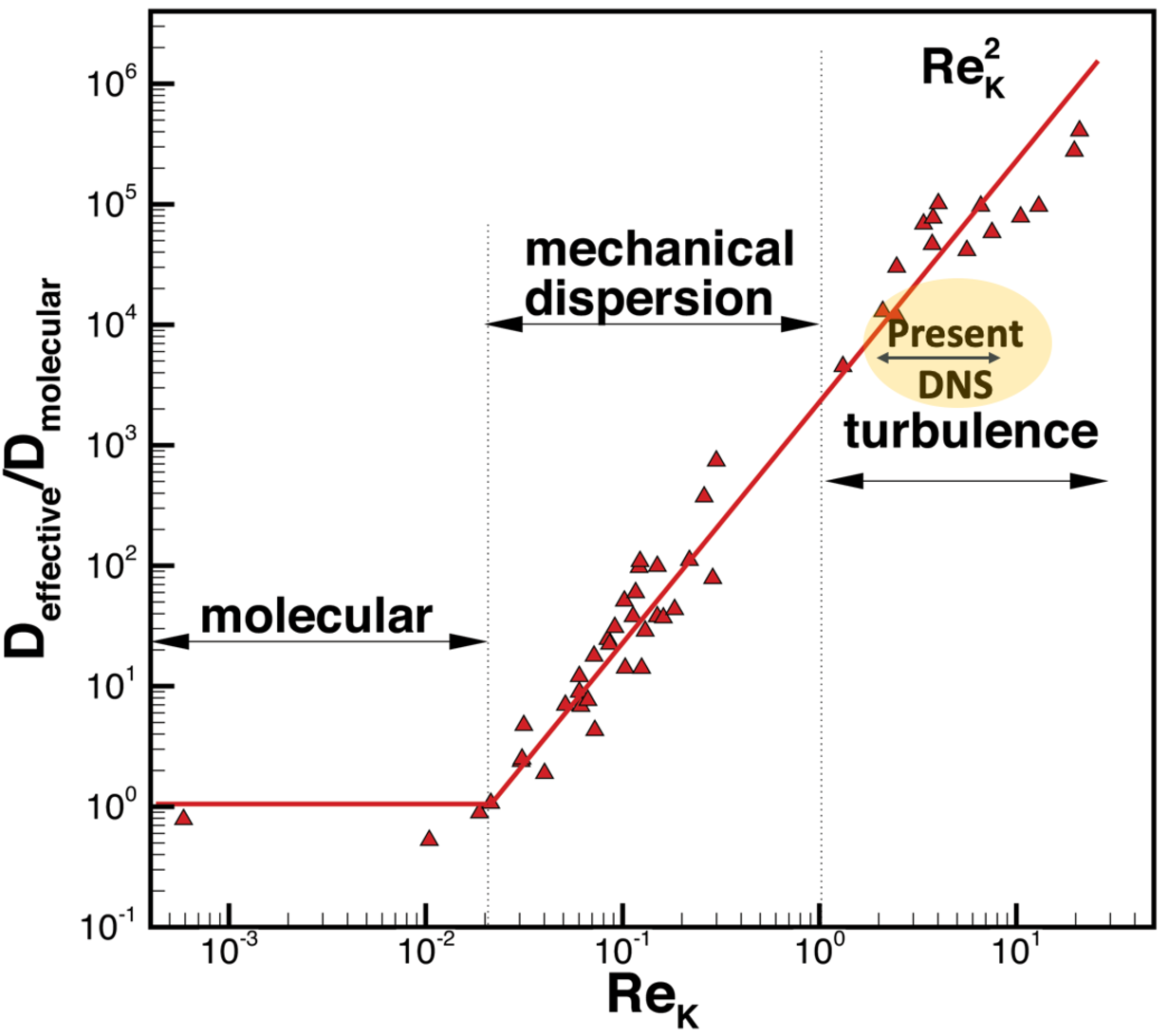}
\caption{\small Effective dispersion coefficient versus $Re_{K}$~(modified based on~\citet{voermans2017variation,grant2018hypor})}
\label{fig:regime_bedschem}
\end{figure}

Mass and momentum transport in turbulent flow over a naturally occurring permeable sediment bed are characterized by bed permeability, $K$ (which depends upon its porosity and average grain size according to the Carman-Kozeny relation), sediment bed arrangement (flat versus complex bedforms), and friction or shear velocity, $u_{\tau}$ (based on equation~\ref{eq:utau}). Permeability Reynolds number, $Re_K = u_{\tau} \sqrt{K}/\nu$, (where $\nu$ is the kinematic viscosity) representing the ratio between the permeability scale ($\sqrt{K}$) to the viscous scale (${\nu/u_{\tau}}$), is typically used for granular beds to identify different flow regimes based on the dominant transport mechanisms across the SWI. For flat beds, three flow regimes (see figure~\ref{fig:regime_bedschem}) characterized by~\citet{voermans2017variation,voermans2018model,grant2018hypor} have been identified:
(i) the molecular regime, $Re_K<0.01$, where the bed is nearly impermeable and the transport is governed by molecular diffusion; (ii) the dispersive regime, $0.01<Re_K<1$, where dispersive transport associated with laminarization of stream turbulence is important; and (iii) the turbulent regime, $Re_K>1$, where turbulence is dominant near the highly permeable interface. 
Based on the data collected for flat beds from several local streams and rivers near Oregon State University by coworkers~\citep{jackson2013fluid,jackson2015flow}, it is found that the gravel grain sizes varied over the range 5--70mm, with friction velocity, $u_{\tau}=0.004$--$0.088~\rm m/s$, and $Re_K=2$--70~(figure~\ref{fig:regime_bedschem}). Free surface flow and waves typically do not affect the hyporheic exchange in natural stream and rivers under subcritical conditions with small Froude numbers.

Few experimental studies have evaluated turbulence characteristics over flat permeable beds \citep{zagni1976channel,zippe1983turbulent,manes2009turbulence,manes2011turbulent,suga2010effects,voermans2017variation, kim2020experimental,rousseau2022experimental}. Bed permeability was found to increase friction coefficient, reduce the wall-blocking effect due to impermeable rough walls, and reduce near-bed anisotropy in turbulence intensities. \citet{manes2009turbulence} studied turbulent flow over uniform cubic pattern of single and multiple layers of spheres at $Re_K$ of 31.2 and 44.6. Permeability was shown to influence flow resistance dramatically, and the conventional assumption of hydraulically-rough regime, wherein the friction factor is dependent upon the relative submergence or the ratio of the roughness size to flow thickness, does not apply to permeable beds. Friction factor was shown to progressively increase with increasing Reynolds number.

\citet{voermans2017variation} studied the influence of different $Re_K$ on the interaction between surface and subsurface flows at the SWI of a synthetic sediment bed composed of randomly-arranged monodispersed spheres using refractive-index matched particle tracking velocimetry. Their experiments covered a wide range of $Re_K=0.36$--$6.3$ and varied the permeability of the beds by investigating three different sphere sizes. The results demonstrated a strong relationship between the structure of the mean and turbulent flow at the SWI and $Re_K$. 
Their data shows that for $Re_K= \mathcal{O}(1-10)$, the turbulence shear penetration depth, a measure of true roughness felt by the flow, normalized by the permeability scale ($\sqrt{K}$) is a non-linear function of $Re_K$, as opposed to a commonly assumed linear relationship for $Re_K < {\mathcal O(100)}$~\citep{ghisalberti2009obstructed,manes2012phenomenological}.
\citet{kim2020experimental} also investigated, through experimental observations at $Re_K=50$, the dynamic interplay between surface and subsurface flow in the presence of smooth  and rough permeable walls, composed of a uniform cubic arrangement of packed spheres. 
They confirmed the existence of amplitude modulation, a phenomenon typically identified in impermeable boundaries, whereby the outer large scales modulate the intensity of the near-wall small scale turbulence.  
They postulated that amplitude modulation of subsurface flow is driven by large-scale pressure fluctuations at the SWI and are generated by the passage of large-scale motions in the log-law region of the surface flow. However, detailed data on the pressure field at the SWI is needed to confirm these findings, a task difficult for experimental measurements, and thus requiring a need for pore-resolved direct simulations.

There have been a few pore-resolved DNS or large-eddy simulations (LES) of turbulent flow over permeable flat beds~\citep{breugem2005direct,
breugem2006influence,kuwata2016lattice,leonardi2018surface,fang2018influence}. However, all of these studies used structured, arranged packings of either non-touching particles or compactly packed particles. Although these numerical studies provide considerable insights into the fundamental aspects of turbulent flow over permeable beds, natural systems involve randomly packed beds with varying particle shapes and sizes. Furthermore, with structured packings such as simple cubic or body-centered cubic, there are open flow pathways that can lead to significant flow penetration and transport, which are not generally present in randomly packed natural systems~\citep{finn2013numerical,finn2013relative, fang2018influence}. At the time of writing, there has only been one pore-resolved DNS study~\citep{shen2020direct} involving randomly packed arrangement of monodispersed spheres with multiple layers at $Re_K=2.62$ and bed porosity of 0.41. They provided significant insights into the flow physics of turbulence over randomly packed beds. To investigate the effect of bed-roughness in regular versus randomly packed spheres, they changed only the top layer of the bed to regularly arranged spheres. More intense mixing was observed near the random interface due to increased Reynolds and form-induced stresses, which resulted in a deeper penetration of turbulence (44\% higher) than the uniform, regularly arranged interface. Although this is one of the first studies of flat beds closely resembling natural systems, the investigation was only carried out at low $Re_K$ that falls in the marginally turbulent regime.  In addition, since only the top layer of the sediment bed was changed from random to arranged, the open flow pathways that are characteristics of arranged packing were potentially absent in their study.

To the authors' best knowledge, pore-resolved DNS of randomly packed flat beds over $Re_K\sim {\mathcal O}(10)$ have not been conducted. This range is of critical importance to stream flows as measurements in local creeks show $Re_K$ values on this order~\citep{jackson2013mean,jackson2015flow} (see figure~\ref{fig:regime_bedschem}). The sweep-ejection events over permeable beds can cause significant spatio-temporal variability in the bed shear stresses and pressure forces on the sediment grains as well as at the sediment water interface. Direct measurements of these quantities in experiments pose significant challenges and hence have not been conducted. The present pore-resolved DNS studies aim to provide detailed data on local distribution of the bed stresses as well as drag and lift forces which can be of importance for incipient motion models. Similarly, models for spatio-temporal pressure variation on the sediment bed can be used as boundary conditions in reach-scale modeling of hyporheic exchange~\citep{chen2018hyporheic}. Thus, conducting a detailed analysis of the data for a range of $Re_K$ representative of stream and creek flows are of direct relevance in modeling transport across the sediment-water interface.

In the present study, pore-resolved DNS of flow over a bed of randomly packed, monodispersed, spherical particles for $Re_K= 2.56$, $5.17$, and $8.94$, representative of transitional to fully turbulent flows are performed. The main goals of this study are to (i) first characterize the nature of the turbulent flow, Reynolds and form-induced stresses, turbulence penetration depths, and sweep-ejection events as a function of $Re_K$, (ii) quantify the spatio-temporal variability of the bed stress and resultant forces on the sediment grains and pressure fluctuations at the SWI using higher-order statistics and propose a model fit for the probability distribution functions that can be used in reduced-order models, and (iii) quantify the contribution of the top sediment layer on the turbulent flow characteristics over permeable beds.

The rest of the paper is arranged as follows. The methodology, flow domain, and simulation parameters are described in section~\ref{sec:setup}. To focus on main results and insights from this work, details on grid refinement and validation studies are presented in the Appendix.  
Details of turbulence structure, mean, turbulent, and dispersive stresses, turbulence penetration depths, quadrant analysis, and the role of the top sediment layer followed by detailed statistics of the bed stress, drag and lift forces on the sediment grains, and pressure variations at the SWI are presented in section~\ref{sec:res} . Importance of the results to hyporheic exchange and transport across the SWI is summarized in section~\ref{sec:conclusions}.

\section {Simulation Setup and Mathematical Formulation}\label{sec:setup}
In this section, the flow domain and parameters for cases studied, numerical approach, grid resolution, and averaging procedure for analysis are described.
\subsection{Simulation domain and parameters}\label{sec:sim_param}
Various non-dimensional parameters relevant to the turbulent flow over a permeable bed, made of monodispersed spherical particles are permeability Reynolds number ($Re_K= u_{\tau} \sqrt{K}/\nu$), the friction or turbulent Reynolds number ($Re_{\tau}= u_{\tau} \delta/\nu$), the roughness Reynolds number $D^{+}= D_p u_{\tau}/\nu$, the bulk Reynolds number ($Re_b = \delta U_b/\nu$), the ratio of sediment depth to the free-stream height ($H_s/\delta$), the ratio of the sediment grain diameter to the free-stream height ($D_p/\delta$), bed porosity ($\phi$), type of particle packing (random versus arranged), and the domain lengths in the streamwise and spanwise directions normalized by the free-stream height ($L_x/\delta$, $L_z/\delta$). Here, $u_{\tau}$ is the friction velocity, $U_b$ is the bulk velocity, $K$ is the bed permeability and $\nu$ is the kinematic viscosity. For monodispersed, spherical  particles, the of size of the roughness element, $k_s$, scales with the permeability ($k_s/\sqrt{K}\approx 9$)~\citep{wilson2008grain,voermans2017variation}). It should be noted that for a given mono-dispersed, randomly packed bed of certain porosity and flow rate, only one of the non-dimensional Reynolds numbers is an independent parameter. 

\begin{figure}
   \centering
   \subfigure[Domain schematic]{
   \includegraphics[width=5cm,height=4cm,keepaspectratio]{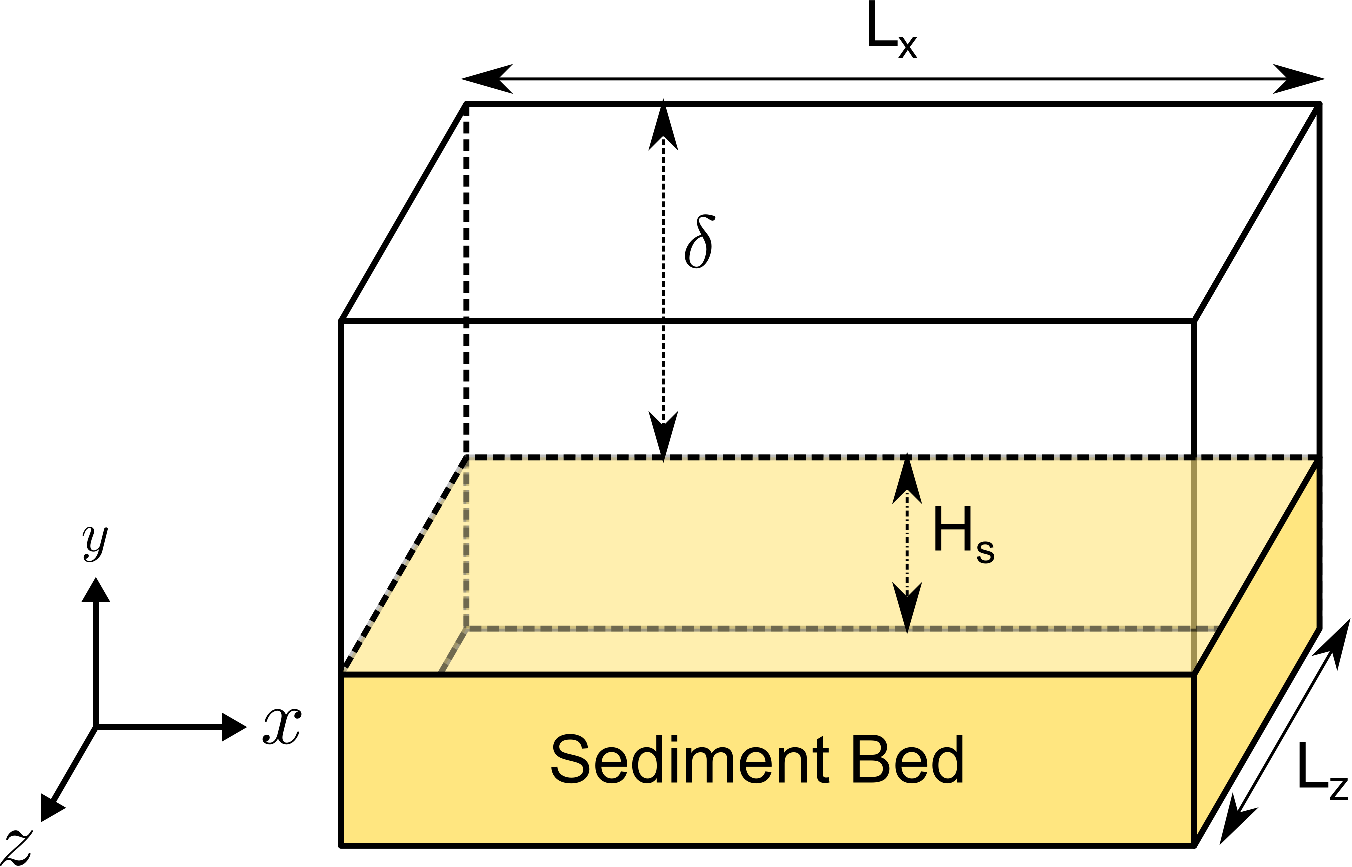}}
    \subfigure[Bed porosity]{
   \includegraphics[width=3.5cm,height=3.8cm,keepaspectratio]{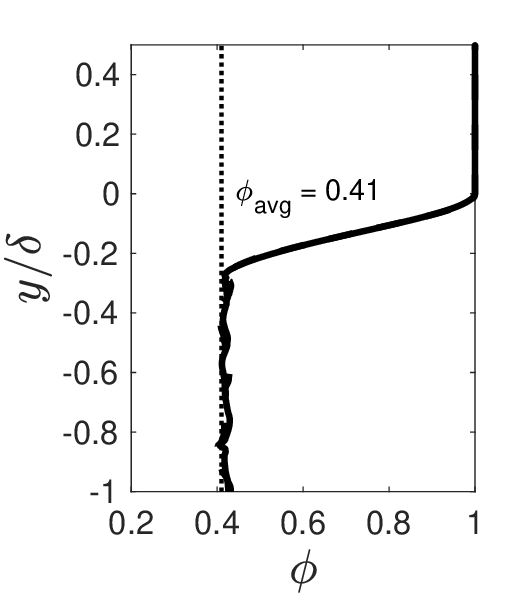}}
   \subfigure[Permeable bed]{
   \includegraphics[width=7cm,height=6cm,keepaspectratio]{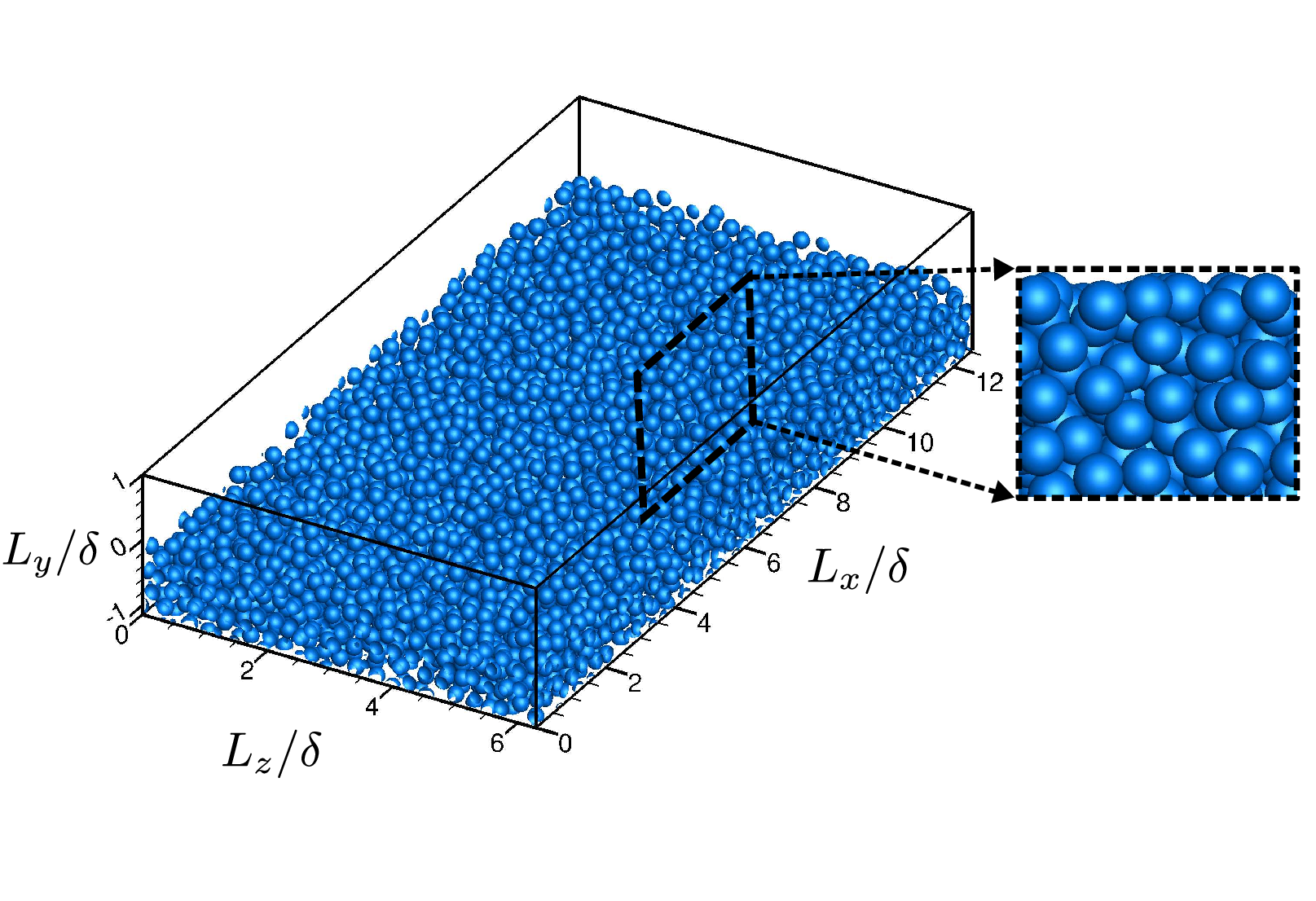}}
    \subfigure[IWM-F]{
    \includegraphics[width=5.5cm,height=6.0cm,keepaspectratio]{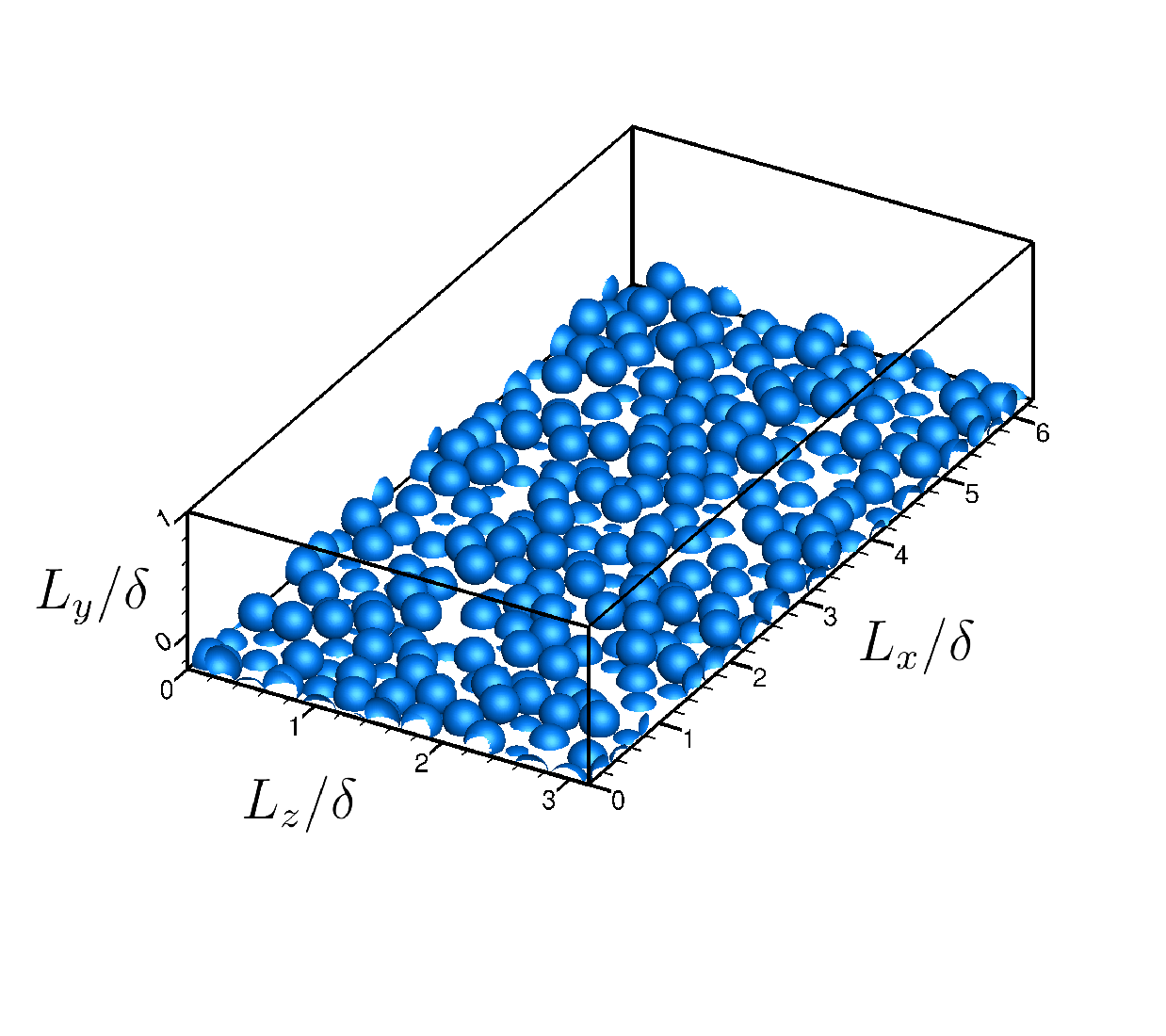}}  
\caption{\small (a) Schematic of the permeable bed, 
(b) porosity profile, 
(c)  permeable bed with randomly packed sediment particles (inset shows close-up view in xy-plane), and (e)  impermeable-wall medium Reynolds number case with full layer of particles. Periodicity is used in the streamwise ($x$) and spanwise ($z$) directions.  Top and bottom surface are defined as slip boundaries for the permeable bed cases, whereas, for the impermeable-wall full layer (IWM-F) case, the bottom surface is a no-slip wall.} 
\label{fig:dom}
\end{figure}

Figure~\ref{fig:dom}a shows the schematic of the sediment bed and the computational domain used in the present study. 
A doubly periodic domain in streamwise ($x-$) and spanwise ($z-$) directions with four layers of randomly packed, monodispersed sediment grains of porosity $\phi = 0.41$ is shown in figure~\ref{fig:dom}c. The random packing of monodispersed, spherical particles is generated using the code developed by~\citet{dye2013description}. The bed porosity profile for low, medium and high permeability Reynolds number cases is shown in figure~\ref{fig:dom}b. In order to quantify the role and influence of the roughness due to only the top layer of the sediment bed on turbulence structure, and bed pressure and shear stresses, a rough impermeable wall configuration is generated as shown in figure~\ref{fig:dom}d. The roughness elements match exactly with the top layer of the porous sediment bed and are placed on top of a no-slip solid wall.

\begin{table}
\begin{center}
\def~{\hphantom{0}}
\begin{tabular}{@{}lc c c c c c c c c c }
Case & Domain &$Re_K$ & $Re_{\tau}$ & $Re_{b}$ & $D^+$  & $\phi$ & $H_s/{\delta}$ & $D_p/{\delta}$ & ${(L_x, L_z)/ \delta}$ \\ 
VV & permeable &2.56 &180  &1,886 &77 &0.41 &1.71 &0.43 &($4\pi$,$2\pi$)   \\
PBL & permeable &2.56 &270  &2,823 &77 &0.41 &1.14 &0.29 &($4\pi$,$2\pi$)   \\
PBM & permeable  &5.17 &545  &5,681 &156 &0.41 &1.14 &0.29 &($2\pi$,$\pi$)  \\
PBH & permeable  &8.94 &943  &9,965 &270 &0.41 &1.14 &0.29 &($2\pi$,$\pi$)  \\
IWM-F & impermeable  &- &545  &5,683 &156 &- &0.29 &0.29 &($2\pi$,$\pi$)  \\
\end{tabular}
\caption{Parameters used in the present pore-resolved DNS: $D_p$ is the sphere diameter, $\delta$ is the free-stream height, $H_s$ is the sediment depth, $\phi$ is the porosity, $L_x$ and $L_z$ are the streamwise and spanwise domain lengths, $Re_K$, $Re_{\tau}$, $Re_b$ and $D^+$ are the permeability, friction, bulk and roughness Reynolds numbers, respectively. $(~)^{+}$ denotes wall units.}
\label{tab:cases1}
\end{center}
\end{table}
Table~\ref{tab:cases1} shows detailed simulation parameters for the cases  used to investigate the structure and dynamics of turbulence over a porous sediment bed. Four permeable bed cases (VV, PBL, PBM, and PBH) are studied by varying the permeability Reynolds number. Case VV is used to verify and validate the DNS simulations of turbulent flow over a sediment bed with experimental data from~\citet{voermans2017variation} as well as simulation data of~\citet{shen2020direct} and its details are given in Appendix C. The free-stream height for the VV case is based on the DNS study of~\citet{shen2020direct}. 

Cases PBL, PBM, and PBH correspond to low (2.56), medium (5.17), and high (8.94) permeability Reynolds numbers, and are used to investigate the influence of $Re_K$ over the turbulent flow regime shown in figure~\ref{fig:regime_bedschem}. Finally, case IWM-F is an impermeable-wall case with only the top layer of the PBM sediment bed used as roughness elements over a no-slip surface. The free-stream height, $\delta$, for the PBL, PBM, PBH and IWM-F cases is set be 3.5$D_{p}$ and is similar to the experimental domains of~\citet{voermans2017variation, manes2009turbulence}. 
The length of the streamwise and spanwise domains is based on the DNS of smooth channel turbulent flow~\citep{moser1999direct}. Moreover, since roughness is expected to break the long, elongated streamwise flow structures commonly observed in smooth channel flow, the domain size used is sufficient to impose the periodicity condition, which was also confirmed by evaluating integral length scales (see Appendix A).

\subsection{Numerical method}\label{sec:numerical_method}
The numerical approach is based on a fictitious  domain  method  to  handle  arbitrary  shaped  immersed  objects without  requiring  the  need  for  body-fitted  grids~\citep{apte2009frs}. Cartesian grids are used in the entire simulation domain, including both fluid and solid phases. An additional body force is imposed on the solid part to enforce the rigidity constraint and satisfy the no-slip boundary condition. The absence of highly skewed unstructured mesh at the bead surface has been shown to accelerate the convergence and lower the uncertainty~\citep{finn2013relative}. 
The following governing equations are solved over the entire domain, including the region within the solid bed, and a rigidity constraint force, $\bf f$, that is non-zero only in the solid region is applied to enforce the no-slip condition on the immersed object boundaries. The governing equations are
	\begin{align}
		\nabla\cdot{\bf u} &= 0, \label{eq:NSa} \\
		\rho \bigg[\frac{\partial {\bf u}}{\partial t} + \left({\bf u}\cdot \nabla\right) {\bf u}	\bigg] &= 
		-\nabla p + \mu \nabla^2{\bf u} + {\mathbf f} \:, 
  		\label{eq:NSb}
	\end{align}
\noindent{}where $\bf u$ is the velocity vector (with components given by ${\bf u}=(u,v,w)$, $\rho$ the fluid density, $\mu$ the fluid dynamic viscosity, and $p$ the pressure. A fully parallel, structured, collocated grid, finite volume solver has been developed and thoroughly verified and validated for a range of test cases including flow over a cylinder and sphere for different Reynolds  numbers, flow over touching spheres at different orientations, flow developed by an oscillating cylinder, among others.
The details of the algorithm as well as very detailed verification and validation studies have been published elsewhere~\citep{apte2009frs,finn2013numerical,finn2013relative}. The solver was used to perform direct one-to-one comparison with a body-fitted solver with known second-order accuracy for steady inertial, unsteady inertial, and turbulent flow through porous media~\citep{finn2013relative} to show very good predictive capability. It has also been recently used for direct numerical simulations of oscillatory, turbulent flow over a sediment layer~\citep{ghodke2016dns,ghodke2018roughness}, and pore-resolved simulations of turbulent flow within a porous unit cell with face-centered cubic packing~\citep{he2018angular,he2019characteristics}.

\subsection{Averaging}\label{sec:DA_proc}
Since the flow properties are highly spatially heterogeneous near rough sediment bed boundary, time averaging followed by spatial averaging is applied. Flow statistics such as Reynolds stresses, form-induced disturbances, shear stress and pressure fluctuations, among others are computed using the time-space averaging.
This consecutive time-space averaging involves Reynolds decomposition ($\psi = \overline{\psi} + \psi^{\prime}$) accompanied by spatial decomposition of the time averaged variable $\overline{\psi} = \langle{\overline{\psi}}\rangle + \widetilde{\overline \psi}$~\citep{nikora2007double,nikora2013spatially}.
Here $\psi$ represents an instantaneous flow variable,  $\overline{\psi}$ is its temporal average, $\psi^{\prime} = \psi - \overline{\psi}$  is the instantaneous turbulent fluctuation. The angular brackets, $\langle\rangle$, denote spatial averaging operator. The quantity $\widetilde{\overline \psi}$ known as the form-induced or dispersive disturbance in space is defined as  $\widetilde{\overline \psi} =  \overline{\psi} - \langle\overline{\psi}\rangle $. This represents the deviation of the time-averaged variable, $\overline{\psi}$, from its spatially averaged value, $\langle\overline{\psi}\rangle$. ~\citet{nikora2013spatially} proposed to denote the form-induced disturbance quantity as $\widetilde{\overline \psi}$, whereby a horizontal overbar is added, to emphasize that time-averaging has been done prior to spatial-averaging. This modified notation is adopted in this work. The original notation of this quantity proposed in~\citet{nikora2007double} was without the overbar, $\widetilde{\psi}$, and has been used in the literature published in this field~\citep{voermans2017variation,fang2018influence,shen2020direct}.

Quantities are averaged over the fluid domain, giving the intrinsic spatial average, of the time averaged variable, $\langle{\overline\psi}\rangle = 1/V_f \int_{V_f} \overline \psi dV$, where $V_f$ is the volume occupied by the fluid. In other words, while calculating the volume average of variables in the particle bed regions, only the portion of the volume occupied by the fluid is taken into account. The representative averaging-volume lengths used for spatial averaging in the streamwise and spanwise directions are the same as the grid resolution in those directions. Since the grids are uniform in $x$ and $z$, this implies that the averaging volumes have the same lengths in the streamwise and spanwise direction. However, in the bed-normal direction, a variable volume averaging is used. In the boundary-layer region, especially near the crest of the bed where steep gradients in flow quantities are present, thin-volumes are used for averaging, whereas, deeper inside the bed thicker averaging volumes are used as described in detail in Appendix B.

\subsection{Grid resolution and flow setup}\label{sec:grids}
Table~\ref{tab:grid} gives the details of grid resolution used for all cases.
The grid resolutions required for these configurations are based on two main considerations: (i) minimum bed-normal grid resolution near the bed to capture the bed shear stress, and (ii) minimum resolution required to capture all details of the flow over spherical particles. Since the intensity of turbulence and mean flow penetration into the bed is expected to reduce further deep into the bed, finer grid resolutions are used in the bed-normal region in the top layer compared to other layers.

For DNS of channel flows, the bed-normal grid resolution in wall units should be $\Delta y^{+}<1$, in order to accurately capture the bed shear stress in the turbulent flow. The grid resolutions in the streamwise and spanwise directions are typically 3--4 times coarser, following the smooth channel flow simulations by~\citet{moser1999direct}. Note that, the roughness features and permeability are known to break the elongated flow structures along the streamwise direction in smooth walls, reducing the anisotropy in the near-bed region~\citep{ghodke2016dns}. To capture the inertial flow features within the pore and around spherical particles, grid refinement studies were conducted on flow over a single sphere at different Reynolds numbers representative of the cases studied here (details in Appendix A). Accordingly, roughly 90 (PBL), 180 (PBM), and 548 (PBH) grid points are used in the bed-normal direction in a region covering the {\it top layer} and extending slightly into the free-stream. In the $x$ and $z$ directions, a uniform grid with a minimum of  26 (PBL), 38 (PBM), and 40 (PBH) grid points per-sediment grain are used to resolve the bed geometry. Effect of uniform, but non-cubic grids within the sediment bed was thoroughly evaluated by comparing drag coefficients to those obtained from cubic grids over a single and a layer of particles to show no discernible differences over the range of Reynolds numbers studied here (see appendix A).

Below the top layer, the bed-normal resolution is slowly reduced and nearly uniform grid in all directions is used deep inside the bed, as the frictional Reynolds number in the bottom layers of sediment decreases significantly. From the crest of the top sediment layer, the grid is stretched, coarsening it gradually towards the top of the channel using a standard hyperbolic tangent function~\citep{moser1999direct}. 
Based on these grid resolutions, the total grid count for the PBL case is $\sim$ 232 million cells, for the PBM case is $\sim$ 200 million cells, and for the PBH case is $\sim$ 428 million cells as given in Table~\ref{tab:grid}. 
\begin{table}
\begin{center}
\def~{\hphantom{0}}
\begin{tabular}{@{}lc c c c c c c}
Case  &$N_x \times N_y \times N_z$ & \multicolumn {3}{c}{Bed-Normal Grid Distribution} &($\Delta x^{+}, \Delta y^{+}, \Delta z^{+}) $\\ 
&& Channel region &  Top layer & Bottom layers & \\
VV &$768 \times 288 \times 384$ &96 &86 &106  &(2.94, 0.95, 2.94)   \\
PBL &$1152 \times 350\times 576$  &150 &90 &110  &(2.94, 0.95, 2.94)   \\
PBM &$846 \times 530 \times 448$   &184 &180 &166   &(4.01, 0.95, 3.8)  \\
PBH &$882 \times 1082 \times 448$  &342 &548 &192   &(6.74, 0.55, 6.63)  \\
IWM-F &$846 \times 364 \times 448$   &184 &180 &-   &(4.01, 0.95, 3.8)  \\
\end{tabular}
\caption{Grid parameters used in the present DNS. $(~)^{+}$ denotes wall units. Details on the grid transition regions are given in appendix B.}
\label{tab:grid}
\end{center}
\end{table}

The flow in the simulations is driven by constant mass flow rate. A target mass flow rate is adjusted until the friction velocity, $u_{\tau}$, which results in the required $Re_K$ is obtained. $Re_{\tau}$ is then calculated based on the free-stream height $\delta$. 
\citet{pokrajac2006definition} noted that there is lack of a general definition of $u_{\tau}$ applicable to the boundary layers with variable shear stress where the roughness height is comparable with the boundary layer thickness. Accordingly, $u_{\tau}$ can be specified based on (i) bed shear stress, (ii) total fluid shear stress based on the roughness crest, (iii) fluid shear stress extrapolated to the zero-displacement plane, or (iv) shear stress obtained by fitting data to log-law.~\citet{pokrajac2006definition} proposed to use $u_{\tau}$ based on the the fluid shear stress at the roughness crest, to obtain least ambiguous definition. Since the present work builds upon the experimental data of~\citet{voermans2017variation}, the friction velocity, $u_{\tau}$, is calculated from the maximum value of the time-space averaged total fluid stress which happens to be very close to the sediment crest. The friction velocity is then based on the sum of the viscous, turbulent, and the form-induced shear stresses~\citep{nikora2004velocity,voermans2017variation}, 
\begin{align}
     \tau(y) &= \rho\nu\del(\phi\langle\overline{u}\rangle)/\del{y} - \rho\phi\langle\overline{u^{\prime}v^{\prime}}\rangle - \rho\phi\langle\widetilde{\closure{u}} \widetilde{\closure{v}}\rangle.
    \label{eq:utau}
\end{align}

Following smooth wall DNS studies by~\citet{moser1999direct}, between 20-25 flow-through times (computed as the length over average bulk velocity $L_x/U_{b}$) are needed for the turbulent flow to reach stationary state. Once a stationary flow field is obtained, computations were performed for an additional time period of $T = 25 \delta/u_{\tau}$ to collect single-point and two-points statistics. For the PBL case, where the domain in the streamwise and spanwise directions is twice as that used in the PBM and PBH cases, the stats were collected for over a time period $T = 13 \delta/u_{\tau}$. The flow statistics were monitored to obtain statistically stationary values over the above averaging periods. 

\section{Results}\label{sec:res}
The main results for the different cases studied in this work are discussed. The Reynolds and form-induced stresses are first compared for different Reynolds numbers (section~\ref{sec:stresses}). The structure of turbulence is visualized using vorticity contours followed by a detailed quadrant analysis describing the sweep and ejection events (section~\ref{sec:quadrant}). Variation of the turbulence penetration depth, interfacial mixing length, and similarity relations are investigated as a function of $Re_K$ (section~\ref{sec:pen_comp_rek}). Next, the role of the top layer of the sediment is quantified by comparing the permeable bed statistics with an impermeable rough wall with roughness equivalent to the permeable bed (section~\ref{sec:role_top_layer}). Finally, probability distribution function (PDF) of the viscous and pressure components of the normalized bed shear stress are presented in section~\ref{sec:pdf_bedshear_rekcomp} followed by the pressure fluctuations at the sediment-water interface in section~\ref{sec:presfluc_rekcomp}.

\subsection{Reynolds and form-induced stresses} 
\label{sec:stresses}

\begin{figure}
   \centering
   \subfigure[]{
   \includegraphics[width=3.6cm,height=7.3cm,keepaspectratio]{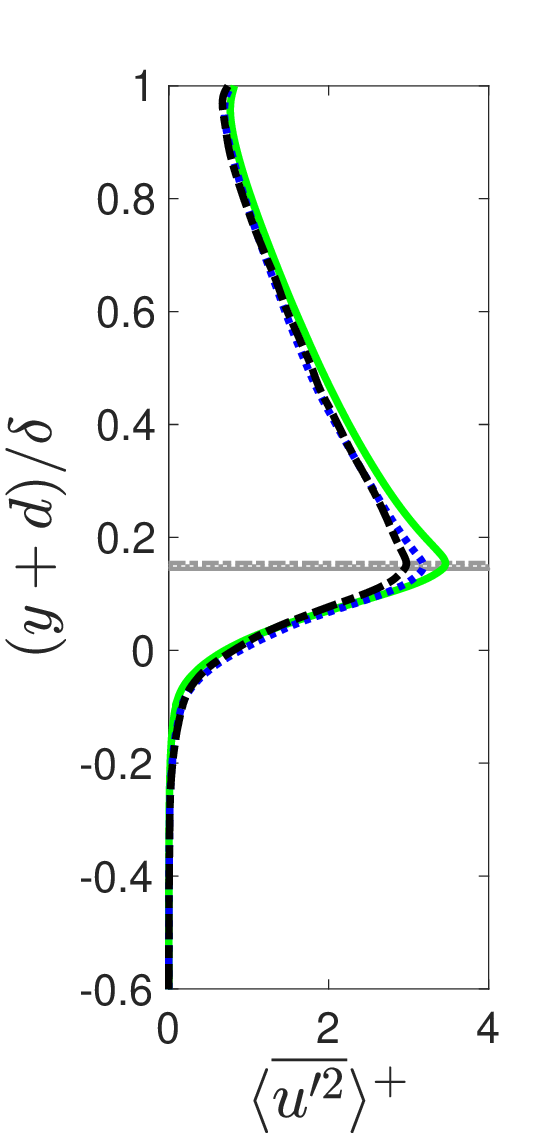}}
   \subfigure[]{
   \includegraphics[width=3.0cm,height=7.3cm,keepaspectratio]{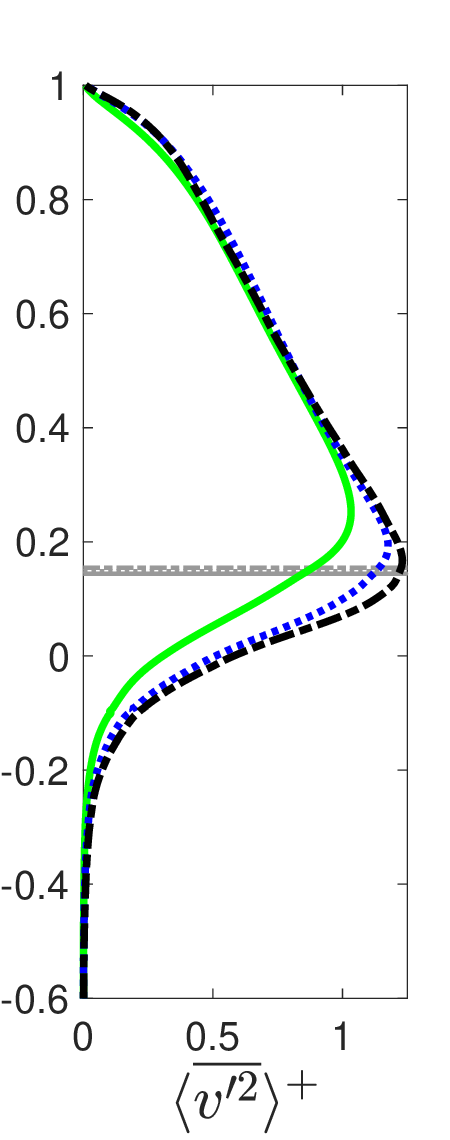}}
   \subfigure[]{
   \includegraphics[width=3.0cm,height=7.3cm,keepaspectratio]{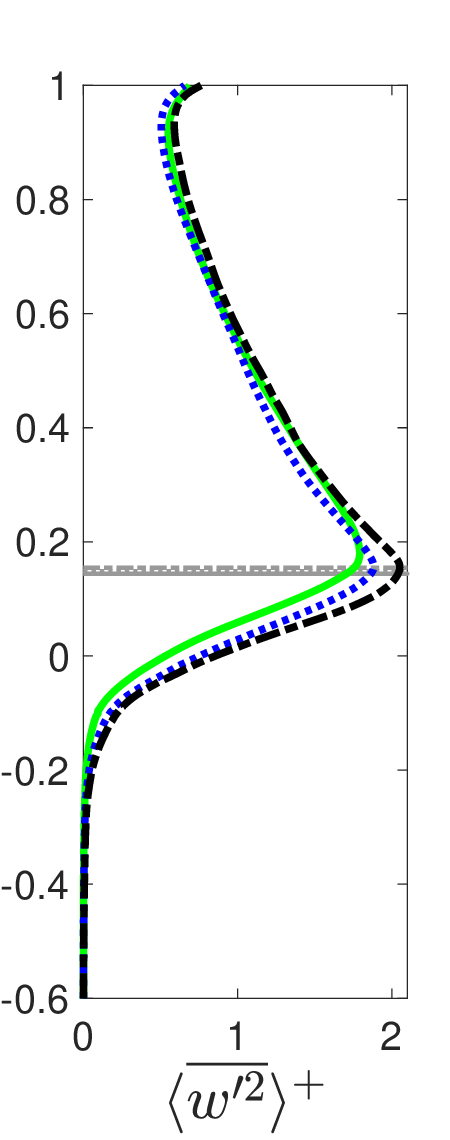}}
   \subfigure[]{
   \includegraphics[width=3.0cm,height=7.3cm,keepaspectratio]{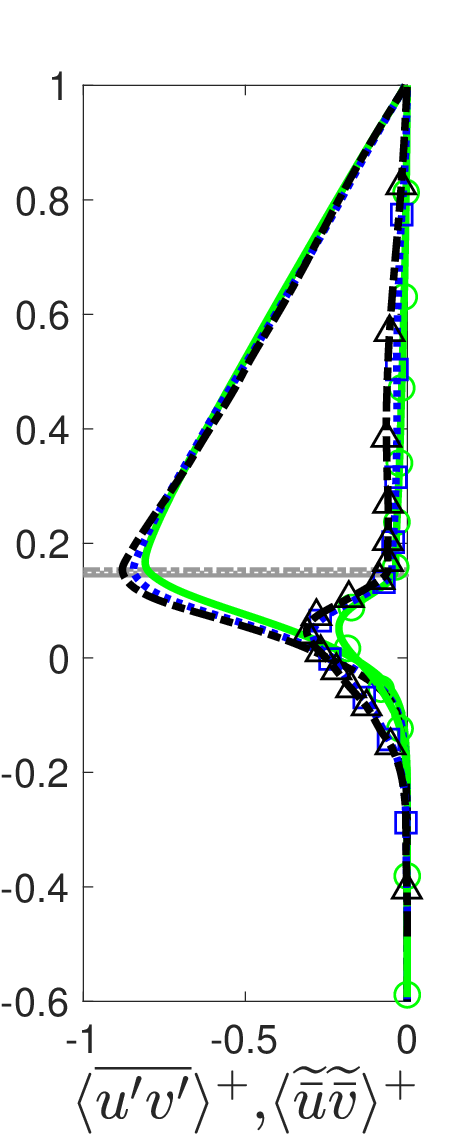}}
    \subfigure[]{
   \includegraphics[width=3.6cm,height=7.3cm,keepaspectratio]{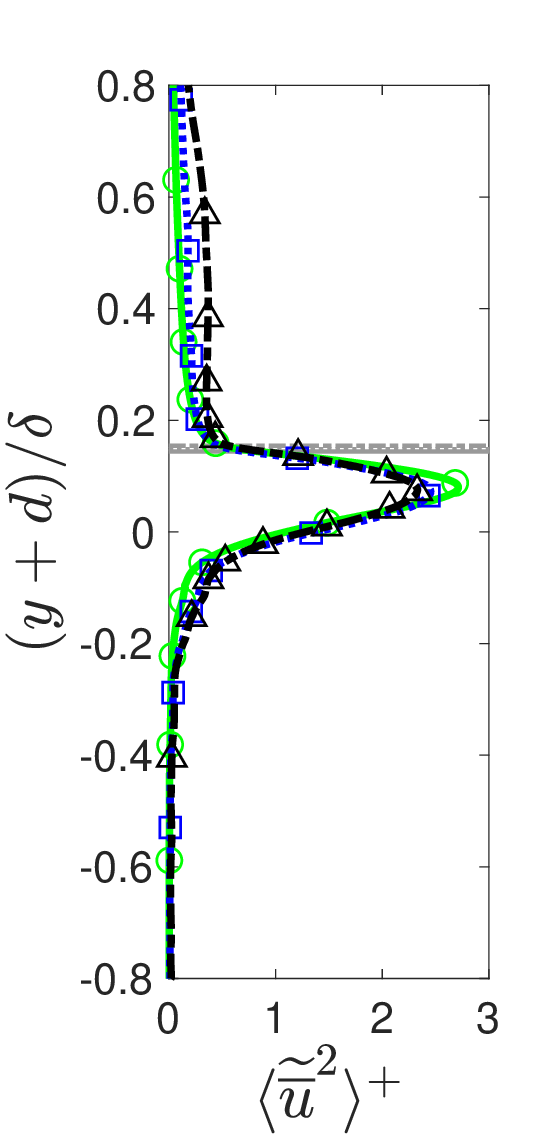}}
    \subfigure[]{
   \includegraphics[width=3.0cm,height=7.3cm,keepaspectratio]{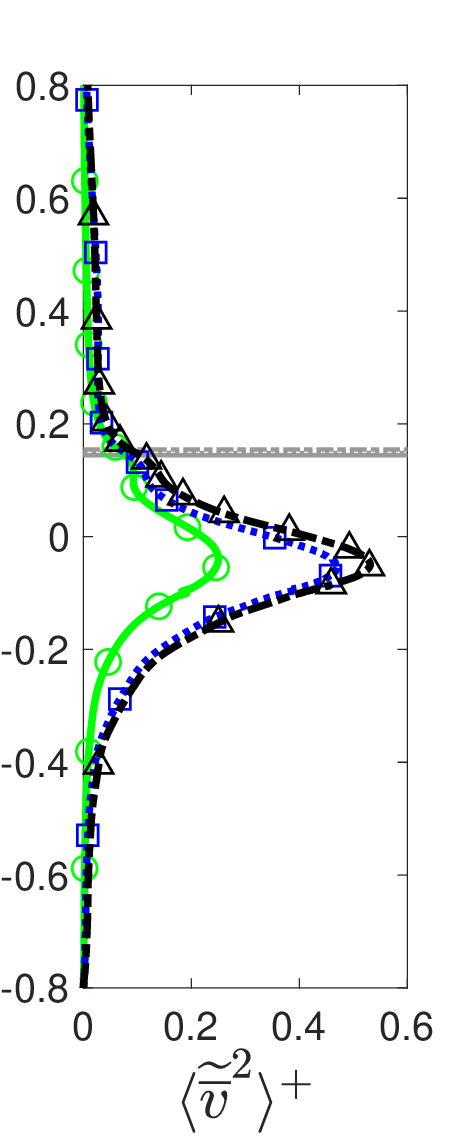}}
    \subfigure[]{
   \includegraphics[width=3.0cm,height=7.3cm,keepaspectratio]{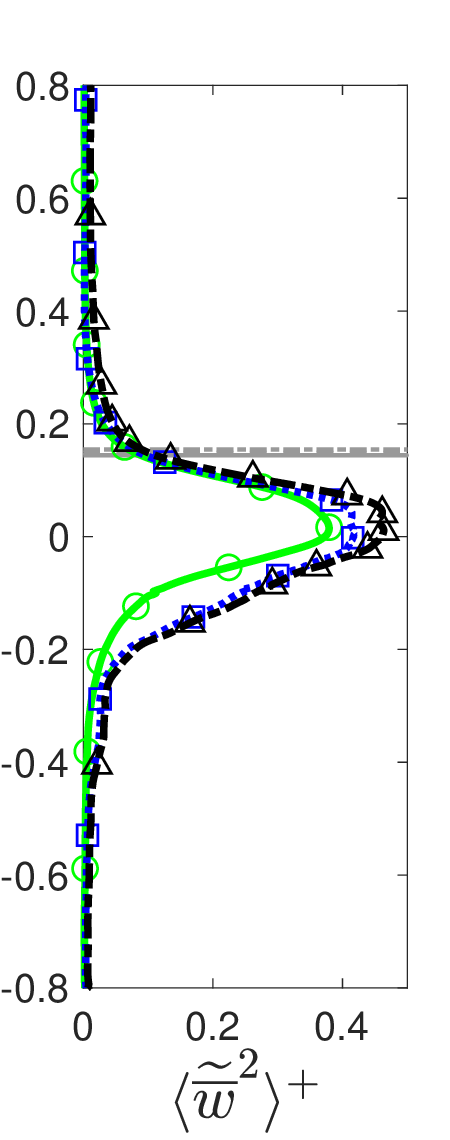}}
    \subfigure[]{
   \includegraphics[width=3.0cm,height=7.3cm,keepaspectratio]{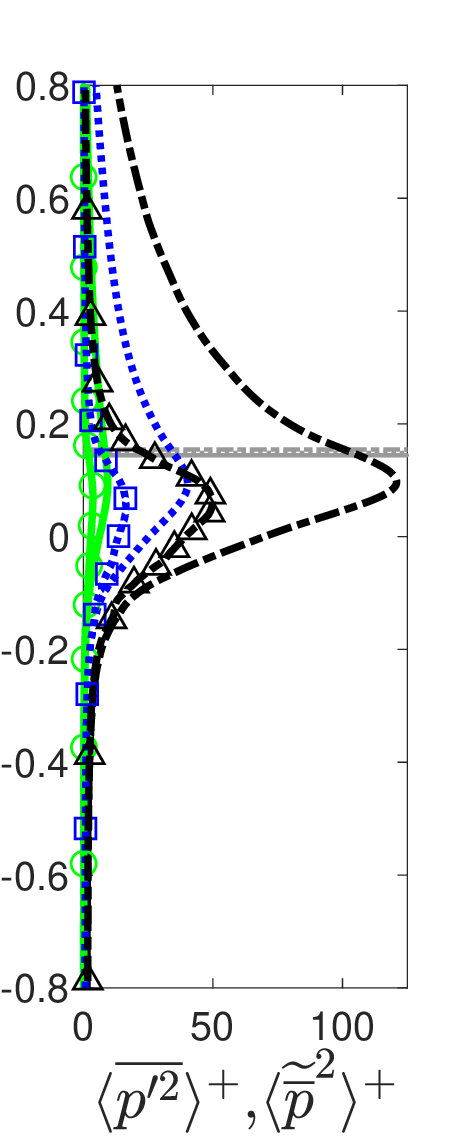}}
\caption{\small Comparison of Reynolds (lines) and form-induced (lines and symbols) stress profiles for PBL (\greenline, \greencirclesolidline), PBM (\bluedottedline, \bluerectangledottedline), and PBH (\blkdashdotline, \blacktriangledashdotline) cases: (a-c) streamwise, bed-normal, and spanwise components of spatially averaged Reynolds stress tensor, (d) spatially averaged shear Reynolds stress and shear form-induced stress, (e-g) streamwise, bed-normal, and spanwise components of form-induced stress, and (h) mean-square pressure fluctuations and form-induced pressure disturbances. Horizontal lines show the crest of sediment bed for PBL (\grayline), PBM (\graydottedline), and PBH (\graydashdotline) cases. Pressure is normalized by $\rho u_{\tau}^2$.  }
\label{fig:reys_fis_rekcomp}
\end{figure}

Figure~\ref{fig:reys_fis_rekcomp} shows the bed-normal variation of all components of Reynolds and form-induced stresses for the PBL, PBM, and PBH cases. The zero-displacement plane (see Appendix D for details) defines the sediment-water interface (SWI) in this study and is used as a virtual origin while comparing and contrasting the primary and secondary statistics. The variables are normalized by $u_{\tau}^2$ (pressure by $\rho u_{\tau}^2$), and $y$ is shifted by $d$, the zero-displacement thickness, and then normalized by $\delta$, effectively making virtual origin the same for all the cases. This implies that SWI planes for all cases align. It is observed that the magnitude of the form-induced stresses is generally smaller than the Reynolds stresses. The location of the peaks are also different, with the form induced stresses peaking below the sediment-water interface, whereas the Reynolds stresses peak close to the bed crest.

The profiles of streamwise ($x$--direction), $\langle\overline{u^{\prime2}}\rangle^{+}$, and bed-normal ($y$--direction) Reynolds stress, $\langle\overline{v^{\prime2}}\rangle^{+}$, (figure~\ref{fig:reys_fis_rekcomp}a,b) exhibit similarity for $(y+d)/\delta >= 0.4$, substantiating the wall similarity hypothesis reported by~\citet{raupach1991rough}, \citet{breugem2006influence}, and \citet{fang2018influence}. Near the bed, the streamwise stress decreases and bed-normal and spanwise ($z$--direction, figure~\ref{fig:reys_fis_rekcomp}c) stresses increase with increasing Reynolds number, suggesting weakening of the wall-blocking  effect with increase in non-dimensional bed permeability. Moreover, not only does the flow penetrate deeper inside the bed (quantified in section~\ref{sec:pen_comp_rek}) but the intensity also increases with increasing $Re_K$ as seen from the bed-normal component of the stress. However, this comes at the expense of loss in intensity in the streamwise direction. 
The peak values and their locations for the Reynolds stresses are shown in table~\ref{tab:reys_fis_rekcomp}. The peak values of $\langle\overline{u^{\prime2}}\rangle^{+}$ are larger than $\langle\overline{v^{\prime2}}\rangle^{+}$ for all Reynolds numbers, similar to turbulent flows over smooth or rough impermeable wall cases. The location of peak in $\langle\overline{u^{\prime2}}\rangle^{+}$ and $\langle\overline{w^{\prime2}}\rangle^{+}$ is close to the crest for all $Re_K$, whereas that in $\langle\overline{v^{\prime2}}\rangle^{+}$ shifts downward starting from above the crest level for the PBL case and moving close to the crest for the PBH case. The Reynolds shear stress shown in figure~\ref{fig:reys_fis_rekcomp}d also peaks near the crest, increases with $Re_K$, and shows similarity in the outer region.

\begin{table}
  \begin{center}
\def~{\hphantom{0}}
\begin{tabular}{@{}lc c c c c }
Case & $\langle\overline{u^{\prime2}}\rangle^{+}$ & $\langle\overline{v^{\prime2}}\rangle^{+}$ &
$\langle\overline{w^{\prime2}}\rangle^{+}$ & $\langle\overline{u^{\prime}v^{\prime}}\rangle^{+}$ & $\langle\overline{p^{\prime2}}\rangle^{+}$ \\ 
PBL &3.46 (0.15) &1.03 (0.25) &1.79 (0.18)  &0.80 (0.16) &9.41 (0.094) \\ 
PBM &3.19 (0.15) &1.18 (0.19) &1.88 (0.16) &0.84 (0.15) &40.25  (0.095)\\ 
PBH &2.97 (0.15) &1.23 (0.16) &2.05 (0.16) &0.88 (0.15) &120.88 (0.097)\\ 
\hline 
Case & $\langle\widetilde{\overline u}^2\rangle^{+}$ & $\langle\widetilde{\overline v}^2\rangle^{+}$ & $\langle\widetilde{\overline w}^2\rangle^{+}$ & $\langle\widetilde{\closure u}\widetilde{\closure v}\rangle^{+}$ & $\langle\widetilde{\overline p}^2\rangle^{+}$  \\ 
PBL &2.71 (0.08) &0.25 (-0.04) &0.38 (0.014) &0.21 (0.05) &3.7 (0.07)  \\ 
PBM &2.45 (0.07) &0.47 (-0.05) &0.42 (0.01) &0.29 (0.04) &16.5 (0.063)  \\ 
PBH &2.33 (0.07) &0.53 (-0.05) &0.46 (0.009) &0.31 (0.04) &49.9 (0.06)  \\ 
\end{tabular}
\caption{The peak value and location [$(y+d)/\delta$] given in brackets, of Reynolds stresses and form-induced stresses for the PBL, PBM, and PBH cases. The peak values of Reynolds and form-induced stresses are normalized by $u_{\tau}^2$ (pressure by $\rho u_{\tau}^2$).}
\label{tab:reys_fis_rekcomp}
 \end{center}
\end{table}

The bed-normal variation of the form-induced stresses for the PBL, PBM and PBH cases are shown in figure~\ref{fig:reys_fis_rekcomp}d--g. The influence of $Re_K$ is much more pronounced on the form-induced stresses. While the magnitudes for the streamwise stresses are comparable, the bed-normal and spanwise stresses are much smaller than the corresponding Reynolds stresses. More importantly, the peak values occur significantly below sediment crest. Inside the bed, $\langle\widetilde{\overline u}^2\rangle^{+}$ decreases with $Re_K$, while both $\langle\widetilde{\overline v}^2\rangle^{+}$ and $\langle\widetilde{\overline w}^2\rangle^{+}$ values increase with  $Re_K$ mimicking the trend with Reynolds stresses. However, in contrast to $\langle\overline{v^{\prime2}}\rangle^{+}$, the peaks in $\langle\widetilde{\overline v}^2\rangle^{+}$ occur at a similar location (table~\ref{tab:reys_fis_rekcomp}) for all Reynolds numbers. This suggests that the penetration depth of $\langle\widetilde{\overline v}^2\rangle^{+}$ is independent of $Re_K$ and is dependent on the local porosity distribution in the top layer of the bed, which is similar for the three $Re_K$ cases studied in this work. 
It is interesting to note that, the values of $\langle\widetilde{\overline w}^2\rangle^{+}$ are larger than the $\langle\widetilde{\overline v}^2\rangle^{+}$ at lower $Re_K$, but are comparable and slightly smaller than $\langle\widetilde{\overline v}^2\rangle^{+}$ at higher $Re_K$. 
This may be attributed to the increased flow penetration, causing the bed-normal form-induced stresses peak further below the bed crest. 

The turbulent fluctuations and form-induced disturbances in pressure, shown in figure~\ref{fig:reys_fis_rekcomp}h, exhibit a very strong correlation with $Re_K$. Again, the form-induced pressure disturbances are generally much smaller than the turbulent fluctuations. There is more than a ten-fold increase in the peak value between the PBL and PBH cases for both the turbulent fluctuations and formed induced disturbances in pressure.
The location of the peak in form-induced disturbances as well as turbulent fluctuations occur just below the crest and is almost the same for all $Re_K$ studied. The fluctuations quickly decay for $(y+d)/\delta < -0.3$, indicating that a significant magnitude contribution comes from the top layer of the sediment bed.
This suggests that the local protrusions of partially exposed sediment particles in the top layer and resultant stagnation flow are responsible in altering the flow structures that in turn produce larger form-induced disturbances and pressure fluctuations. The increased pressure disturbances due to the bed roughness elements can lead to enhanced mass transport rates at the sediment-water interfaces with potentially reduced residence times for pollutants and contaminants.

\subsection{Turbulence structure and quadrant analysis} 
\label{sec:quadrant}
\begin{figure}
   \centering
   \subfigure[]{
   \includegraphics[width=6cm,height=6cm,keepaspectratio]{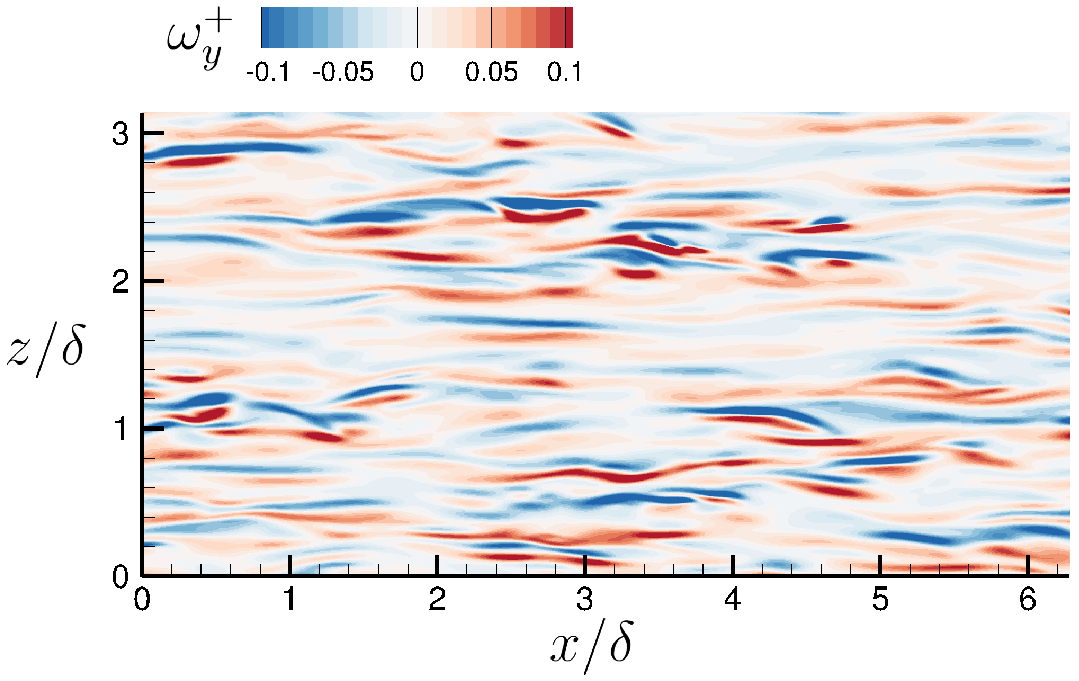}}
   \subfigure[]{
   \includegraphics[width=6cm,height=6cm,keepaspectratio]{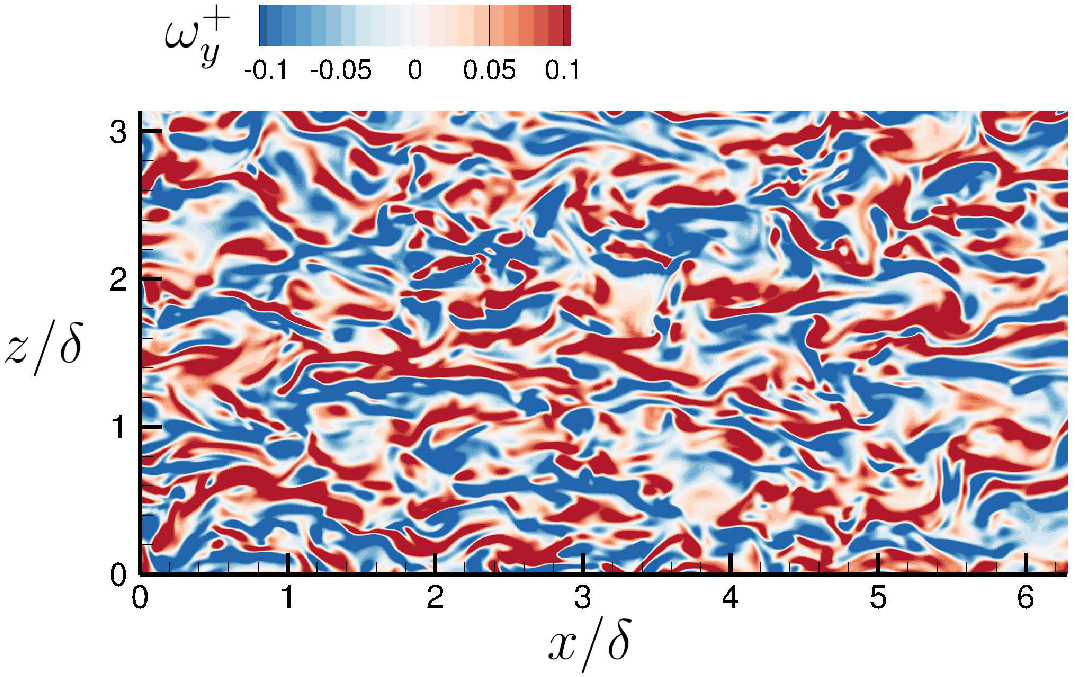}}
   \subfigure[]{
   \includegraphics[width=6cm,height=6cm,keepaspectratio]{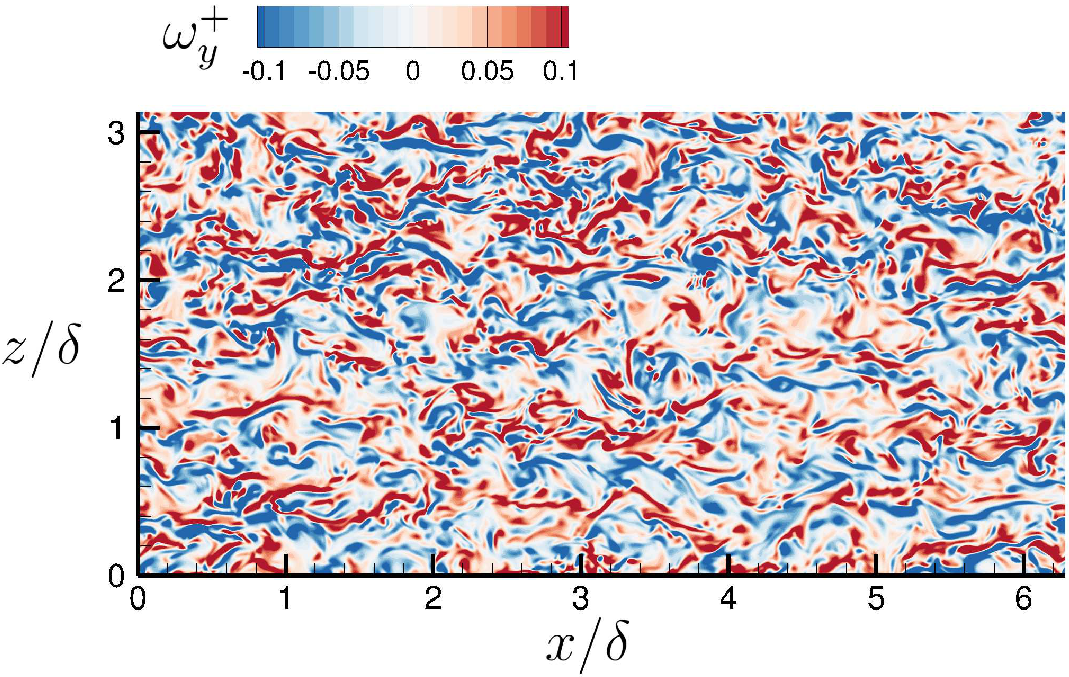}}
    \subfigure[]{
   \includegraphics[width=6cm,height=6cm,keepaspectratio]{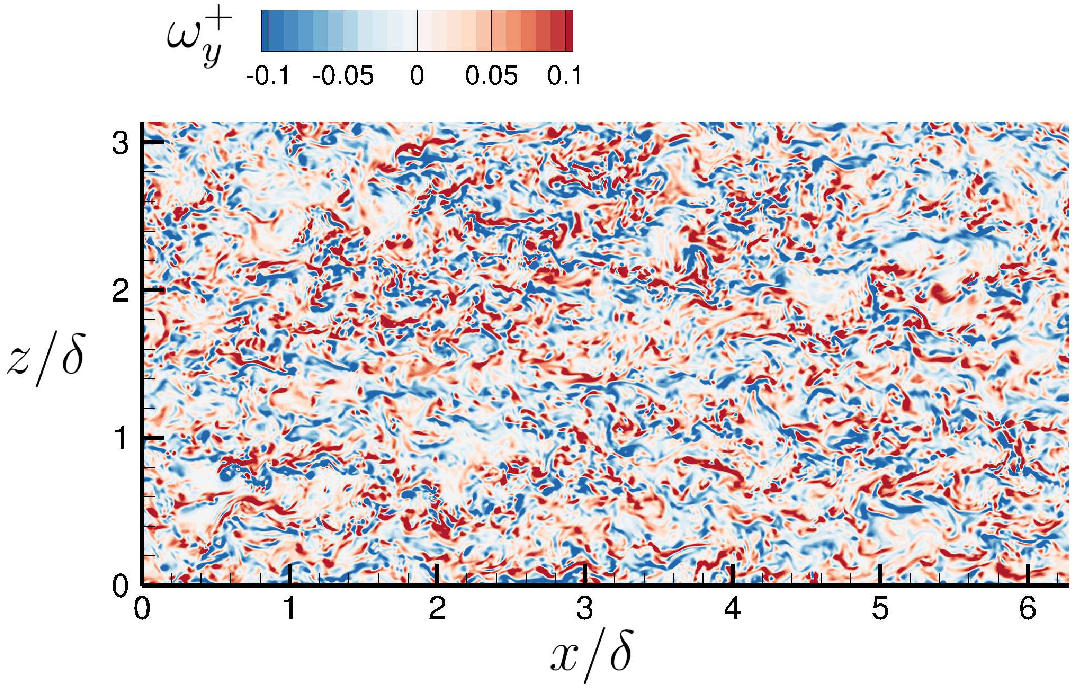}}
   \vspace{-3mm}
\caption{\small Contours of bed-normal vorticity, $\omega_{y}^{+}$, normalized by $u_{\tau}^2/\nu$ at $y/\delta = 0.005$, between  $0 < x/\delta <6.28$ and $0 < z/\delta <3.14$ for  (a) smooth-wall case simulated at $Re_{\tau} = 270$, (b) PBL case, (c) PBM  case and  (d) PBH case.} 
\label{fig:vor_rekcomp}
\end{figure}

Distinct variations in the characteristics of primary turbulence structure are first shown in this section followed by the quadrant analysis. Contours of instantaneous bed-normal vorticity, $\omega_{y}^{+}$ =  $\omega_{y} \nu / u_{\tau}^2 $, just above the crest ($y/\delta = 0.005$) are shown in figure~\ref{fig:vor_rekcomp}. Results from a simulated smooth wall case at $Re_{\tau} = 270$ (same $Re_{\tau}$ as the PBL case) are also shown in figure~\ref{fig:vor_rekcomp}a for comparison. In the smooth wall case, distinct long elongated streaky structures, which are a result of low and high speed streaks generated quasi-streamwise vortices, are visible. The influence of strong mean gradient and an impenetrable smooth wall results in these long streaky structures~\citep{lee1990structure}. In the low Reynolds number permeable bed case (PBL), shown in figure~\ref{fig:vor_rekcomp}b, the starting of the breakdown of these structures can be seen. The roughness and permeability of the bed help in the breakdown. Although the long elongated streaks in the PBL case are shortened due to roughness, at this $Re_K$ the flow anisotropy is somewhat retained. With further increase in Reynolds number, the streaks are broken down even more and in the PBH case (figure~\ref{fig:vor_rekcomp}d), the streaky structures are significantly less pronounced with the flow becoming more intermittent. As $Re_K$ increases, weakening of the wall-blocking effect due strong bed-normal velocities prevents the formation of these long streaky structures and lead to reduction in flow anisotropy.

\begin{figure}
   \centering
    \subfigure[]{
   \includegraphics[width=3.7cm,height=3.7cm,keepaspectratio]{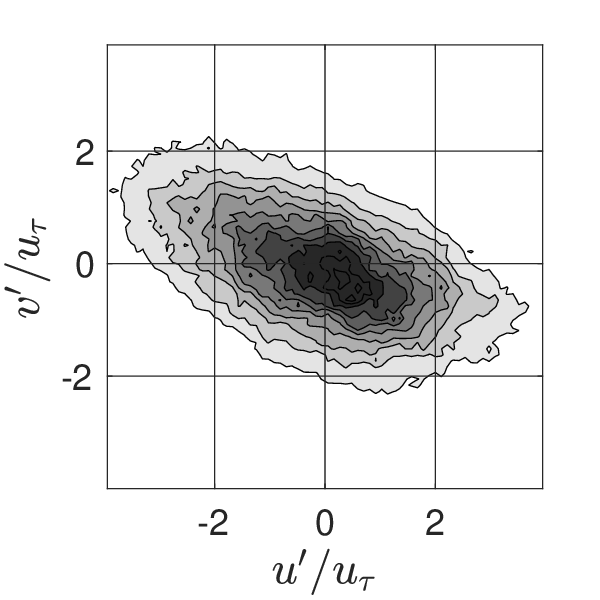}}
   \subfigure[]{
   \includegraphics[width=3.7cm,height=3.7cm,keepaspectratio]{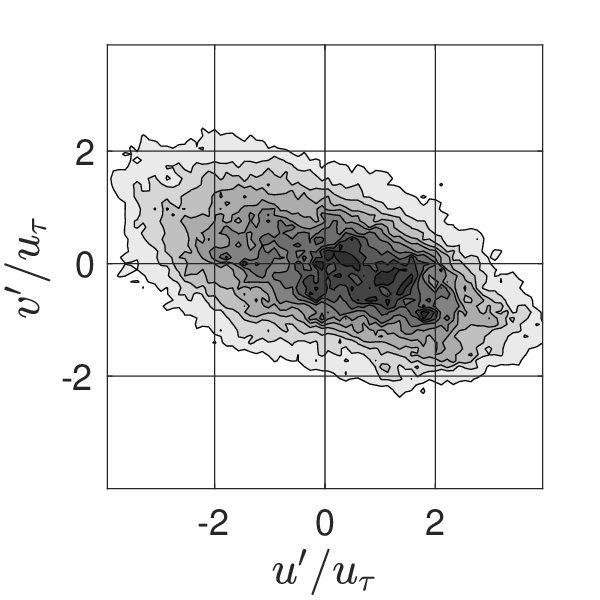}}
   \subfigure[]{
   \includegraphics[width=4.3cm,height=3.7cm,keepaspectratio]{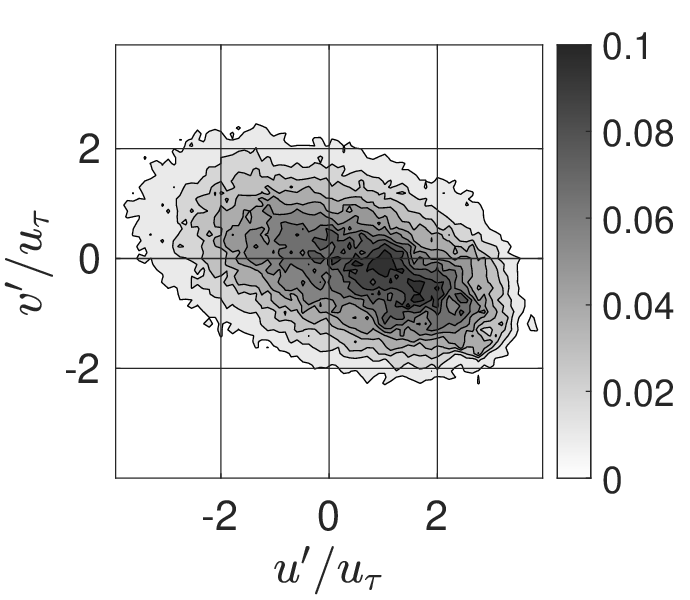}}
    \subfigure[]{
   \includegraphics[width=3.7cm,height=3.7cm,keepaspectratio]{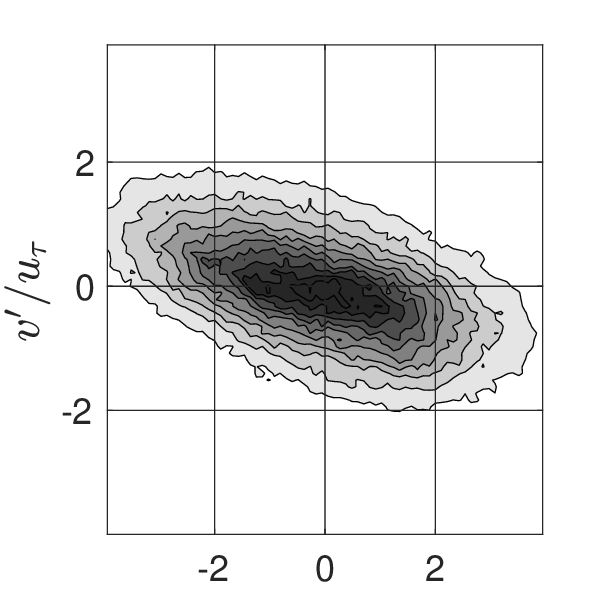}}
   \subfigure[]{
   \includegraphics[width=3.7cm,height=4.0cm,keepaspectratio]{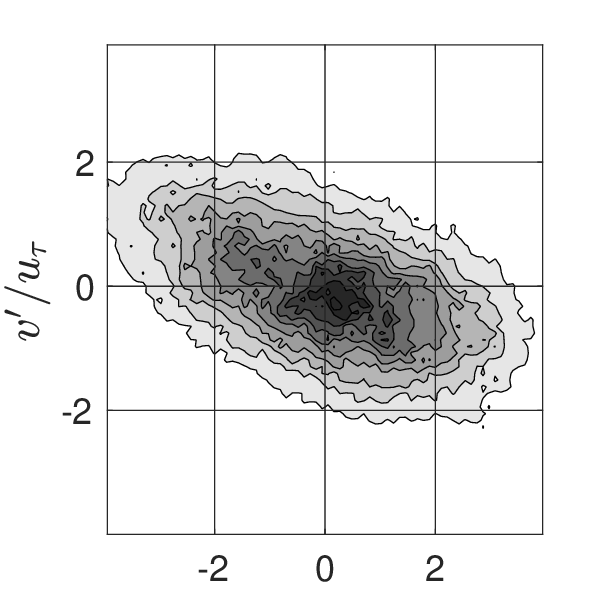}}
   \subfigure[]{
   \includegraphics[width=4.3cm,height=4.0cm,keepaspectratio]{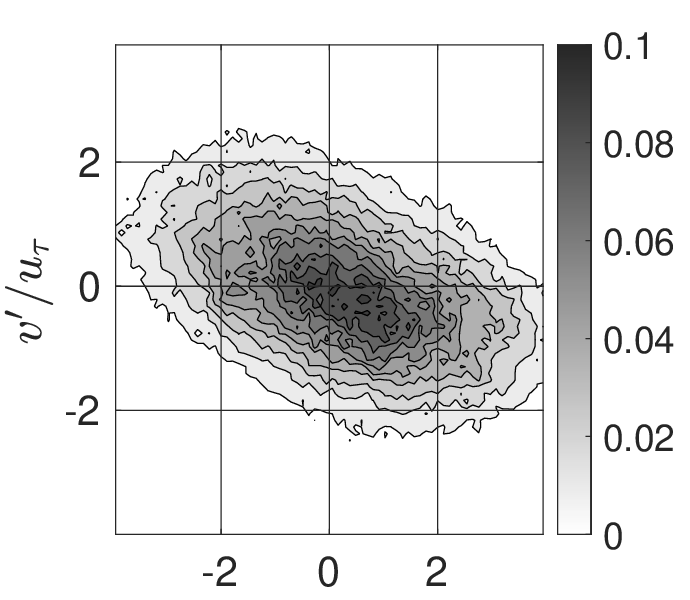}}
    \subfigure[]{
   \includegraphics[width=3.7cm,height=4.0cm,keepaspectratio]{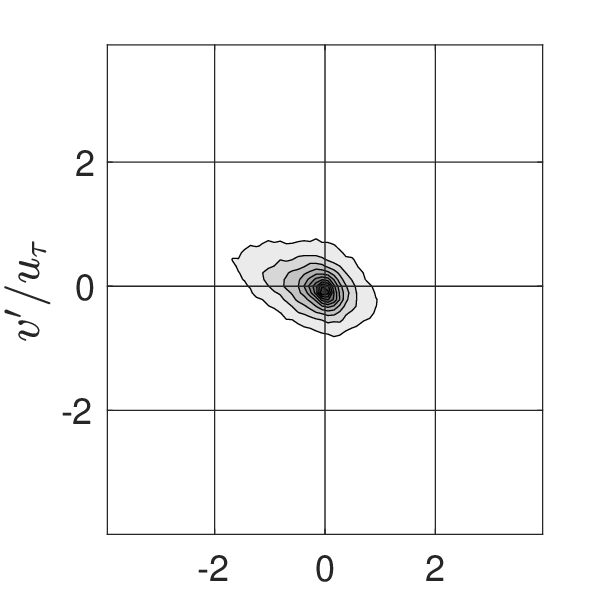}}
   \subfigure[]{
   \includegraphics[width=3.7cm,height=4.0cm,keepaspectratio]{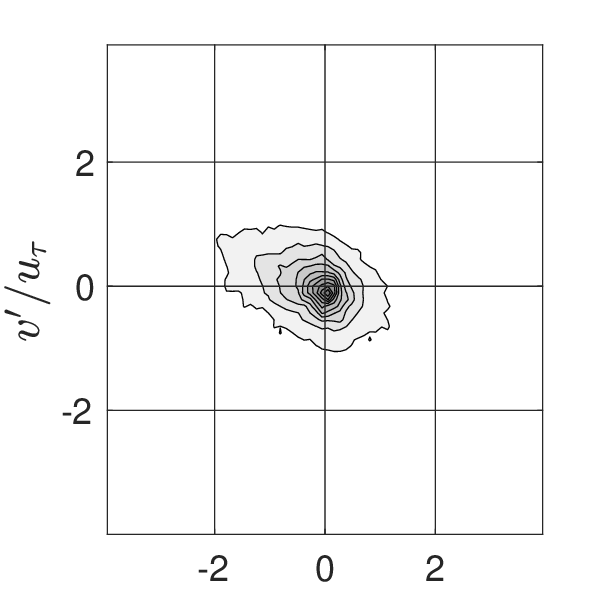}}
   \subfigure[]{
   \includegraphics[width=4.3cm,height=4.0cm,keepaspectratio]{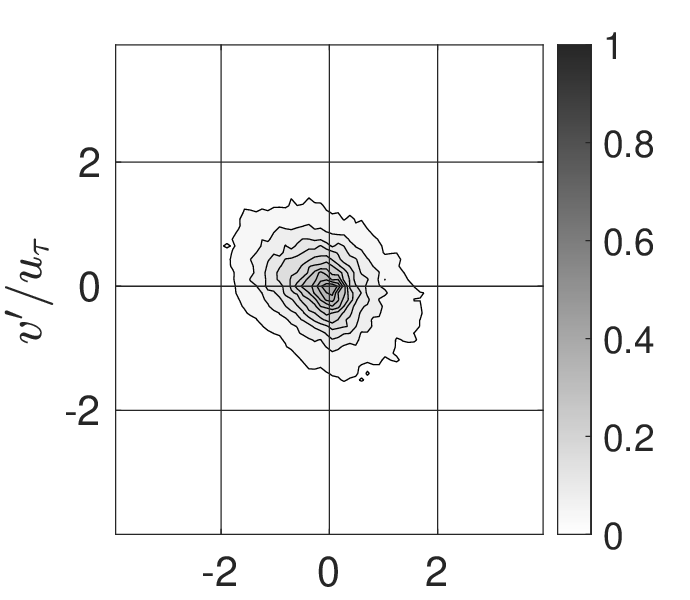}} 
   \subfigure[]{
   \includegraphics[width=3.7cm,height=4.0cm,keepaspectratio]{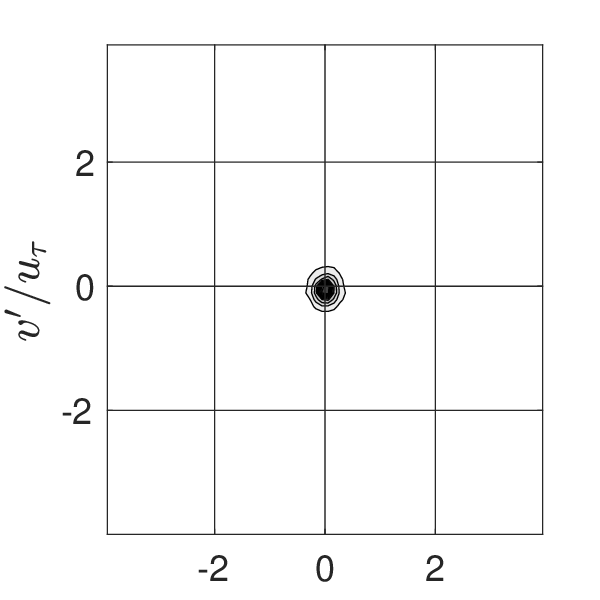}}
   \subfigure[]{
   \includegraphics[width=3.7cm,height=4.0cm,keepaspectratio]{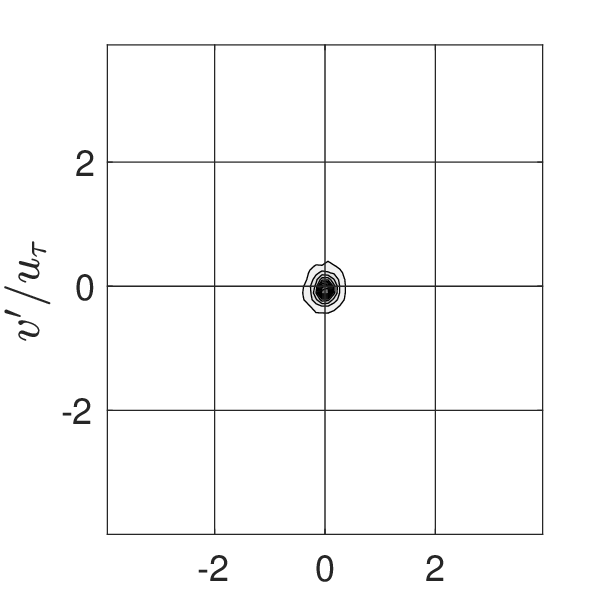}}
   \subfigure[]{
   \includegraphics[width=4.3cm,height=4.0cm,keepaspectratio]{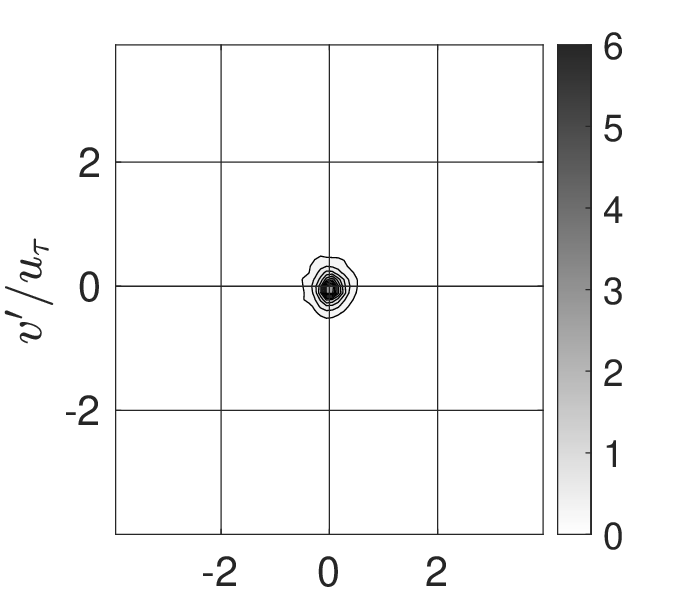}}
\caption{\small Quadrant analysis of probability distribution functions of turbulent velocity fluctuations for the PBL (left panel), PBM (middle panel), and PBH (right panel) cases at different elevations: $y/\delta$ ($y/D_{p}$) of (a--c) 0.143~(0.5), (d--f) 0~(0), (g--i) -0.143~(-0.5), (j--l) -0.286~(-1). The velocities have been normalized by $u_{\tau}$.}
\label{fig:jpdf_rekcomp}
\end{figure}

Quadrant analysis is performed to understand the influence of near bed flow structures on Reynolds stress. Joint probability distribution functions (PDFs) for turbulent velocity fluctuations, $u^{\prime}$ and $v^{\prime}$ calculated at different elevations from the sediment bed are shown in figure~\ref{fig:jpdf_rekcomp}.
The product of $u^{\prime} v^{\prime}$ is negative in the second and fourth quadrants, representing turbulence production. 
The second quadrant, where $u^{\prime} < 0, v^{\prime} > 0$, corresponds to an ejection event whereas, the fourth quadrant, where $u^{\prime} > 0, v^{\prime} < 0$, corresponds to a sweep event. 
Figures~\ref{fig:jpdf_rekcomp}(a-l) show the correlation for different $Re_K$ at four bed-normal elevations within one particle diameter below and above the crest. Figures~\ref{fig:jpdf_rekcomp}a--c are one diameter above the crest in the free-stream, figures~\ref{fig:jpdf_rekcomp}d--f are at the bed crest, figures~\ref{fig:jpdf_rekcomp}g--i are half-way into the top layer of the bed, and figures~\ref{fig:jpdf_rekcomp}j--l are at the bottom of the top layer ($y/D_p = -1$), respectively.

Inside the bed (figures~\ref{fig:jpdf_rekcomp}j--l), at the bottom of the top layer, the probability for ejection and sweep events is small and nearly equal for all three $Re_K$. This suggests that the turbulent structures lose both their directional bias and strength, becoming nearly isotropic in nature below the top layer. At half the distance into the top layer,  figures~\ref{fig:jpdf_rekcomp}g--i, it can be seen that for the PBL and PBM cases, the ejection events are slightly more dominant, transporting fluid away from the bed; however, for the PBH case, both the ejection and sweep events are equally dominant indicating that at higher Reynolds number, sweep events are enhanced, increasing transport of momentum towards the bed. At the bed crest, figures~\ref{fig:jpdf_rekcomp}d--f, both ejection and sweep events become dominant over a greater range of velocity fluctuations showcasing the interaction between fluid flow and permeable sediment bed. For lower $Re_K$, the joint PDFs are more concentrated in a narrow band, indicating large fluctuations in $u^{\prime}$ are associated with smaller excursions in $v^{\prime}$. With increase in Reynolds number, the PDFs appear more diffused highlighting increase in bed-normal velocity fluctuations, and reduction in anisotropy structures at the SWI, which is confirmed from the vorticity contours discussed earlier. Further away from the bed at $y/D_{p}=0.5$ as shown in figures~\ref{fig:jpdf_rekcomp}a--c, the ejection and sweep events again decrease in intensities compared to those at the SWI.

\subsection{Penetration depths, mixing length, and similarity relations}
\label{sec:pen_comp_rek}
Passage of turbulent structures over the sediment bed and resulting sweep events were shown to penetrate into the sediment bed in section~\ref{sec:quadrant}. These sweep events induce momentum fluxes that can be on the order of the mean bed shear stress. The penetration of turbulence increases the flow resistance and effective roughness.  
 The depth of turbulent shear penetration is associated to the characteristic size of the turbulent eddy across the sediment-water interface (SWI). {Knowing how these scales are related to the permeability ($\sqrt{K}$) or the mean particle size ($D_p$) at different $Re_K$ is important for reduced order models of turbulent momentum and mass transport across the interface.}

\begin{figure}
   \centering
    \subfigure[]{
   \includegraphics[width=6cm,height=4cm,keepaspectratio]{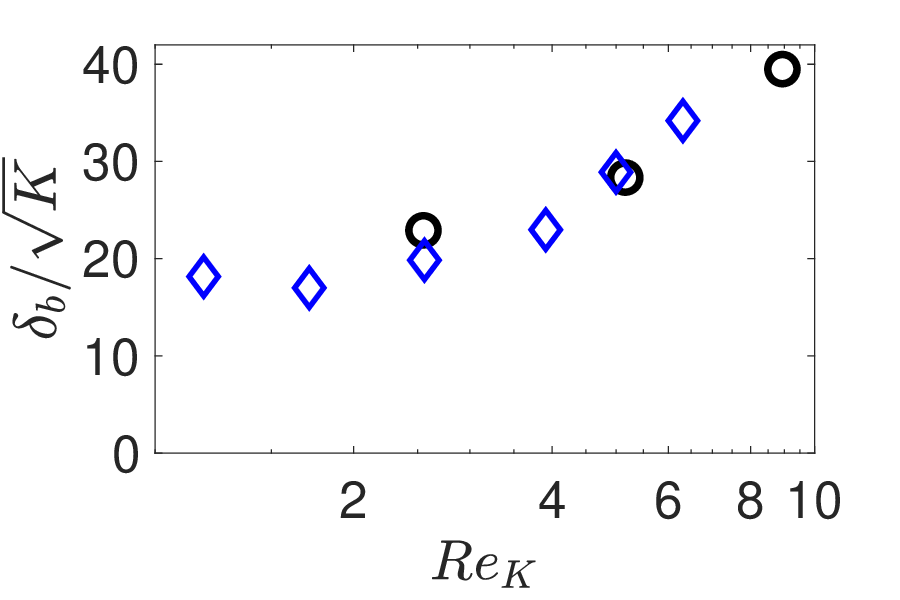}}
   \subfigure[]{
   \includegraphics[width=6cm,height=4cm,keepaspectratio]{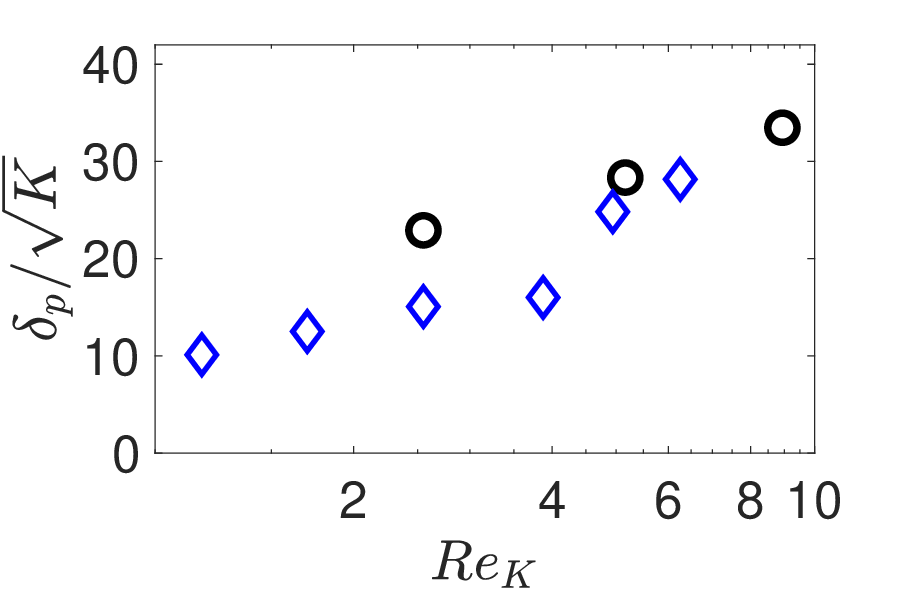}}
   \subfigure[]{
   \includegraphics[width=6cm,height=4cm,keepaspectratio]{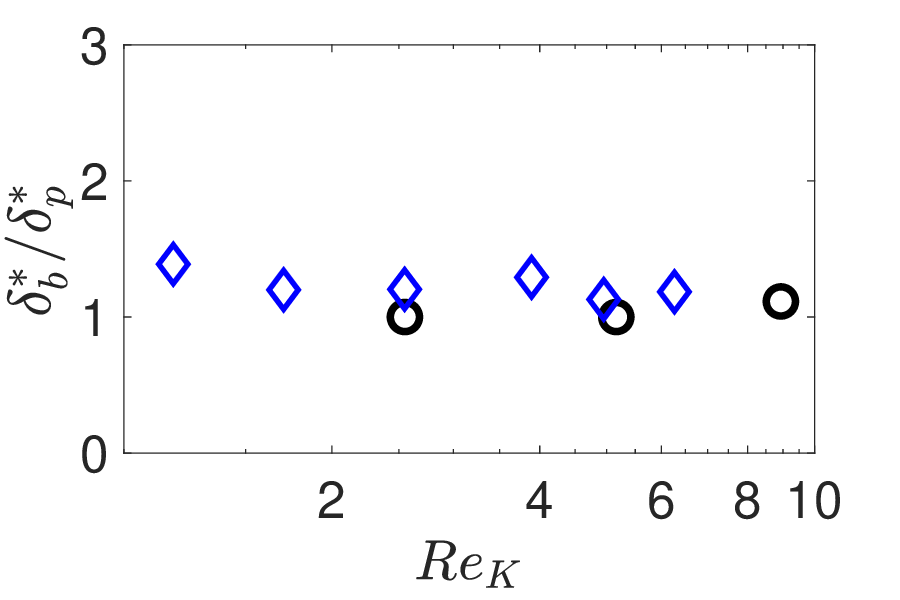}}
    \subfigure[]{
   \includegraphics[width=6cm,height=4cm,keepaspectratio]{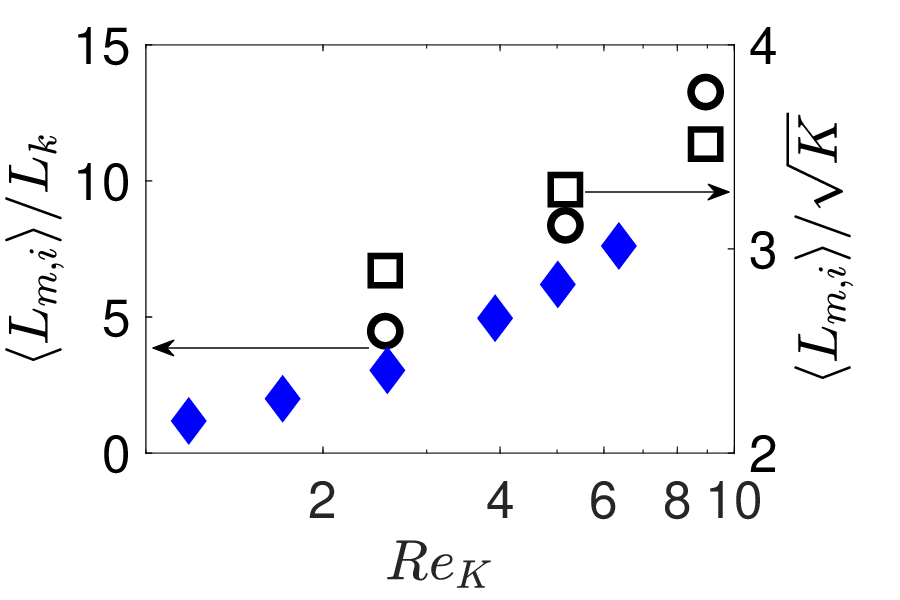}}
\caption{\small  (a) Brinkman layer thickness, $\delta_{b}$, normalised by $\sqrt{K}$; ~\citet{voermans2017variation} (\bluediamond), DNS (\blkcircle), (b) turbulent shear stress penetration, $\delta_{p}$, normalised by $\sqrt{K}$; ,~\citet{voermans2017variation} (\bluediamond), DNS (\blkcircle), (c) ratio of the flow penetration depth, $\delta_{b}^*$, to the shear stress penetration depth, $\delta_{p}^*$, measured relative to the crest; ~\citet{voermans2017variation} (\bluediamond), DNS (\blkcircle), and (d) Left axis: Ratio of interfacial mixing-length to the Kolmogorov length scale; ~\citet{voermans2018model} (\bluefilldiamond), DNS(\blkcircle), and Right axis: interfacial mixing-length normalised by $\sqrt{K}$; DNS (\blksquare ). }  
\label{fig:bk_ts}
\end{figure}

The mean flow penetration depth, $\delta_{b}$, known as the Brinkman layer thickness, is calculated from the mean velocity profiles below the sediment crest. Deep inside the bed, the mean velocity reaches a constant value (Darcy velocity), which is denoted as $U_{p}$. The Brinkman layer thickness is then calculated by measuring the the vertical distance from the SWI ($y = -d$) to a location inside the bed, where the difference between the local mean velocity ($\langle{\overline{u}}\rangle(y)$) and Darcy velocity ($U_{p}$) has decayed to 1\% of the velocity value at the SWI ($U_{i}$), i.e $\langle{\overline{u}}\rangle(y)_{y+d=-\delta_{b}} $= $0.01(U_{i} - U_{p}) + U_{p}$ (where subscript `$i$' indicates the SWI location). The corresponding value of the Brinkman layer thickness measured from the crest of the bed is $\delta_b^{*} = \delta_b+d$. 
Figure~\ref{fig:bk_ts}a shows strong correlation of the normalized Brinkman layer thickness with permeability Reynolds number. Similar trends are observed in experimental data of~\citet{voermans2017variation} despite the fact that the actual position and size of the particles in the random arrangement used in the present study are different than those in the experiments.

The turbulent shear stress penetration is defined as the depth at which the Reynolds stress is 1\% of its value at the SWI, i.e. $\delta_p=\langle\overline{u^{\prime}v^{\prime}}\rangle_{y=-\delta_{p}}$ = $0.01\langle\overline{u^{\prime}v^{\prime}}\rangle_{i}$, and the value measured from the crest of the bed is $\delta_p^{*} = \delta_p+d$. Following the work of~\citet{ghisalberti2009obstructed} on obstructed shear flows,~\citet{manes2012phenomenological} defined the penetration depth from the crest of the sediment ($\delta_p^{*}$), and showed that it is proportional to the drag length scale, i.e., $\delta_p^{*}\sim f(C_d a)^{-1}$, where $C_d$ is the drag coefficient of the medium, and $a$ is a length scale obtained based on the frontal area per unit volume of the solid medium. For monodispersed spheres, $a$ is proportional to the particle size which in turn is related to the permeability. Thus, using a drag-force balance,~\citet{manes2012phenomenological} argued for a linear relation between $\delta_p^{*}$ and $\sqrt{K}$. However, 
figure~\ref{fig:bk_ts}b shows that the normalized $\delta_p$ is a function of $Re_K$. Both the mean flow ($\delta_b$) and turbulent shear stress penetration ($\delta_p$) show non-linear correlation with the permeability, and increase with increasing $Re_K$. 
A deterministic relation is observed for the ratio, $\delta_b^{*}/\delta_p^{*}$, with the permeability Reynolds number as the ratio approaches a constant value of $1.1$ as shown in figure~\ref{fig:bk_ts}c.

The dominant scale of the turbulent structures at the interface is affected by the interstitial spacing within the pore of which the permeability is a geometric measure, but it does not introduce any physical flow dependent measure. Hence, quantifying the dependence of interfacial mixing length, $\langle L_{m,i}\rangle = (\langle\overline{u^{\prime}v^{\prime}}\rangle_i / (\partial_y \langle \overline{u}\rangle)_i^2)^{1/2}$, on the permeability Reynolds number is important. The mixing length can be thought of as a representative length scale of the turbulent eddies at the SWI responsible for turbulent transport of mass and momentum. It is on the order of the bed permeability and is greater than the Kolmogorov length scale ($L_k = (\nu^3/\varepsilon)^{1/4}$, where $\varepsilon \approx u^2_{\tau}/\langle L_{m,i}\rangle$ is dissipation rate of the turbulent kinetic energy~\citep{tennekes1972first}. Figure~\ref{fig:bk_ts}d shows the interfacial mixing length, $\langle L_{m,i}\rangle$, normalized by the Kolmogorov length scale ($L_{k}$) as well as normalized by the permeability ($\sqrt{K}$). The Brinkman layer thickness, the turbulent shear stress penetration, and the mixing length at the interface show very similar dependence on $Re_K$ over the Reynolds numbers studied, suggesting that that the mixing length is a relevant  characteristic scale for transport of momentum and mass.

Flows over highly permeable boundaries share statistical similarities over different types of permeable geometries as shown by~\citet{ghisalberti2009obstructed}.  Asymptotic values predicted from present simulations for $\sigma_{v,c}/\sigma_{u,c} \sim 0.6$, $\sigma_{v,c}/u_{\tau} \sim 1.1$, $\sigma_{u,c}/u_{\tau} \sim 1.8$, and $U_c/u_{\tau} \sim 2.6$ match well with those observed by~\citet{voermans2017variation}. Here. $\sigma_{u,c}= \langle\overline{u^{\prime2}}\rangle^{1/2}_{c}$, and $\sigma_{v}= \langle\overline{v^{\prime2}}\rangle^{1/2}_{c}$, and  $U_c$ is the mean velocity at the crest. The data are normalized by $-\langle\overline{u^{\prime}v^{\prime}}\rangle_c^{1/2}$, where the subscript `$c$' indicates that these similarity relations are defined at the crest of the sediment bed, $y = 0$. 
The successful comparison of various turbulent quantities with the experimental work of~\citet{voermans2017variation,voermans2018model} has important implications for flows over monodispersed sediment beds in general. \citet{voermans2017variation} studied a range of $Re_K$ between ($1-6.3$) by varying the permeability, through the use of medium and large diameter particles, along with flow rates. In the present DNS, the particle sizes and permeability are kept constant and $Re_K$ is varied by changing the bulk flow rate. 
Geometrical features of the sediment beds, such as size of particles, hence the permeability, local porosity variations, and tortuosity, between the experiments and the DNS are very different. The fact that the DNS predictions follow closely with the experimental data suggests that, for turbulent flow over randomly arranged monodispersed sediment beds, the actual location of sediment particles in the bed has little influence on the turbulence statistics. 
However, matching the $Re_K$ is a necessary but not a sufficient condition. A random as opposed to structured arrangement of monodispersed particles in the top layer is also important to achieve statistical similarity and is investigated below. 
 
\subsection{Role of the top layer of the sediment bed}\label{sec:role_top_layer}

To quantify the influence of the top layer of the sediment bed on flow turbulence and statistics at the SWI, a rough impermeable wall case is investigated by matching the entire top layer of the permeable bed at medium Reynolds number (PBM). The top layer of the sediment is placed over a no-slip wall as shown in figure~\ref{fig:dom}d corresponding to the case IWM-F in table~\ref{tab:cases1}. 

\begin{figure}
   \centering
   \subfigure[]{
   \includegraphics[width=3.6cm,height=7.3cm,keepaspectratio]{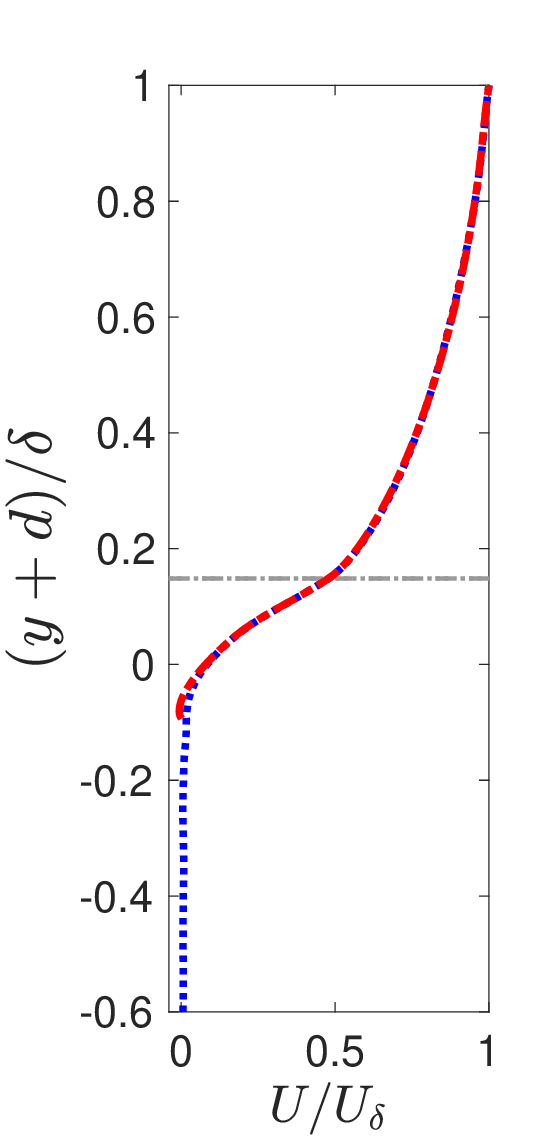}}
   \subfigure[]{
   \includegraphics[width=3.0cm,height=7.3cm,keepaspectratio]{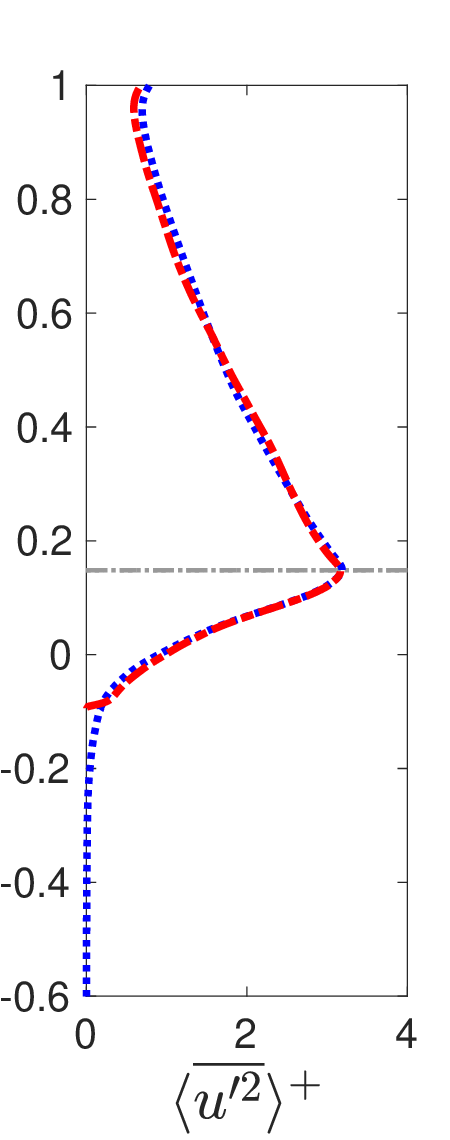}}
   \subfigure[]{
   \includegraphics[width=3.0cm,height=7.3cm,keepaspectratio]{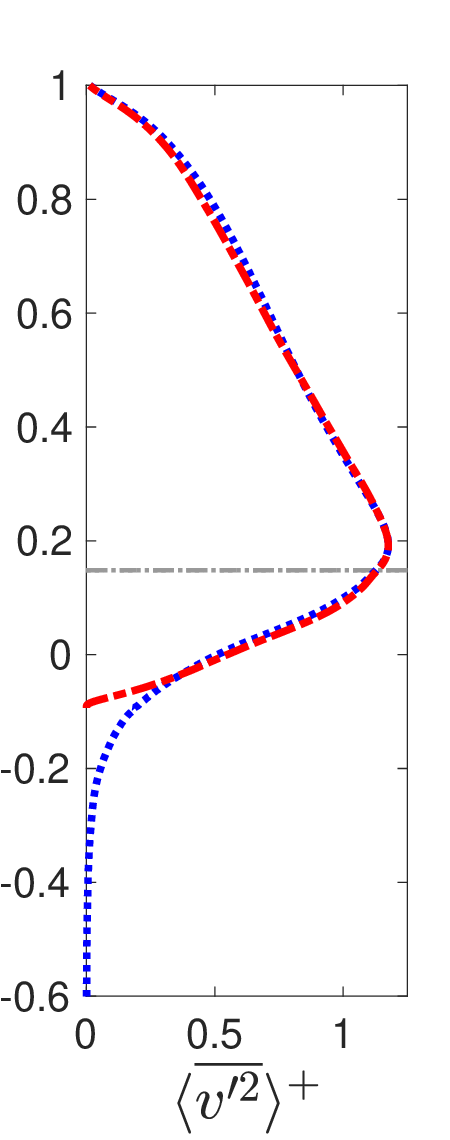}}
    \subfigure[]{
   \includegraphics[width=3.0cm,height=7.3cm,keepaspectratio]{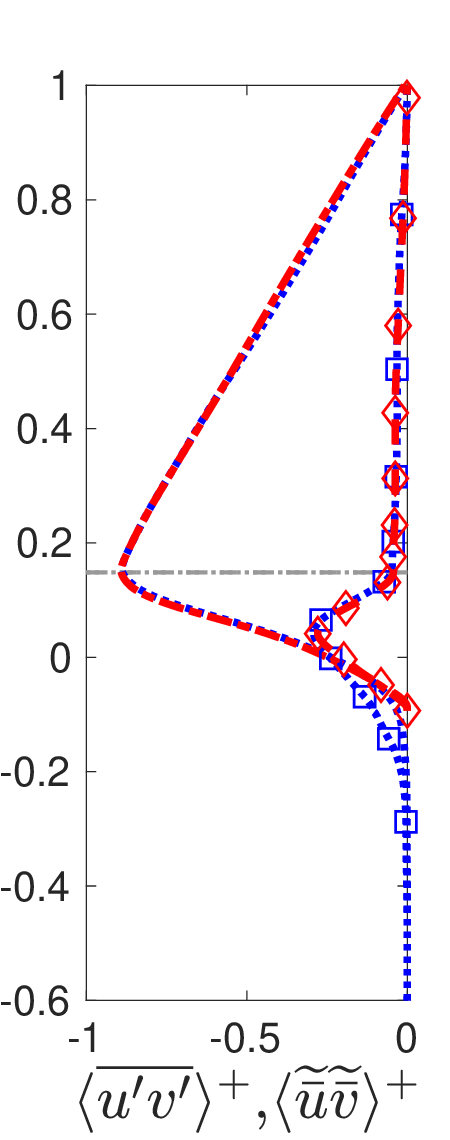}}
    \subfigure[]{
   \includegraphics[width=3.6cm,height=7.3cm,keepaspectratio]{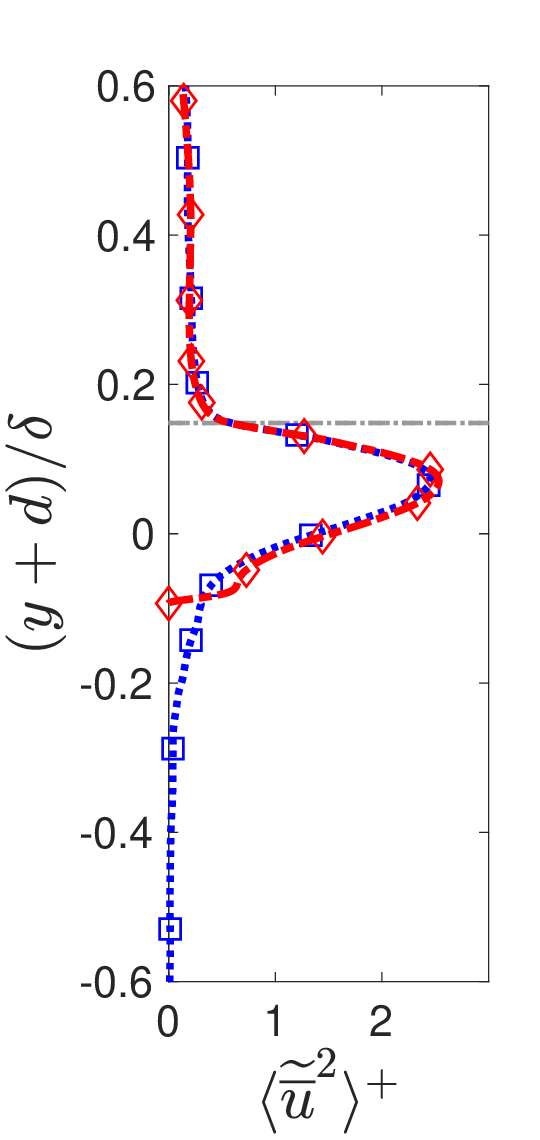}}
   \subfigure[]{
   \includegraphics[width=3.0cm,height=7.3cm,keepaspectratio]{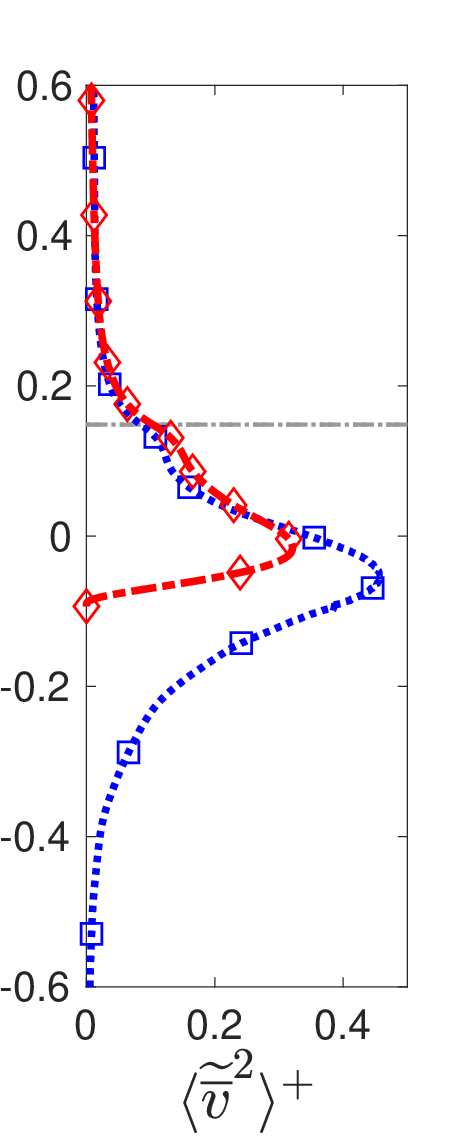}}
   \subfigure[]{
   \includegraphics[width=3.0cm,height=7.3cm,keepaspectratio]{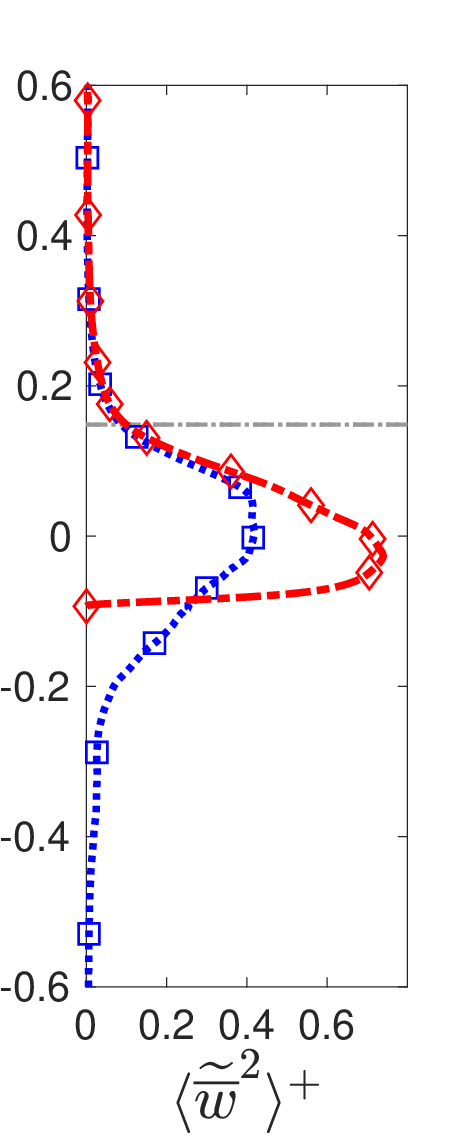}}
   \subfigure[]{
   \includegraphics[width=3.0cm,height=7.3cm,keepaspectratio]{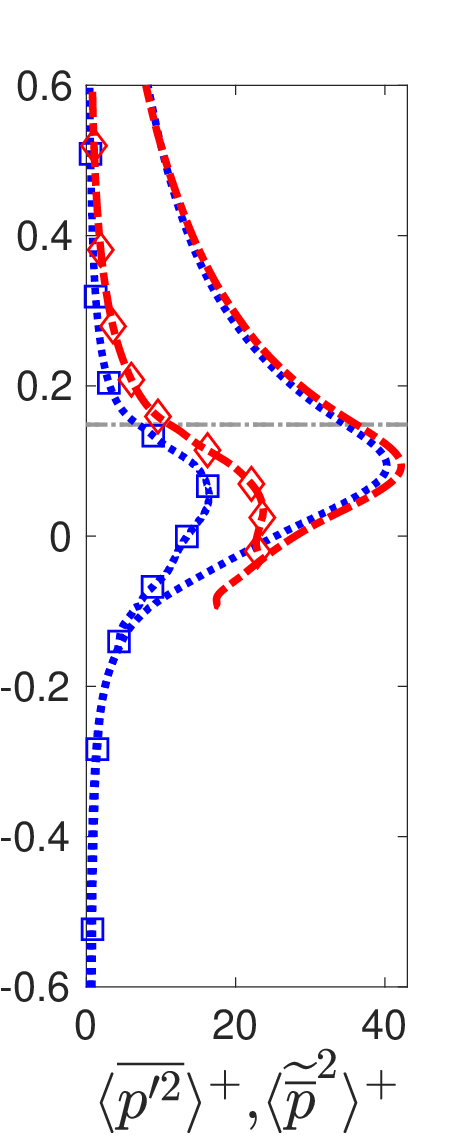}}
\caption{\small Comparison of the mean velocity, Reynolds (lines) and form-induced (lines and symbols) stress profiles for PBM (\bluedottedline, \bluerectangledottedline) and IWM-F (\reddashdotline,\reddiamonddashdotline) cases: (a) mean velocity, (b-c) streamwise, and bed-normal components of spatially-averaged Reynolds stress tensor, (d) spatially averaged Reynolds stress and shear form-induced stress,  (e-g) streamwise, bed-normal and spanwise components of form induced stresses, and (h) mean-square pressure fluctuations and form-induced pressure disturbances. The crest lines for PBM and IWM-F cases overlap and are shown by the  horizontal lines (\graydottedline, \graydashline). Pressure is normalized by $\rho u_{\tau}^2$.}
\label{fig:reys_fis_pbm_iwm}
\end{figure}

Figures~\ref{fig:reys_fis_pbm_iwm}a--h compare the bed-normal variation of mean velocity, Reynolds and form-induced stresses, and pressure disturbances for the PBM and impermeable wall IWM-F cases. The mean velocity, Reynolds stress, and turbulent pressure fluctuation profiles for both the cases overlap each other with no noticeable difference due to the presence of an underlying solid wall in IWM-F. The majority of high-magnitude bed-normal fluctuations are restricted to the top layer of the bed for both cases. The presence of a solid wall underneath the full layer of spherical roughness elements in IWM-F has minimal influence on both the magnitude and penetration of the mean flow and turbulent fluctuations. This is because the full layer of roughness elements creates pockets underneath where the flow can penetrate. Since the turbulent kinetic energy within this layer is still small, the flow characteristics and momentum transport mechanisms resemble that of a permeable bed. Similar behavior was observed and reported by~\citet{manes2009turbulence} in their experimental work.

As in the permeable bed cases, the form-induced stresses (figures~\ref{fig:reys_fis_pbm_iwm}e--g) have lower magnitudes compared to their corresponding Reynolds stresses and the peak values occur significantly below the sediment crest even for IWM-F case. While the peak value for streamwise component ($\langle\widetilde{\overline u}^2\rangle^{+}$) is well captured, the peaks in bed-normal ($\langle\widetilde{\overline v}^2\rangle^{+}$) and spanwise ($\langle\widetilde{\overline w}^2\rangle^{+}$) components show differences between the two cases. Deeper penetration and higher magnitude of form-induced normal stresses are observed in PBM compared to the IWM-F case. The additional layers of sediment grains underneath the top layer in PBM provide connected pathways for the bed-induced flow disturbances to penetrate deeper with a gradual loss in intensity. While in IWM-F, the underlying solid wall blocks the flow penetration and redistributes the stresses tangentially into the spanwise direction, which is evident by larger peak magnitude of $\langle\widetilde{\overline w}^2\rangle^{+}$ compared to $\langle\widetilde{\overline v}^2\rangle^{+}$.

The form-induced pressure disturbances ($\langle\widetilde{\overline p}^2\rangle^{+}$), shown in figure~\ref{fig:reys_fis_pbm_iwm}h, penetrate deeper into the bed in PBM,  resulting in a reduction of its peak value compared to the IWM-F case. The wall blocking effect in IWM-F results in pressure disturbances extending above the crest level, up to about $\left ({y+d}\right)/{\delta} <0.3$. The turbulent pressure fluctuations, however, are well captured and are typically much larger than the form-induced pressure disturbances. Presence of the high magnitude pressure fluctuations only in the top layer is an important observation provided by the present DNS data, as it showcases that inclusion of the effect of a single layer of roughness elements with random arrangement for reach-scale hyporheic exchange models can potentially better capture the turbulent fluctuations.

\subsection{Stress and force statistics on the particle bed}\label{sec:pdf_bedshear_rekcomp}

Direct measurements of shear stress or the drag and lift forces on sediment grains in a laboratory or in the field are challenging. The present pore-resolved simulations provide access to the spatio-temporal variations in these variables. Specifically, knowing higher order statistics of bed shear stress as a function of Reynolds number can critically influence reduced order models for mass and momentum transport that are based on the friction velocity ($u_{\tau}$). Furthermore, incipient motion, sliding, rolling and saltation, driving the bedload transport are modeled based on the bed shear stress exceeding a critical value. Higher-order statistics of bed shear stress are important in stochastic modeling of incipient motion~\citep{ghodke2018spatio}. Motion and rearrangement of sediments can alter local bed porosity and effective permeability, directly impacting hyporheic exchange. The DNS data are used to compute probability distribution functions (pdfs) and statistics of the local variation of net bed shear stress on particle surfaces as well as the net drag and lift forces on the particle bed at different $Re_K$.

The net stress ($\bm{\tau}^t = {\bm \tau}^v - p{\mathbf I}$) on the particle surface includes contribution from the viscous (skin-friction, ${\bm \tau}^v$) as well as pressure (form, $p{\mathbf I}$) stresses. The normal and tangential components of the net stress can be directly evaluated on the particle surface from the velocity and the pressure fields and then transformed into the Cartesian frame using the surface normal vector to obtain the streamwise (${\tau}_x^t$), the bed-normal (${\tau}_y^t$), and the spanwise (${\tau}_z^t$) components, respectively.  The local distribution of the net stress on the particle surface can be integrated over its surface area to obtain the net force on the particle,
\begin{equation}
{\mathbf F} = \int_{\Gamma} {\bm{\tau}}^t\cdot {\mathbf n}~d\Gamma \equiv \int_{\Gamma} \left({\bm\tau}^v - p{\mathbf I}\right) \cdot {\mathbf n}~d\Gamma,
\end{equation}
where $d\Gamma$ is the unit surface area of the particle, $\mathbf n$ is the particle surface normal vector. Accordingly, statistics of the local distribution of the net stress as well as the drag and lift force components are evaluated.

\subsubsection{Local distribution of net stress on the particle surface}

\begin{figure}
   \centering
   \subfigure[]{
  \includegraphics[width=4cm,height=4cm,keepaspectratio]{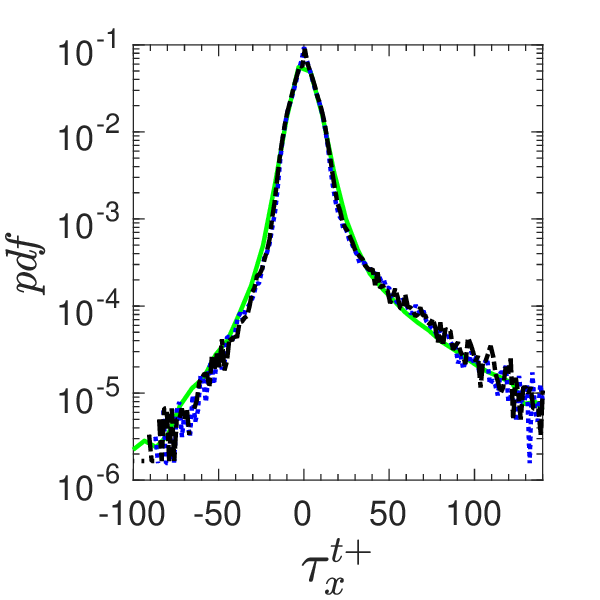}}
  \subfigure[]{
  \includegraphics[width=4cm,height=4cm,keepaspectratio]{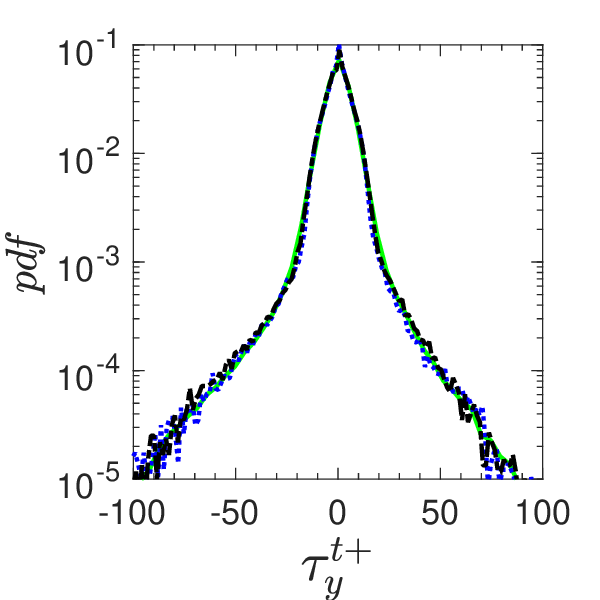}}
  \subfigure[]{
   \includegraphics[width=4cm,height=4cm,keepaspectratio]{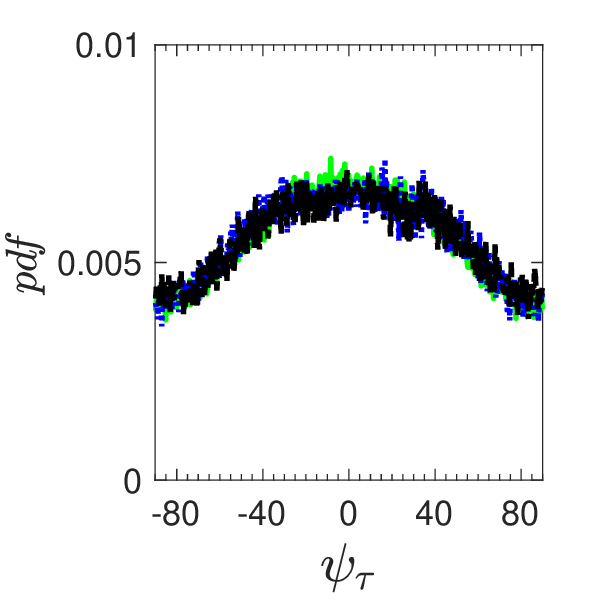}}
\caption{\small PDFs of the net bed stress components (a) streamwise ($\tau_{x}^t$) and (b) bed-normal ($\tau_{y}^t$). Superscript $(.)^+$ denotes normalized quantity by the total bed stress in the $x$ direction ($\tau_w^{t}$). Also shown is the yaw angle $\psi_{\tau}$ based on the viscous components of the stresses (c). Legend: PBL (\greenline), PBM (\bluedottedline), and PBH (\blkdashdotline) cases.}
\label{fig:pdf_bssaplus}
\end{figure}

The probability distribution functions of the streamwise ($\tau^t_x$) and bed-normal ($\tau^t_y$) net local stress normalized by the total stress in the $x$--direction integrated over all particles in the bed ($\tau_w^{t}$), are shown in figures~\ref{fig:pdf_bssaplus}a-b. It is found that these distributions collapse nicely for all Reynolds numbers, suggesting that they are independent of $Re_K$. The dominance of positively skewed streamwise shear stress ($\tau_x^t$) events is clearly visible by this normalization. Positive skewness in the PDFs is associated with the exposed protrusions of the spherical roughness elements to the free-stream, as instantaneous streamwise velocity generally increases in the bed-normal direction. For wall-bounded flows, the probability of negative wall shear stress is generally linked to the near-wall low and high speed regions, however, with protrusions from the rough sediment bed, these structures are destroyed as was shown in section~\ref{sec:quadrant}. The probability of negative $\tau_{x}^{t+}$ fluctuations is then highly associated with trough regions between the roughness elements and are representative of reverse flow behind the exposed sediment particles. As the net bed stress increases with increase in $Re_K$, the probability of extreme streamwise stress events increases. The PDFs of bed-normal ($\tau^{t+}_y$) and spanwise ($\tau^{t+}_z$, not shown) stress are more symmetric due to the absence of directional influence of a strong mean flow gradient.

The relationship between the two viscous components of shear stress can be understood by their yaw angle, $\psi_{\tau} = {\rm atan}(\tau_{z}^v(t)/\tau_{x}^v(t))$. \citet{jeon1999space} reported that the shear-stress yaw angles for smooth walls are within the range of range of $-45$ to $45$ degrees, indicating that events with large values of $\tau_{x}^v$ are associated with small $\tau_{z}^v$. However, in flow over rough permeable beds, probability of yaw angles above $45$ degrees is much higher as seen in figure~\ref{fig:pdf_bssaplus}c. This shows that comparable magnitudes of $\tau_{x}^v$ and $\tau_{z}^v$ occur more frequently in flows over sediment beds. However, the yaw angle is independent of $Re_K$, suggesting that it is influenced more by the roughness distribution of the bed rather than the flow. The roughness elements reduce the directional bias of near bed vortex streaks, resulting in more isotropic vorticity fields, wherein the probability of large scale fluctuations occurring simultaneously in both components of shear stress increases. 

\begin{figure}
   \centering
  \subfigure[]{
  \includegraphics[width=6cm,height=4cm,keepaspectratio]{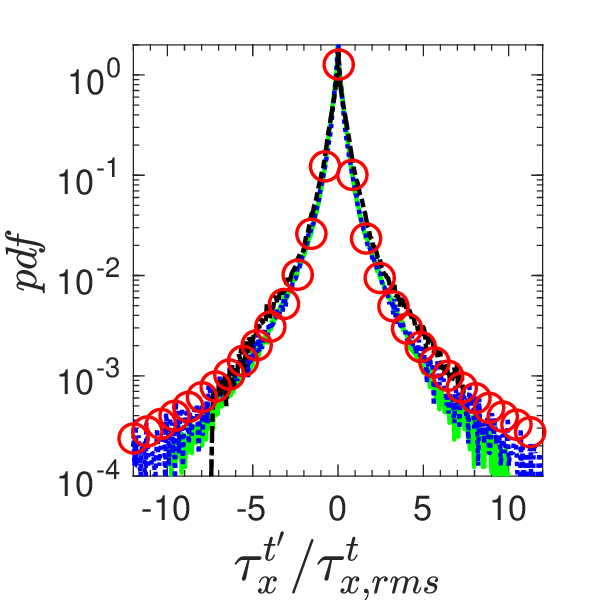}}
   \subfigure[]{
  \includegraphics[width=6cm,height=4cm,keepaspectratio]{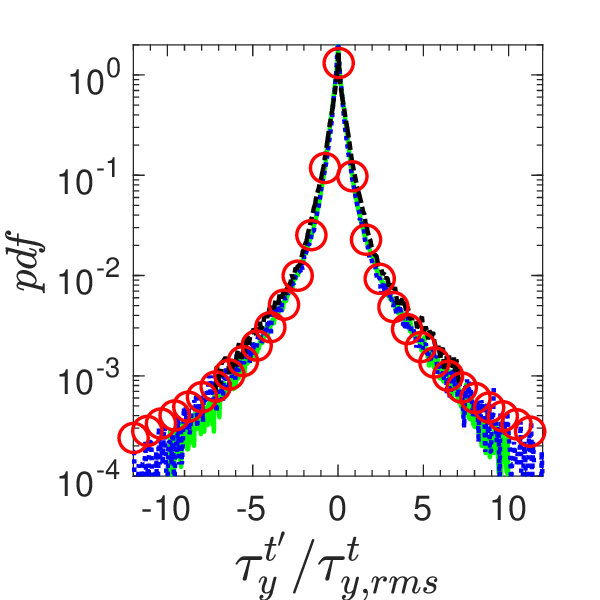}}
\subfigure[]{
  \includegraphics[width=6cm,height=4cm,keepaspectratio]{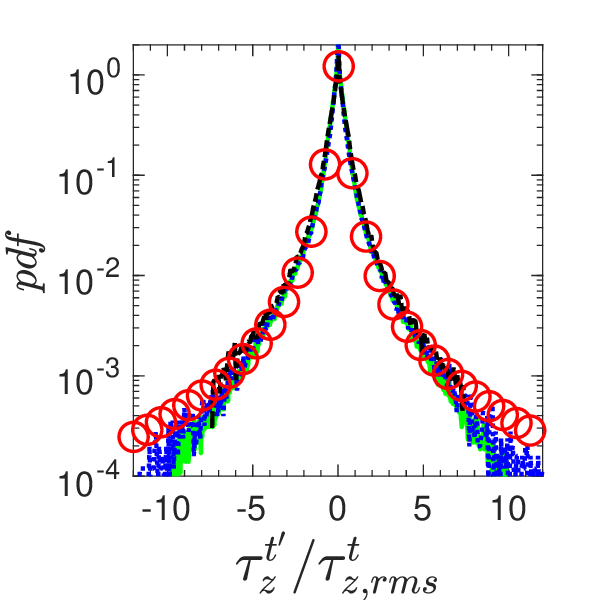}}
   \caption{\small PDFs of net stress fluctuations normalized by their root-mean-square values for (a) streamwise ($\tau^{t\prime}_{x}$), (b) bed-normal  ($\tau^{t\prime}_{y}$), and (c) spanwise ($\tau^{t\prime}_{z}$) components for PBL (\greenline), PBM (\bluedottedline), and PBH (\blkdashdotline) cases. Also shown is a non-Gaussian $t$-location scale fit (\redcircle).} 
\label{fig:pdf_bssafluc}
\end{figure}

As the normalized PDFs of the local distribution of the net bed stress collapse for different $Re_K$, it is conjectured that a simple deterministic fit to the PDFs of fluctuations in stress is possible and is investigated. The PDFs of the local distribution of the net stress fluctuations on the particle surface normalized by their rms values are shown in figure~\ref{fig:pdf_bssafluc}a--c. The higher order statistics for the net stress are shown in the table~\ref{tab:bss_stats_rekcomp}. The mean (not shown) and standard deviation of shear stresses increase with $Re_K$ suggesting higher probability of extreme events for larger $Re_K$. The fluctuation PDFs are symmetric, but non-Gaussian, and show peaky distribution with heavy tails and high Kurtosis values given in table~\ref{tab:bss_stats_rekcomp}. A $t$-location model fit based on the variance, zero skewness, and shape factor determined by excess kurtosis represents the distributions very well.

\begin{table}
\begin{center}
\def~{\hphantom{0}}
\begin{tabular}{@{}l |c c c | c c c | c c c }
\multicolumn{1}{c}{}&\multicolumn{3}{c}{$\tau^{t^{\prime}}_x$}& \multicolumn{3}{c}{$\tau^{t^{\prime}}_y$}&\multicolumn{3}{c}{$\tau^{t^{\prime}}_z$} \\ \hline
Case & $\hat{\sigma}(\cdot)$ & $Sk(\cdot)$ & $Ku(\cdot)$ 
     & $\hat{\sigma}(\cdot)$ & $Sk(\cdot)$ & $Ku(\cdot)$
     & $\hat{\sigma}(\cdot)$ & $Sk(\cdot)$ & $Ku(\cdot)$\\ 
PBL &3.3e-1 &-1.28e-1 &26.51 &3.2e-1 &7.86e-2 &26.67 &3.0e-1 &8.0e-3 &22.75\\ 
PBM &1.38 &-5.16e-1 &19.92  &1.41 &1.57e-1 &20.95 &1.29 &-7.89e-2 &19.5 \\ 
PBH &2.7 &1.65e-1 &14.86 &2.79 &1.3e-1 &15.85 &2.71 &-8.1e-2 &15.2 \\ 
\end{tabular}
\caption{Higher order statistics for streamwise ($\tau_x^{t^\prime}$), bed-normal ($\tau_y^{t^\prime}$), and spanwise ($\tau_z^{t^\prime}$) stress fluctuations showing the standard deviation $\hat{\sigma}(\cdot)$, skewness $Sk(\cdot)$, and kurtosis $Ku(\cdot)$.}
\label{tab:bss_stats_rekcomp}
\end{center}
\end{table}

\begin{figure}
   \centering
   \subfigure[]{
   \includegraphics[width=4cm,height=4cm,keepaspectratio]{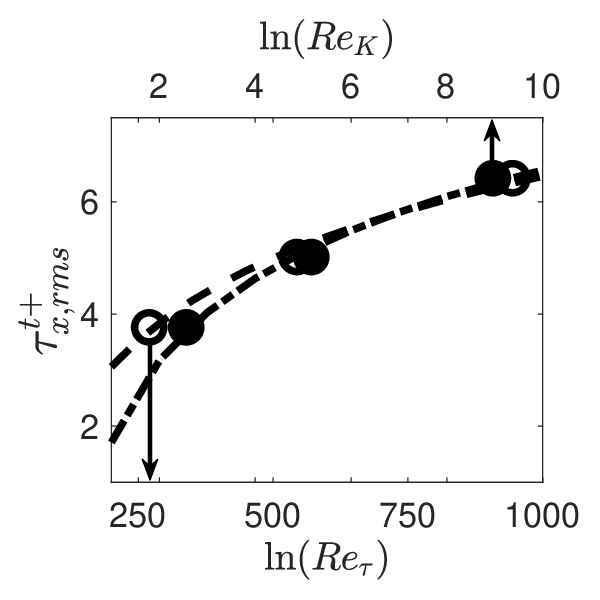}}
    \subfigure[]{
   \includegraphics[width=4cm,height=4cm,keepaspectratio]{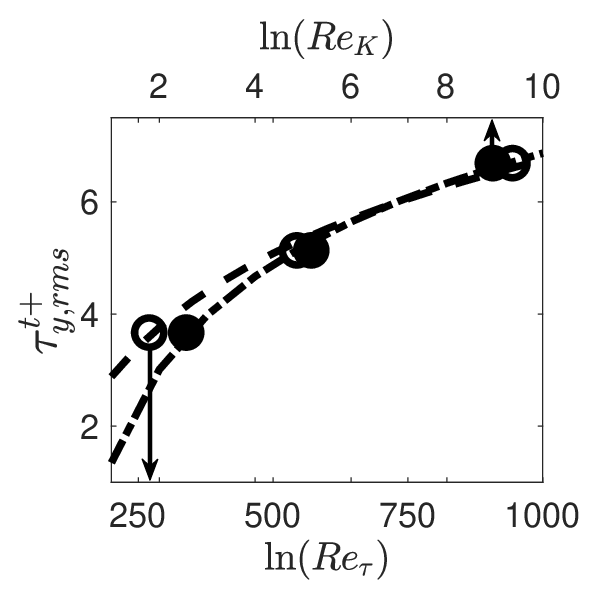}}
    \subfigure[]{
   \includegraphics[width=4cm,height=4cm,keepaspectratio]{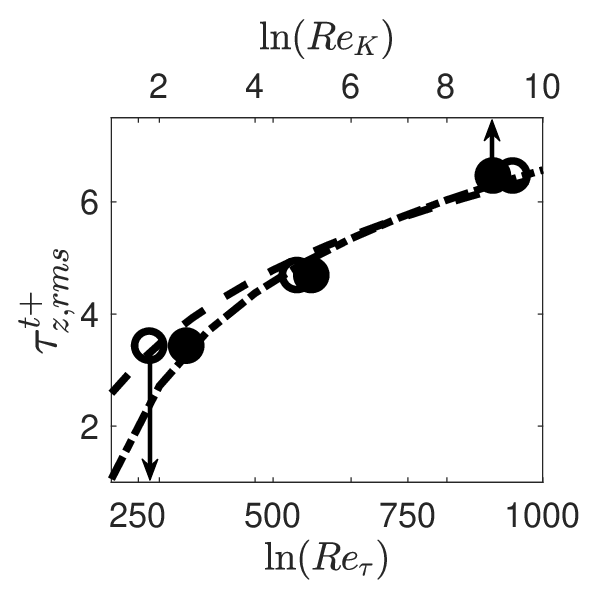}}
\caption{\small Variation of the root-mean squared fluctuations of the net bed stress normalized by the total bed stress ($\tau_w^t$) with the friction and permeability Reynolds numbers: (a) $\tau^{+}_{x,rms}$, (b) $\tau^{+}_{y,rms}, $ and (c) $\tau^{+}_{z,rms}$. Function of $Re_{\tau}$: DNS data (\blkcircle), fitted logarithmic correlation from DNS data for permeable sediment beds \textbf{\blkdashline}, and Function of $Re_{K}$: DNS data (\blkfillcircle), fitted logarithmic correlation from DNS data for permeable sediment beds \textbf{\blkdashdotline}.}
\label{fig:taurms_retau_corre}
\end{figure}

For smooth wall-bounded flows, the root mean-squared fluctuations of streamwise stress follow a logarithmic correlation as proposed by~\citet{orlu2011fluctuating}. Accordingly, a logarithmic correlation between the root mean-squared fluctuations and the friction Reynolds number is also assumed in the present permeable bed DNS,
\begin{eqnarray}
     \tau^{t+}_{x,rms} = \tau_{x,rms}^t/\tau_{w}^t &=&2.10\ln Re_{\tau} - 8.11
    \label{eq:tauyxrms_corr_1}, \\
    \tau^{t+}_{y,rms} = \tau_{y,rms}^t/\tau_{w}^t &=&2.4\ln Re_{\tau} - 9.83
    \label{eq:tauyyrms_corr_1}, \\
     \tau^{t+}_{z,rms} = \tau_{z,rms}^t/\tau_{w}^t &=& 2.4\ln Re_{\tau} - 10.10,
    \label{eq:tauyzrms_corr_1}
\end{eqnarray}
and is shown in figures~\ref{fig:taurms_retau_corre}a--c. Here, the root mean-squared fluctuations are obtained by computing the net stress fluctuations on particle surface over the entire bed and then time-averaging. For the present monodispersed bed, the friction  ($Re_{\tau}$) and permeability Reynolds numbers ($Re_K$) are related to each other, and thus the above relation can also be plotted in terms of the permeability Reynolds number as shown.

The logarithmic dependence of shear stress fluctuations together with a symmetric, non-Gaussian distribution for local stress fluctuations, is an important result, as in the field measurements for natural stream or river bed studies, typically the friction velocity, $u_\tau$, is measured~\citep{jackson2013fluid,jackson2015flow} to compute $Re_{\tau}$. 
Equations~\ref{eq:tauyxrms_corr_1}--\ref{eq:tauyzrms_corr_1} can then be used to evaluate the bed stress variability for different $Re_K$.  Together with the non-Gaussian model distribution for the PDFs of stress fluctuations, a stochastic approach for mobilization and incipient motion of sediment grain can be developed.

\subsubsection{Net drag and lift force distribution}

The local stress distributions on particle surface are integrated over each individual particle and then time averaged to obtain the mean and fluctuating drag and lift forces. Percentage contributions of the viscous and pressure stresses to the average drag and lift forces in the full bed as well as just the top layer of the bed are given in table~\ref{tab:force}. Majority of the contribution to the drag and lift forces comes from pressure distribution. 
It was also found that the top layer of the bed results in average forces that are 3--4 times the average in the full bed indicating that a significant contribution to the lift and drag force comes from the top layer of the bed.

\begin{table}
\begin{center}
\def~{\hphantom{0}}
\begin{tabular}{@{}cl | c c c c c | c c c c c }
\multicolumn{1}{c}{} &\multicolumn{1}{c}{}  &\multicolumn{5}{c}{Drag force} & \multicolumn{5}{c}{Lift force} \\ \hline
\multicolumn{1}{c}{}& Case &\%$F_d^v$ & \%$F_d^p$ & ${\hat \sigma}(\cdot)$& $Sk(\cdot)$& $Ku(\cdot)$& \%$F_{\ell}^v$ &  \%$F_{\ell}^p$ &${\hat\sigma}(\cdot)$& $Sk(\cdot)$& $Ku(\cdot)$\\ 
&PBL & 39.4 & 60.6 &0.36 & -0.44& 14.84 &0.9 & 99.1 &0.31 & 0.35 &12.09\\
Full Bed&PBM & 23.7 & 76.3 &1.59 & -0.88 & 14.49& 9.9 & 90.1 &1.26& -0.15 &11.15 \\
&PBH & 16.7 & 83.25 &2.82 &0.46 &11.20 & 3.2& 96.8 &2.83 &0.58 & 8.46\\ \hline
& PBL & 36.8 & 63.2 &0.6 &-0.35 &6.79  &20.1 & 79.9 &0.62 &-3.0e-4 &4.10 \\
Top Layer &PBM & 22.15 &77.85 &2.53 &-0.63 &5.35  &19.3&80.7 &2.5 &0.11 &3.47  \\
&PBH & 13.4 & 86.6 &4.50 &-0.18 &4.42 & 4.6 &95.4 &4.99  &-0.08 &3.81 \\ \hline
\end{tabular}
\caption{Drag and lift force statistics for the full bed and the top layer showing the percentage contributions of viscous ($F_d^v$, $F_{\ell}^p$) and pressure ($F_d^p$, $F_{\ell}^p$) components to the mean force, and the standard deviation ($\sigma$), skewness ($Sk$), and kurtosis ($Ku$) of the fluctuations in the force. } 
\label{tab:force}
\end{center}
\end{table}

\begin{figure}
   \centering
  \subfigure[]{
  \includegraphics[width=6cm,height=4cm,keepaspectratio]{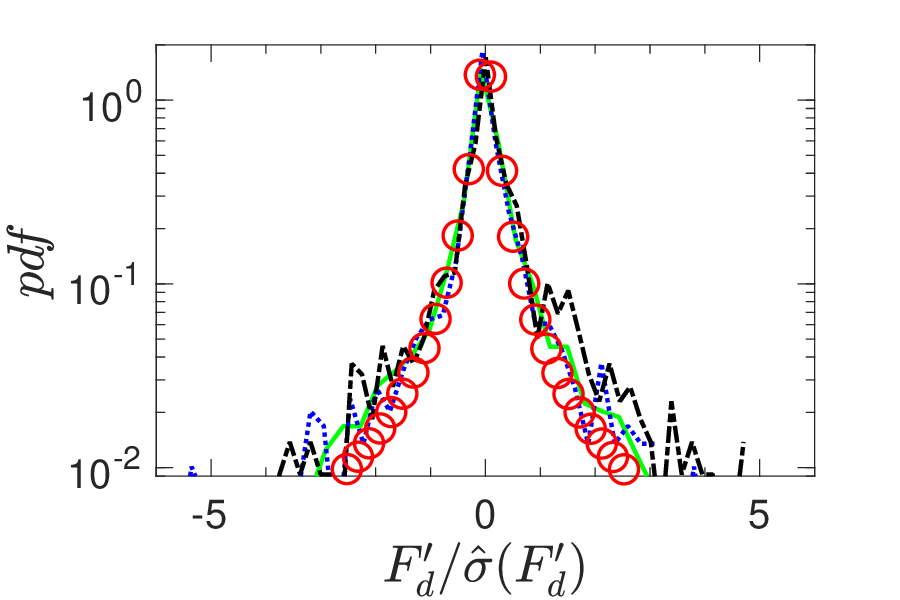}}
 \subfigure[]{
  \includegraphics[width=6cm,height=4cm,keepaspectratio]{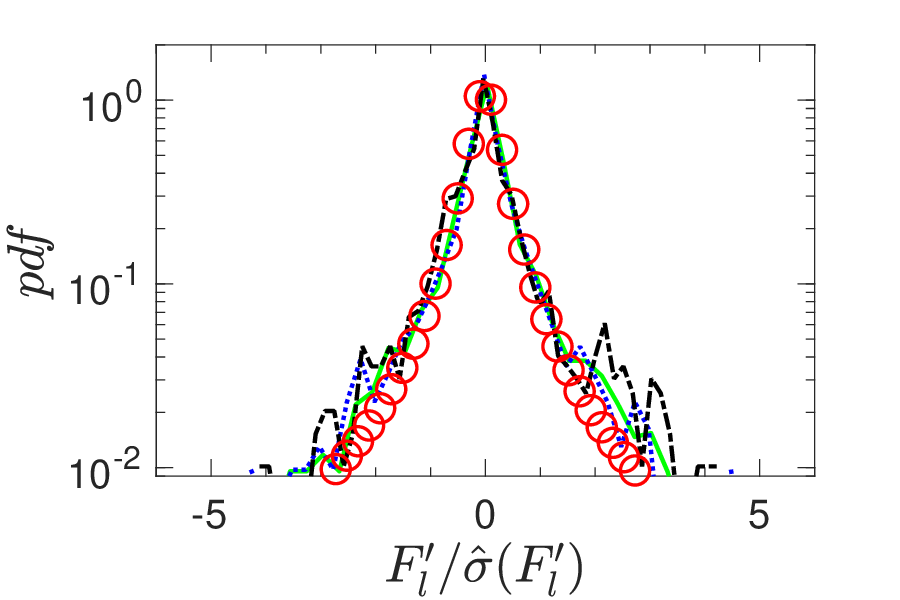}}
    \subfigure[]{
  \includegraphics[width=6cm,height=4cm,keepaspectratio]{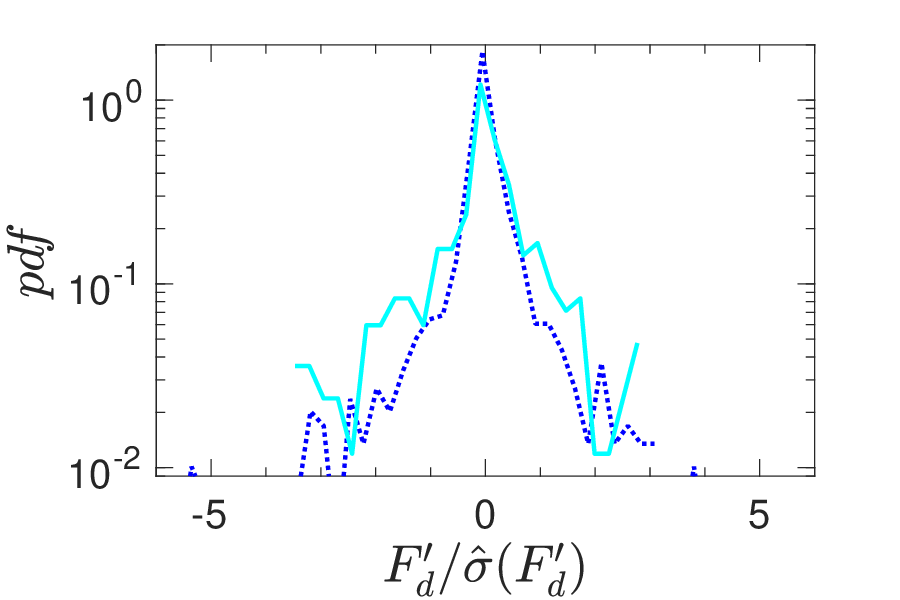}}
   \subfigure[]{
  \includegraphics[width=6cm,height=4cm,keepaspectratio]{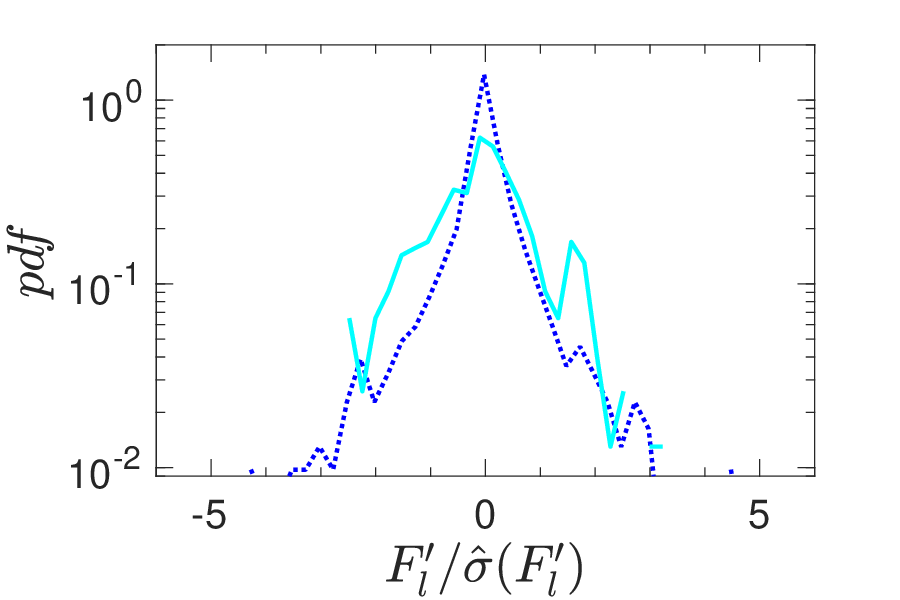}}
   \caption{\small PDFs of drag and lift force fluctuations (a,c) drag force and (b,d) lift force normalized by their respective standard deviation values. Top panel shows permeable bed results for PBL (\greenline), PBM (\bluedottedline), and PBH (\blkdashdotline) cases. Also shown is a $t$-location scale model fit (\redcircle). Bottom panel compares the values for the full permeable bed (PBM, \bluedottedline) and the top layer of the permeable wall (PBMTL, \cyanline) for the medium $Re_K$. }
\label{fig:pdf_force}
\end{figure}

The probability distribution functions of the  fluctuations of drag ($F^{\prime}_d$,  $x-$component) and lift forces ($F^{\prime}_{\ell}$, $y-$component) on all particles within the bed normalized by their respective standard deviation values are shown in figures~\ref{fig:pdf_force}a,b. Similar to the local stress distributions, the drag and lift force fluctuations also exhibit a symmetric, non-Gaussian distribution with heavy tails. Higher-order statistics of the forces (see table~\ref{tab:force}) indicate minimal skewness, and very high kurtosis suggesting probability of extreme forces due to turbulence. A model $t$-location fit function based on the variance and excess kurtosis data fits well for all Reynolds numbers.

Finally, to assess the contribution of the top layer on force statistics, figures~\ref{fig:pdf_force}c,d compare the PDFs of the drag and lift forces for PBM full bed as well as only the top layer of the bed, referred to as the PBMTL data. Note that for the PBMTL data, force fluctuations are normalized by the total force only in the top layer. A close match suggests that the top layer of the bed contributes to the majority of the net drag and lift force fluctuations in the bed.
This has implications for large scale Reynolds-averaged Navier-Stokes modeling where a low-Reynolds number model is used to estimate shear stress on the bottom solid wall of the domain. Including roughness effects through a single layer of sediments may help improve prediction of reduced-order models.

\subsection{Pressure distributions at SWI}\label{sec:presfluc_rekcomp}
Pressure fluctuations at the SWI play a critical role in hyporheic transport even for a flat bed. Specifically, pressure fluctuations due to turbulence are conjectured to have significant impact on mass transport within the hyporheic zone as it can directly influence the residence times through turbulent advection. Reach-scale modeling of hyporheic zone transport typically use a one-way coupling approach, wherein the pressure fields obtained from the streamflow calculations are used as boundary conditions for a separate mass-transport computation within the hyporheic zone using Darcy-flow like models~\citep{chen2020modeling}. These studies have shown that better characterization of the pressure distributions at the SWI can have a significant impact on predicting transport. Specifically, using the present DNS data, the variation of pressure at the SWI with Reynolds number is quantified. 

\begin{figure}
   \centering
    \subfigure[]{
   \includegraphics[width=4cm,height=4cm,keepaspectratio]{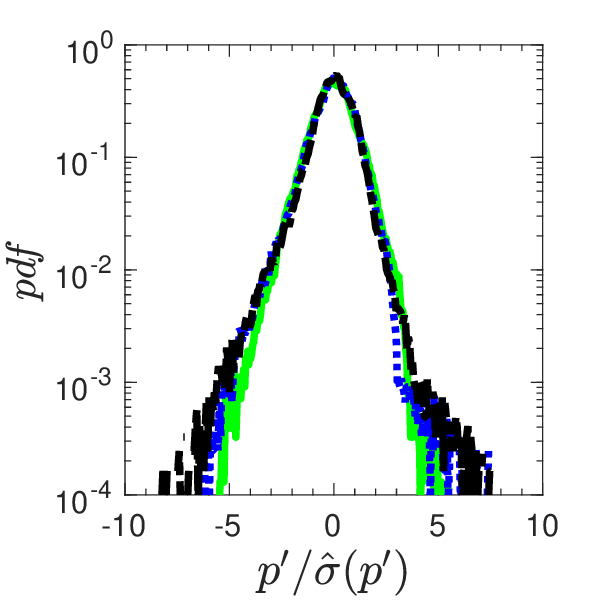}}
   \subfigure[]{
   \includegraphics[width=4cm,height=4cm,keepaspectratio]{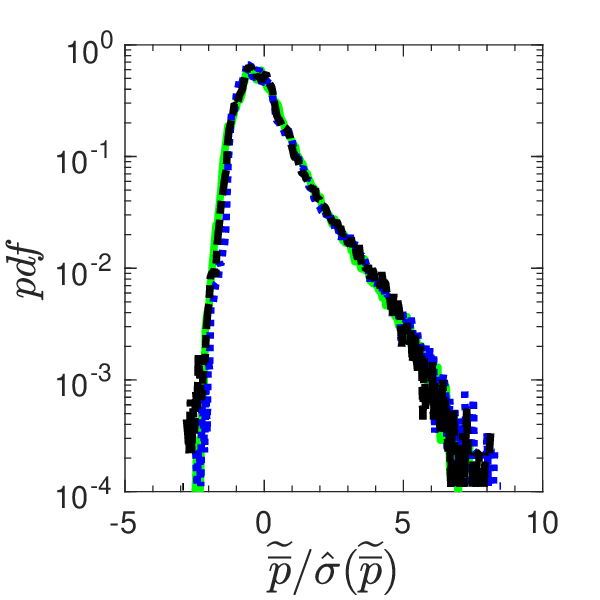}}
   \subfigure[]{
   \includegraphics[width=4cm,height=4cm,keepaspectratio]{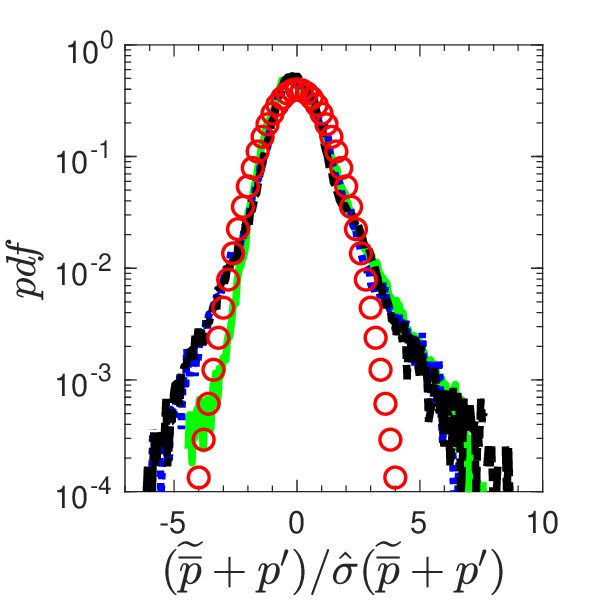}}
    \subfigure[]{
   \includegraphics[width=4cm,height=4cm,keepaspectratio]{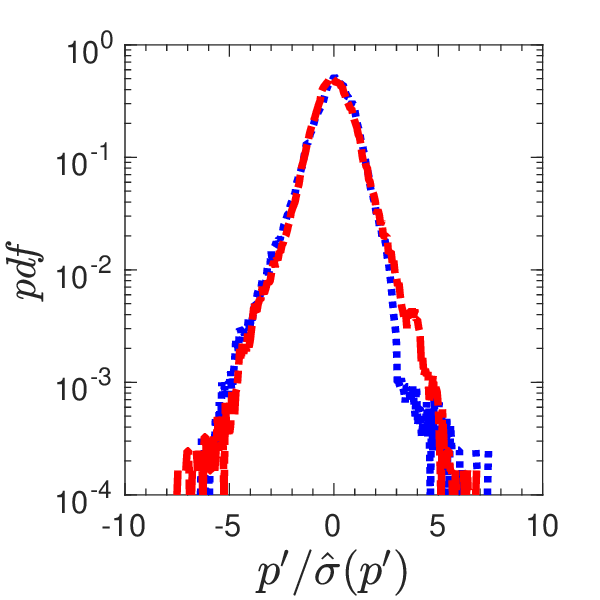}}
   \subfigure[]{
   \includegraphics[width=4cm,height=4cm,keepaspectratio]{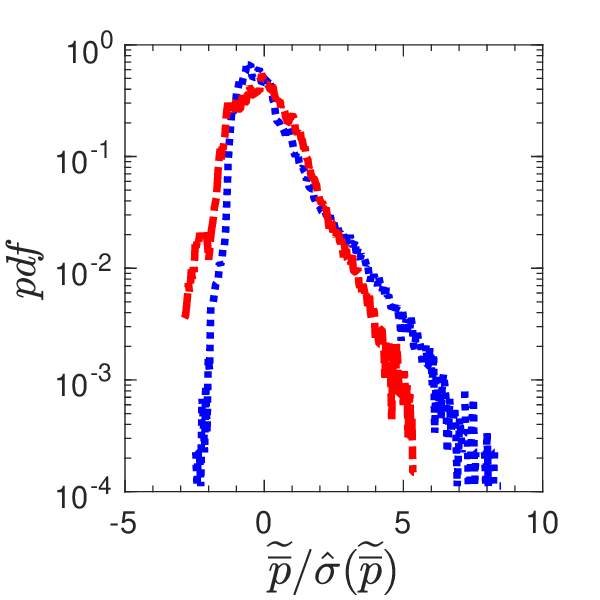}}
   \subfigure[]{
   \includegraphics[width=4cm,height=4cm,keepaspectratio]{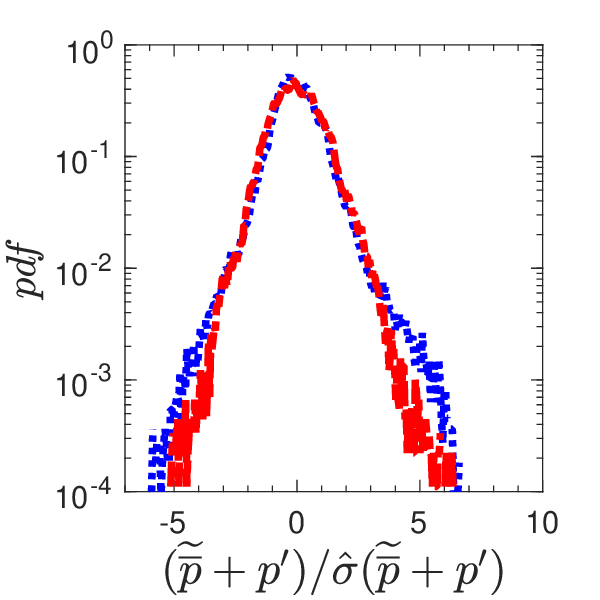}}
\caption{\small PDFs of (a,d) turbulent pressure fluctuations ($p^{\prime}$), (b,e) form-induced pressure disturbances ($\widetilde{\overline p}$) (c,f) sum of $p^{\prime}$ and $\widetilde{\overline p}$. Top panel shows data for the permeable bed cases  PBL (\greenline), PBM (\bluedottedline), and PBH (\blkdashdotline),  and a Gaussian model fit (\redcircle). Bottom panel compares the PDFs for the full permeable bed (PBM) and the impermeable rough wall (IWF-M, \reddashdotline). }
\label{fig:prespdfs_rekcomp}
\end{figure}

PDFs of pressure fluctuations and disturbances normalized by their respective standard deviations and averaged over multiple flow through times for the PBL, PBM and PBH cases are compared at their respective zero-displacement planes ($y=-d$) or the sediment-water interface. Figures~\ref{fig:prespdfs_rekcomp}(a,b) show the pdf distributions for the normalized turbulent pressure fluctuations, $p^{\prime}$, and normalized form-induced pressure disturbances, $\widetilde{\overline p}$. The turbulent fluctuations exhibit close to a normal distribution, whereas the form-induced pressure disturbances have skewed distributions with longer positive tails. This is attributed to the roughness protrusions that create positive pressure stagnation regions. Figure~\ref{fig:prespdfs_rekcomp}c shows the pdfs of the sum of the normalized turbulent and form-induced, $p^{\prime} + \widetilde{\overline p}$, pressure PDFs for the three cases. The pressure sum PDFs for all cases are statistically similar and symmetric, slightly heavy-tailed than a Gaussian, however, the Gaussian distribution nearly captures the pressure data within $\pm 3 \hat{\sigma}$. This result suggests that the pressure behavior inside the bed can be approximated with a simpler Gaussian distribution across a range of permeability Reynolds numbers typical of natural stream or river beds. Table~\ref{tab:prespdfs_stats} lists all higher order statistics for turbulent fluctuations, form-induced disturbances, and their sum. The skewness and kurtosis for $\widetilde{\overline p}$ is higher than $p^{\prime}$ in alignment with the skewed distribution. The mean and variance values for the two are roughly of the same order at the SWI. However, the peak in turbulent pressure fluctuations is much larger than the peak in the form-induced disturbances as seen from the bed-normal variations shown earlier in figure~\ref{fig:reys_fis_rekcomp}h.

\begin{table}
\begin{center}
\def~{\hphantom{0}}
\begin{tabular}{@{}l| c c c | c c c | c c c }
\multicolumn{1}{c}{}&\multicolumn{3}{c}{$p^{\prime}$} & \multicolumn{3}{c}{$\widetilde{\overline p}$} & \multicolumn{3}{c}{$\widetilde{\overline p}+p^{\prime}$} \\ \hline
Case &  $\hat{\sigma}(\cdot)$ &  $Sk(\cdot)$ & $Ku(\cdot)$ & $\hat{\sigma}(\cdot)$ &  $Sk(\cdot)$ & $Ku(\cdot)$ & $\hat{\sigma}(\cdot)$ &  $Sk(\cdot)$ & $Ku(\cdot)$\\ 
 PBL & 1.12 & -1.84e-1 & 4.22 & 1.60 & 1.70 & 8.04 & 1.86 & 1.09 & 6.76 \\ 
 PBM & 2.98 & -5.15e-1 & 5.68 &  2.38 & 2.07 & 9.76 &  3.68 & 0.48 & 6.65 \\
 PBH & 6.11 & -5.66e-1 & 7.52& 4.27 & 1.87 & 8.77& 7.54 & 0.68 & 8.52\\
\end{tabular}
\caption{Higher order statistics for turbulent fluctuations and form induced pressure disturbances, showing the standard deviation $\hat{\sigma}(\cdot)$, skewness $Sk(\cdot)$, and kurtosis $Ku(\cdot)$. }
\label{tab:prespdfs_stats}
\end{center}
\end{table}

The PDFs of normalized pressure fluctuations and disturbances for the  PBM and IWM-F cases are also compared at their respective zero-displacement planes ($y=-d$) or SWI in figures~\ref{fig:prespdfs_rekcomp}d--f. 
The variance in turbulent pressure fluctuations and form-induced disturbances is generally larger in the IWM-F case than those in the permeable bed case (PBM). The probability of higher negative $\widetilde{\overline p}$ increases in the IWM-F case due to the blocking effect of the impermeable wall. However, the sum of the normalized distributions of $p^{\prime}$ and $\widetilde{\overline p}$ show a reasonable match between the PBM and IWM-F cases as shown in figure~\ref{fig:prespdfs_rekcomp}f. Therefore, the distribution of the sum of normalized pressure fluctuation and disturbances in the top layer of the sediment bed is found to be sufficient to capture the pressure variation at the sediment-water interface.

\section{Conclusions} 
\label{sec:conclusions}
Pore-resolved direct numerical simulations of turbulent flow over a randomly packed, monodispersed bed of spherical particles are performed for three permeability Reynolds numbers, $Re_K= 2.56$,~5.17, and 8.94 ($Re_{\tau} = 270, 545$, and $943$), in the hydrodynamically fully rough regime representative of natural stream systems. A thoroughly validated fictitious domain based numerical approach~\citep{apte2009frs,finn2013relative,ghodke2016dns,ghodke2018roughness,he2019characteristics} is used to conduct these simulations. The numerical computations are first validated against the experimental data by~\citet{voermans2017variation} 
at $Re_K \sim 2.56$, and are then used to investigate different Reynolds numbers. The time-space averaging methodology is used to compute the mean velocity, Reynolds stresses, and form-induced stresses. Differences in the near-bed turbulence structure, statistics of the local distribution of the net bed stress on the sediment grains and the resultant drag and lift forces, pressure distributions at the sediment-water interface, and the contribution of the top layer of sediment grains to turbulence statistics were quantified in detail. The {key findings} of this work are summarized below. 

(i) The peak and significant values of the Reynolds stresses occur in the top layer of the bed for all three $Re_K$ cases; decreasing quickly below one grain diameter from the sediment crest. While peak values in streamwise stress decrease, those in bed-normal and spanwise stresses increase with increasing Reynolds number. Streamwise, bed-normal, and shear Reynolds stresses exhibit similarity in the free-stream region, substantiating the wall similarity hypothesis.

(ii) Form-induced stresses are typically lower in magnitude than their respective Reynolds stress counterparts, with the locations of their peak values occurring further below the crest. For low $Re_K$, the spanwise form-induced stress is typically larger than the bed-normal values, a result similar to a rough, impermeable wall. However, at higher $Re_K$, the bed-normal stresses are comparable to the spanwise stresses due to increased flow penetration.

(iii) Mean flow penetration (Brinkman layer thickness) and shear penetration show non-linear, increasing correlation with $Re_K$, their ratio approaching a constant deterministic value. The length scale of dominant turbulent eddies at the sediment-water interface is better represented by the mixing length obtained from the Reynolds stress and mean flow gradient. The normalized interfacial mixing length at the sediment-water interface increases with $Re_K$ and shows similar behavior as the Brinkman layer thickness and shear penetration depth, suggesting that the mixing length is relevant as the characteristic length scale for transport of momentum and mass across the SWI. Quadrant analysis for turbulent fluctuations showcases domination of ejection and sweep events at the SWI. Within one particle diameter inside the bed, the turbulent structures lose both their directional bias and strength, becoming more isotropic in nature.

(v) To quantify the role of the top layer of the sediment bed on the flow, an impermeable rough wall with the same roughness elements as the top layer was investigated. The mean and Reynolds stress profiles show very little differences between the permeable bed and impermeable rough wall. Form-induced stresses are, however, influenced by the impermeability, redistributing stresses tangential along the wall.

(vi) The form-induced disturbances and the turbulent fluctuations in pressure are strongly dependent on Reynolds number, with a ten-fold increase in the peak value between the lowest and highest $Re_K$ cases studied. This increase is attributed to the nature of the flow over the exposed particles in the top layer. A majority of the high magnitude pressure fluctuations are restricted to the top layer of the bed for all Reynolds numbers. The standardized PDFs of the sum of the pressure fluctuations and form-induced pressure disturbances at the sediment-water interface are statistically similar, symmetric, and collapse for different $Re_K$.

(vii) The PDFs of local distribution of the net bed stress computed directly on the sediment grains and normalized by the total bed stress in the streamwise direction, collapse for all Reynolds numbers. The PDFs of fluctuations in the bed stress normalized by their root-mean-squared values are symmetric and exhibit a peaky, non-Gaussian distribution with heavy tails. A logarithmic correlation between the root mean-squared stress fluctuations and the friction Reynolds number (as well as $Re_K$) is observed, which together with the non-Gaussian distribution for fluctuations in stress can be used to develop mechanistic force balance models for incipient motion of sediment grains.

(viii) The mean and fluctuations in drag and lift forces on the particle are computed by integrating the local bed stress on the particle surface. Majority of the contribution to the drag and lift forces comes from the pressure distribution for all $Re_K$. In addition, the top layer of the bed results in average forces that are 3--4 times the average value in the full bed indicating that a significant contribution to the lift and drag force comes from the top layer of the bed. Fluctuations in drag and lift forces have minimal skewness and high kurtosis indicative of a symmetric, non-Gaussian distribution with heavy tails. A $t$-location shape function model based on the skewness and excess kurtosis data fits well for all $Re_K$. Since the local distribution of the net bed stress and the drag and lift forces on particles are mainly influenced by the top layer of the sediment grain, including the roughness effects through a single layer of randomly arranged sediments can potentially improve reach-scale predictions based on reduced-order models.

\begin{acknowledgments}
\section*{Acknowledgements} 
This work was initiated as part of SKK's internship at Pacific Northwest National Laboratory. Simulations were performed at the Texas Advanced Computing Center's (TACC) Frontera system. Computing resource from Pacific Northwest
National Laboratory’s EMSL (Environmental Molecular Sciences Laboratory) is also acknowledged.
\end{acknowledgments}

\section*{Funding} 
SKK acknowledges support from Pacific Northwest National Laboratory (PNNL) as part of an internship program. SVA and SKK gratefully acknowledge funding from US Depart of Energy, Office of Basic Energy Sciences (Geosciences) under award number DE-SC0021626 as well as US National Science Foundation award \#205324. The computing resources used were made available under NSF's Leadership Resources Allocation (LRAC) award. XH and TDS acknowledge funding from the DOE Office of Biological and Environmental Research, Subsurface Biogeochemical Research program, through the PNNL Subsurface Science Scientific Focus Area project (http://sbrsfa.pnnl.gov/). 

\section*{Declaration of Interests} 
The authors report no conflict of interest.

\section*{APPENDIX A. Grid refinement, integral scales, and domain sizes} 
The solver has been thoroughly verified and validated for a range of cases~\citep{AptePatankar} and has also been used for large-scale, parallel simulations of oscillatory flow over a layer of sediment particles~\citep{ghodke2016dns,ghodke2018roughness} and flow through porous media~\citep{finn2013relative,he2019characteristics}.

\begin{table}
\begin{center}
\def~{\hphantom{0}}
\begin{tabular}{@{}l c c c}
$Re_D$& 50 & 150 & 350 \\ 
& \multicolumn{3}{c}{\underline{Drag Coefficient, $C_D$}} \\ 
Isotropic, $D_p/\Delta y = 100$ & 1.54 & 0.9 & 0.65 \\ 
$\Delta x =  \Delta z =  3 \Delta y$ & 1.58 & 0.91 & 0.66 \\
$\Delta x = \Delta z =  4\Delta y$ &1.56 & 0.9 & 0.66 \\  
\end{tabular}
\caption{Grid refinement study for flow over an isolated sphere with non-isotropic, rectilinear grids similar to those used in the present study.}
\label{tab:drag}
\end{center}
\end{table}
In turbulent flow over a sediment bed, there is a need to use non-isotropic and high-aspect ratio grids to minimize the total control volumes and yet provide sufficient resolution needed to capture all scales of turbulence. Specifically, for DNS of open channel flows, the resolution near the sediment bed and in the bed-normal direction should be such that $\Delta y^{+} < 1$, where $\Delta y^{+}$ represents resolution in wall units. The code was used to predict flow over an isolated sphere at different Reynolds numbers using isotropic and non-isotropic, rectilinear grids. The drag force was compared with published data~\citep{AptePatankar} and is given in Table~\ref{tab:drag}. It is observed that the high-aspect ratio grids are capable of predicting the drag forces accurately for $Re_D$ up to 350, where $Re_D = UD_p/\nu$ is Reynolds number based on the sphere diameter, $D_p$, and the uniform undisturbed upstream velocity, $U$. Also, the effectiveness of the non-isotropic grids in capturing vortex shedding was verified using the Strouhal number for vortex shedding at $Re_D=350$. The obtained value for Strouhal number was 0.131, which compared reasonably well with the range of values between 0.135--0.14 predicted on finer, isotropic grids in literature~\citep{mittal2008versatile,mittal1999fourier,bagchi2001direct}.

In the present permeable bed cases, the flow velocity near the bed is much smaller than the free-stream velocity and hence the relevant Reynolds number for the spherical roughness elements is the roughness Reynolds number $D^{+}= D_p u_{\tau}/\nu$, where $D_P$ the particle diameter is a measure of the roughness height for monodispersed particle beds. $D^{+}$ values for the PBL, PBM and PBH cases are 77, 156, and 270, respectively, which fall well within the range of Reynolds numbers of the above grid refinement study. The grid resolution used in the present DNS is thus sufficient to capture the inertial flow features around the particle, including flow separation and wakes.

Eulerian two-point auto-correlations are used to compute the integral length scales in streamwise and spanwise directions, and are compared with the domain sizes in those directions. With increase in Reynolds number, the vortical structures are broken down by the roughness elements, and the integral length scales in the streamwise and spanwise directions are expected to decrease. The time averaged Eulerian two-point auto-correlations were computed as
\begin{align}
\nonumber  \rho^E_{ij} \left( |\boldsymbol{s}| \right) &= \dfrac{\overline{\langle u^{\prime}_i(\boldsymbol{x},t) \ u^{\prime}_j(\boldsymbol{x+s},t)\rangle}}{\overline{\langle u^{\prime}_i(\boldsymbol{x},t) \ u^{\prime}_j(\boldsymbol{x},t)\rangle}} , 
    \label{eq:log_eq}
\end{align}
here $\rho^E_{ij}$ is the Eulerian auto-correlation, $\boldsymbol{s}$ represents the set of all possible vector displacements for which the auto-correlation is calculated. The correlations were first computed at 100,000 randomly picked locations $(\boldsymbol{x})$ in the fluid domain at one instant of time and then spatially averaged to obtain the overall representation. This procedure is then repeated over several flow through times to get a temporal average of the spatially averaged values. The Eulerian integral length scales, $\rho^E_{ij}$, are then calculated by integrating the correlations over the respective abscissas and the results are presented in table~\ref{tab:ils}. The length scales for the PBL case are comparable to values obtained by ~\citet{krogstad1994structure,shen2020direct} at similar Reynolds number. The domain sizes for the PBL and PBM cases are approximately 11--13 times the integral length scale in the streamwise and 20--32 times that in the spanwise direction. The domain size for PBH is same as PBM. The domain lengths $L_x \times L_y$ are $12.56\delta\times6.28\delta$ for PBL and $6.28\delta \times3.14\delta$ for PBM are sufficient for the periodicity assumption to obtain mean and turbulence statistics without any domain confinement effects. In comparison, DNS of turbulent flow over rough, impermeable walls by~\citet{ma2021direct} used $4\delta\times 2.4\delta$ for similar Reynolds numbers. 

\begin{table}
\begin{center}
\def~{\hphantom{0}}
\begin{tabular}{@{}lc c c c c c c    }
Case & $L_{11}/\delta$ & $L_{33}/\delta$ &  $L_{11}/D_p$ & $L_{33}/D_p$& $L_x/L_{11}$ & $L_z/L_{33}$ &  \\
PBL & 0.953 & 0.2 & 3.33  & 0.7 & 13.18 & 31.45  \\ 
PBM & 0.558 & 0.151 & 1.95  & 0.53 & 11.25 & 20.79    \\  
\end{tabular}
\caption{Eulerian length scales in the streamwise ($\rho^E_{11}$) and spanwise ($\rho^E_{33}$) directions normalized by the free-stream height, $\delta$, and the particle diameter, $D_P$. Also shown are the streamwise ($L_x$) and spanwise ($L_z$) domain lengths normalized by $L_{11}$ and $L_{33}$, respectively.}
\label{tab:ils}
\end{center}
\end{table}

\section*{APPENDIX B. Variable volume averaging} 
 \citet{pokrajac2015spatial} used a variable volume averaging methodology, to spatially average time-averaged quantities, wherein thinner volumes were used to average the flow in the free-stream and thicker volumes were used to average inside the porous region with a smooth transition in volume height between the two regions. Following a similar approach, in this study thin volumes are used for averaging in the free-stream region near the crest of the bed where steep gradients in flow quantities are present. The averaging volume is gradually coarsened away from this region and deeper inside the bed thicker averaging volumes are used. For averaging purposes the domain in the vertical direction is divided into four regions: (Region 1) uniform averaging-volume height deep inside the bed, (Region 2) a transitioning averaging-volume height between the uniform region below and the top layer of the bed, (Region 3) a refined uniform averaging-volume height in the top layer and the crest region, (Region 4) a transitioning averaging-volume height in the free-stream region. The variable volume averaging approach across the various segments is given as 
\begin{align}
\nonumber  l_y &=  l_1 \eta_1   \qquad \qquad \qquad \qquad \qquad \quad  \textrm{Region 1} \qquad -1.14 \leq y/\delta \leq -0.57 \\
\nonumber  l_y &=  \frac {l_2 \tanh{\left (\gamma_1 \eta_2 \right)}}{\tanh{\left (\gamma_1 \right)}}   \qquad \qquad \qquad \ \ \textrm{Region 2} \qquad -0.57 < y/\delta \leq -0.28 \\
\nonumber  l_y &=  l_3 \eta_3  \qquad \qquad \qquad \qquad \qquad \quad  \textrm{Region 3} \qquad -0.28 < y/\delta \leq 0.031 \\
\nonumber  l_y &=  \frac {l_4 \left (1 - \tanh{\left (\gamma_2 - \gamma_2 \eta_3 \right)} \right )}{\tanh{\left (\gamma_2 \right)}}   \qquad   \textrm{Region 4} \qquad \quad \ 0.031 < y/\delta \leq 1,
\end{align}
where $l_1, l_2, l_3 \text{ and } l_4$ are the vertical heights of each region, $\gamma_1$ and $\gamma_2$ control the rate of transitioning of the averaging volume height in the bed-normal direction, $\eta_1 = \eta/w_1, \eta_2 = (\eta - w_1)/w_2, \eta_3 = (\eta - (w_1 + w_2))/w_3 $, and  $\eta = j/gnj$.  Here, values $w$ are weights based on the ratio of number of assigned volumes for averaging in each region and the total number of volumes, $j$ is the index of the averaging volume, $gnj$ is the  total number of averaging volumes used in the bed normal direction. Typical $\gamma_1$ and $\gamma_2$ values are between $1.5 - 3$ and $0.7-1.3$, respectively. Similar variable volume averaging has been carried out in previous studies by~\citet{karra2022pore,karra2022particle}

\section*{Appendix C: Validation Study}\label{sec:validation}
Pore-resolved direct numerical simulation of turbulent flow over a sediment bed (case VV) was first validated with experimental data of~\citet{voermans2017variation} as well as DNS data of~\citet{shen2020direct}. Permeable bed case with porosity of $0.41$, $Re_K = 2.56$ and $Re_{\tau} \sim 180$ matches with case L12 in~\citet{voermans2017variation}. In the present work, the numerical algorithm developed by~\citet{dye2013description} is used to generate a random distribution of monodispersed spheres for a given porosity. It uses the collective rearrangement algorithm introduced by~\citet{williams2003random}, coupled with a mechanism for controlling the overall system porosity, providing a periodic arrangement in the streamwise and spanwise directions. Although the average porosity in the sediment bed is matched, the actual random configuration of the sediment particles is likely different compared to both the experimental~\citep{voermans2017variation} and published DNS data~\citep{shen2020direct}. 

\begin{figure}
   \centering
   \includegraphics[width=8cm,height=6cm,keepaspectratio]{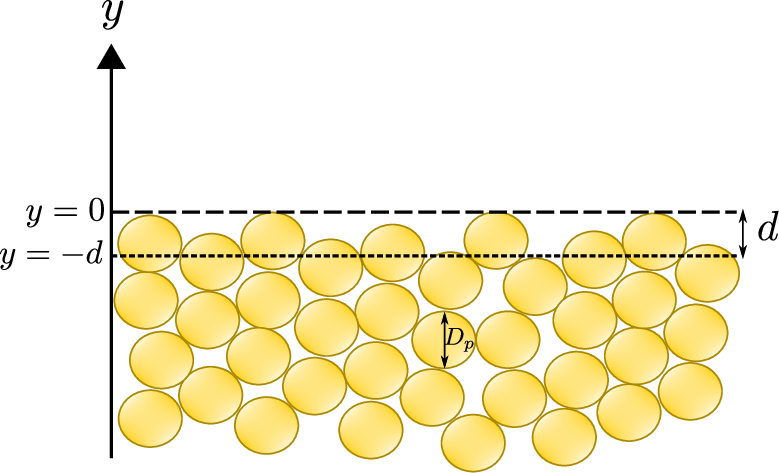}
\caption{\small Schematic showing positions of the sediment crest ($y = 0$), the zero-displacement plane ($y = -d$), and particle diameter ($D_{p}$).}
\label{fig:crestorigin}
\end{figure}

\begin{figure}
   \centering
   \subfigure[]{
   \includegraphics[width=3.1cm,height=7.6cm,keepaspectratio]{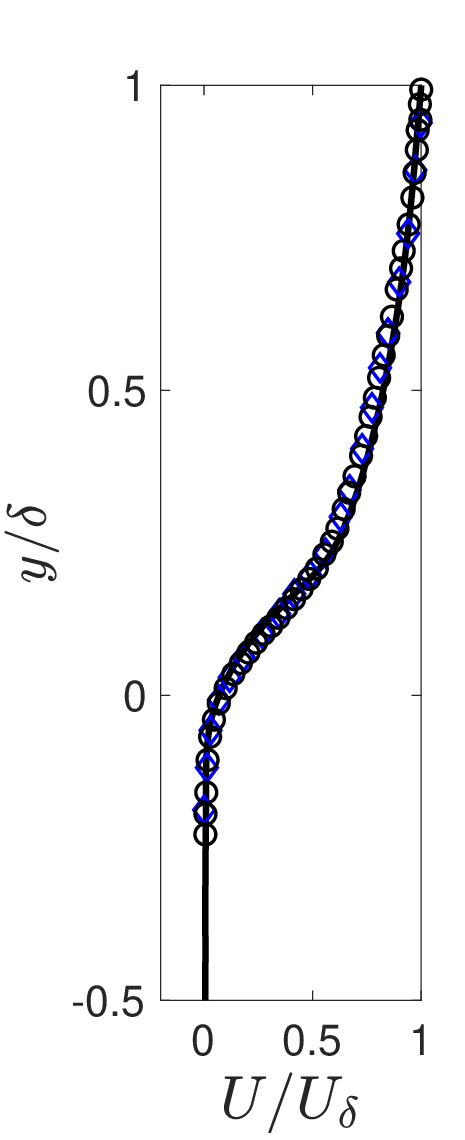}}
    \subfigure[]{
   \includegraphics[width=3.1cm,height=7.6cm,keepaspectratio]{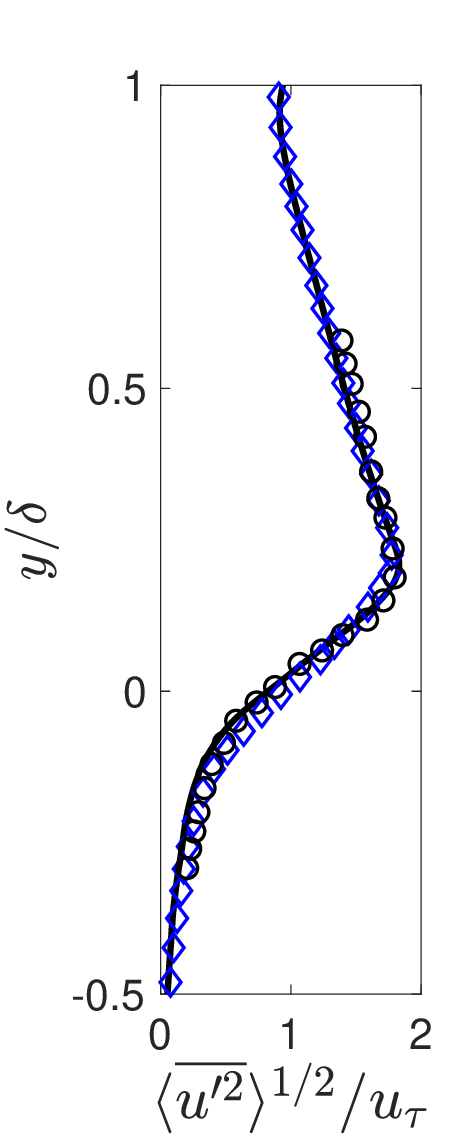}}
   \subfigure[]{
  \includegraphics[width=3.1cm,height=7.6cm,keepaspectratio]{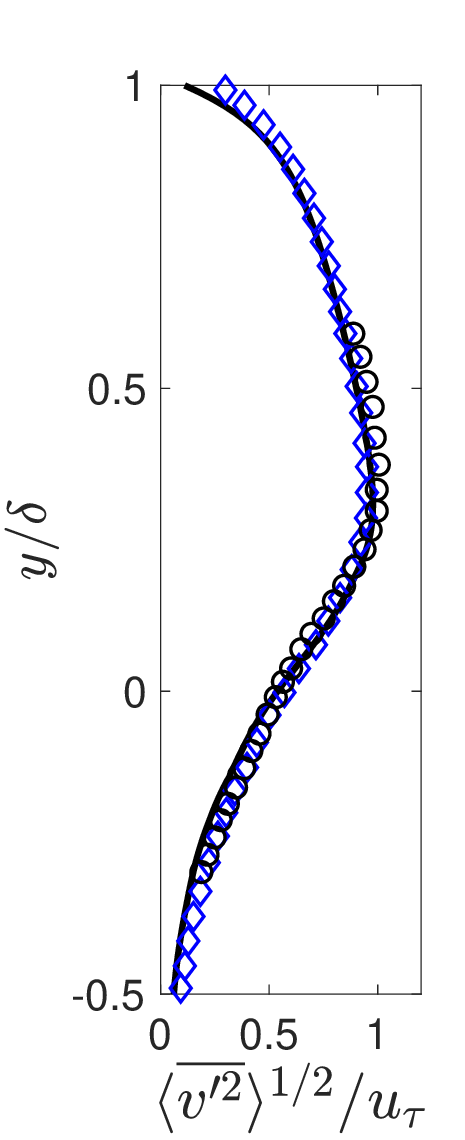}}
  \subfigure[]{
  \includegraphics[width=3.1cm,height=7.6cm,keepaspectratio]{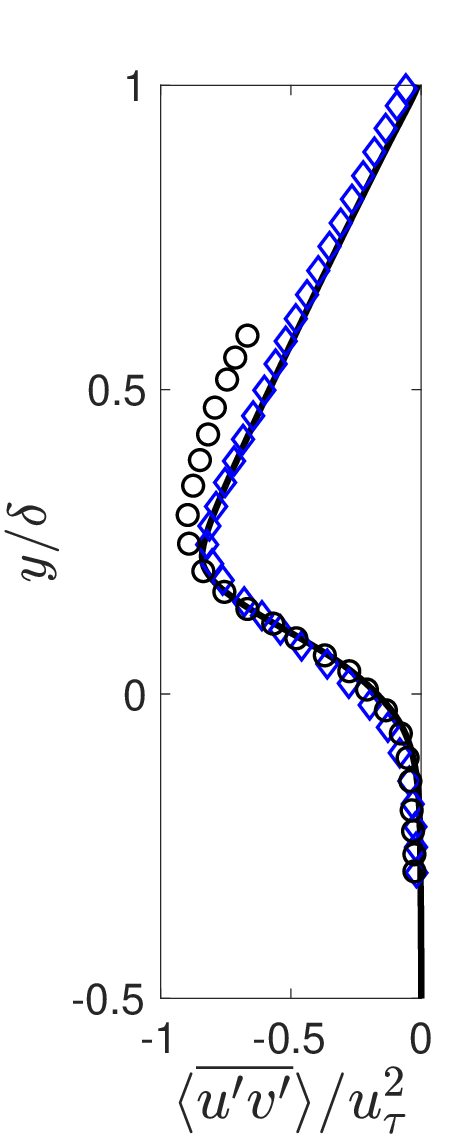}}
\caption{\small Comparison of (a) mean streamwise, velocity and (b) streamwise, (c) wall-normal, and (d) shear components of spatially averaged Reynolds stress tensor. Experimental data by~\citet{voermans2017variation} (\blkcircle), DNS by \citet{shen2020direct} (\bluediamond), present DNS (\blkline). }
\label{fig:vald1}
\end{figure}

\citet{voermans2017variation} defined the origin of the sediment bed to be the inflection point in the porosity profile, that is, where $\partial^2_{yy} \phi=0$. Therefore, in order to facilitate comparison of the current DNS results with the experimental work the origin for case VV is taken to be the inflection point ($\partial^2_{yy} \phi=0$ ) of its porosity profile. Shown also in the schematic in figure~\ref{fig:crestorigin} is the zero-displacement plane, $y = -d$, whose physical meaning and values are given in Appendix D.   
The time-space averaged mean velocity profile normalized by channel free-stream velocity, $U_{\delta}$, is shown in figure~\ref{fig:vald1}a. Excellent agreement is seen between the DNS data, experimental measurements and DNS data from \citet{shen2020direct}. Figures~\ref{fig:vald1}b,~\ref{fig:vald1}c, and~\ref{fig:vald1}d show a comparison of turbulence intensities, namely streamwise, bed-normal and shear stresses. Again very good agreement between DNS and experiment is observed. The slight deviation in Reynolds stress between the DNS and experimental in the outer channel flow region can be attributed to the high measurement uncertainty (between $6-30\%$~\citet{voermans2017variation}) in sampling this variable in the experiment.

\begin{figure}
   \centering
   \subfigure[]{
   \includegraphics[width=3.1cm,height=7.6cm,keepaspectratio]{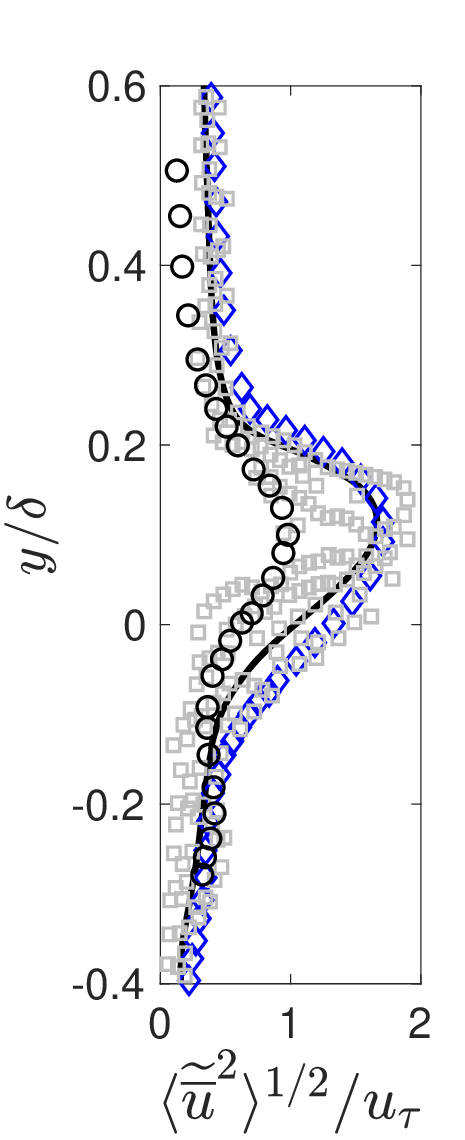}}
    \subfigure[]{
   \includegraphics[width=3.1cm,height=7.6cm,keepaspectratio]{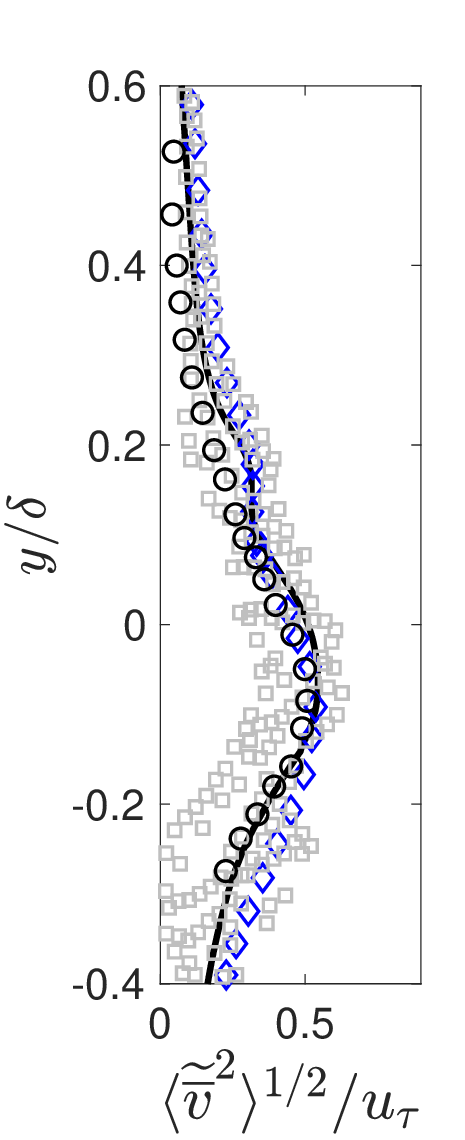}}
   \subfigure[]{
  \includegraphics[width=3.1cm,height=7.6cm,keepaspectratio]{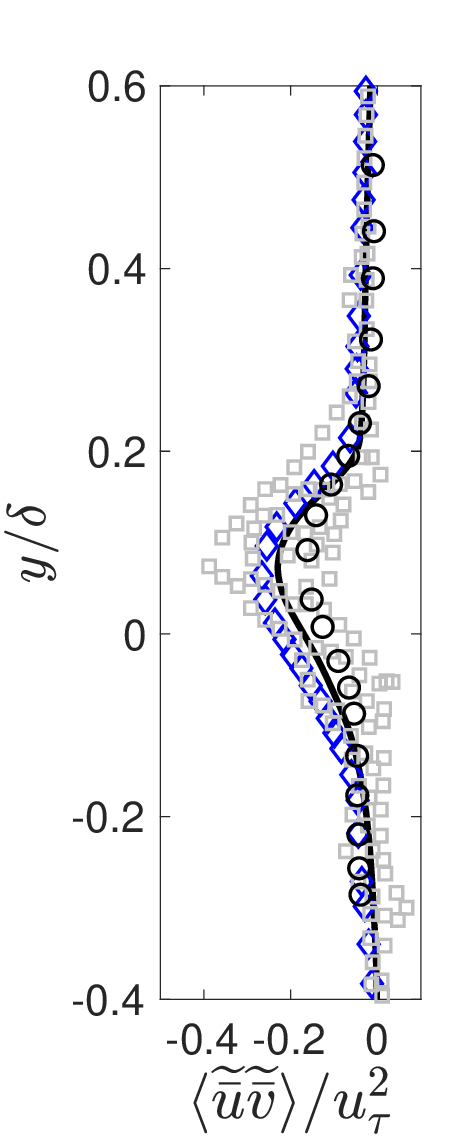}}
\caption{\small Comparison of (a) streamwise, (b) wall-normal, and (c) shear components of form induced stress tensor. Experimental data by~\citet{voermans2017variation} (\blkcircle), emulating experimental sampling (\graysquare), \citet{shen2020direct} (\bluediamond), present DNS (\blkline). }
\label{fig:vald2}
\end{figure}

The normalized form-induced or dispersive stresses are shown in figures~\ref{fig:vald2}a,~\ref{fig:vald2}b, and~\ref{fig:vald2}c. 
The differences between the present DNS and the experimental results can be explained based on the sampling procedures used. Firstly, as mentioned in the previous section, spatial averaging is carried out over an entire $x-z$ volume at a given $y$ location for DNS results. However, for the experimental data, spatial averaging was performed over three different spanwise locations over six different measurements. To quantify the differences in the sampling procedures between the experiments and DNS, the experimental sampling process is replicated in the DNS data whereby spatial averaging is carried out at a few finite uncorrelated spanwise locations and repeated over different streamwise locations. A family of curves, shown by grey squares, indicates the associated uncertainty with the sampling locations of the experimental data. The averaged experimental and DNS data are within this scatter for all streamwise locations. Secondly, it is has been reported in literature~\citep{nikora2002zero, fang2018influence} that the spanwise averaging is highly sensitive to the geometry at the sediment-water interface. For the present DNS, only the mean porosity of the randomly distributed arrangement of monodispersed spherical particles is matched with the experimental geometry. However, the exact sediment-grain distribution in the experiments is unknown and is likely different compared to that used in the DNS. This difference, especially near the top of the bed can also contribute to differences in the form-induced or dispersive stresses.

In spite of the potential differences in the sediment bed distribution between DNS and experimental work, the present results reproduce the mean flow and turbulence stresses observed in the experiment. The form-induced stresses fall within the uncertainty associated with sampling locations in the experiments. In addition, turbulence statistics from the current work are compared with DNS predictions from~\citet{shen2020direct}. Good agreement between the two sets of DNS results is observed. The consistency in predicted results with the published experimental and numerical studies persuasively validates the numerical approach used in this work. 

\section*{D. The log-law and zero-displacement thickness}\label{sec:zero_disp}
In turbulent flows over rough walls and permeable beds, the log-law has the following form
\begin{eqnarray*}
     \dfrac{U(y)}{u_{\tau}} &= \dfrac{1}{\kappa} \log \left (\dfrac{y+d}{y_0}\right), 
    \label{eq:log_eq}
\end{eqnarray*}
\noindent{}where $\kappa$ is the von-K\'{a}rm\'{a}n constant, $d$ is distance between the zero-displacement plane and the top of the sediment crest (see figure~\ref{fig:crestorigin}), and $y_0$ is the equivalent roughness height which is related to the measure of the size of the roughness elements. 

\begin{table}
\begin{center}
\def~{\hphantom{0}}
\begin{tabular}{@{}lc c c c c c c c c  }
Case &  $\kappa$ & $d/{\delta}$ & $d/D_p$ & $d^{+}$ & $y_{0}/\delta$ &$y_{0}/D_p$  & $y_0^{+}$ &$\delta_b/D_p$ &$\delta_p/D_p$  \\
PBL & 0.325 & 0.1686 & 0.59 & 45 & 0.0244 & 0.085 & 6.57 & 0.76 & 0.76  \\ 
PBM & 0.31 & 0.1743 & 0.61 & 95 & 0.0298 & 0.10 & 16.22 & 0.94 & 0.94  \\ 
PBH  & 0.2875 & 0.1829 & 0.64  & 172 & 0.0364 & 0.127 & 34.35 & 1.31 & 1.11 \\ 
\end{tabular}
\caption{The von-K\'{a}rm\'{a}n constant ($\kappa$), zero-displacement thickness ($d$), and equivalent roughness height ($y_0$) normalized by $\delta$, $D_p$ and $\nu/u_{\tau}$. The last two columns shown Brinkman layer thickness $\delta_b$, and shear penetration depth $\delta_p$. $(~)^{+}$ denotes wall units. Results are shown for PBL, PBM, and PBH cases.}
\label{tab:zdp}
\end{center}
\end{table}

Although several techniques have been used to determine these parameters in literature~\citep{raupach1991rough}, the procedure described by~\citet{breugem2006influence} is followed here. First, the extent of the logarithmic layer is determined by plotting $(y+d)^{+}{\partial_{y^{+}} U^{+}}$ against $y^{+}$ for several values of $d$. From the equation of log-law, it is easy to see that the value of $(y+d)^{+}{\partial_{y^{+}} U^{+}}$ is a constant equal to $1/\kappa$ in the logarithmic layer. Therefore, the value of $d$ is the one that gives a horizontal profile in the logarithmic layer. 
The values of $d$, $\kappa$, and $y_0$ determined from a least squares fit of log-law equation to the velocity profile in the logarithmic layer are given in table~\ref{tab:zdp}. 
The von-K\'{a}rm\'{a}n constant ($\kappa$) for the three permeable bed cases is lower than the value of 0.4 for flows over smooth walls. This decrease in $\kappa$ has also been observed in flows over permeable walls by~\citet{suga2010effects,manes2011turbulent,breugem2006influence}. Both the zero-displacement thickness and equivalent roughness height show a dependency on $Re_K$ and increase with increasing Reynolds number. Their values in wall units, $d^+$ and $y_0^+$, compare reasonably well with the studies by~\citet{manes2011turbulent,suga2010effects,breugem2006influence}. 
 
From~\citet{nikuradse1933stromungsgesetze} experiments of flows over impermeable fully rough walls, the ratio of $y_0/D_p$ was found be approximately $1/30$. The values of $y_0/D_p$ for the PBL, PBM, and PBH cases given in table~\ref{tab:zdp} are approximately 2-4 times larger than the value observed in fully rough walls. Importantly, they show a correlation with $Re_K$, due to the influence of the bed permeability. \citet{hinze1975turbulence} (p. 637)  report that for fully rough impermeable walls the ratio of $d/D_p$ is approximately $0.3$. This ratio for the three permeable bed cases is roughly 2 times larger as shown in table~\ref{tab:zdp}  and also shows a dependency on $Re_K$. Therefore, the permeable bed cases in the current study show the characteristics of a fully rough wall regime (based on their roughness Reynolds number ($D^+>70$) as shown in table~\ref{tab:cases1}) influenced by permeability. \citet{breugem2006influence} in their study for cases with $1 < Re_K <10$ found the values of $y_0/D_p$ and $d/D_p$ to be orders of magnitude greater than $1/30$ and $0.3$ respectively as their $D^+$ values were $<7$ which meant that the effects of surface roughness were negligible.

\bibliographystyle{jfm}
\bibliography{hyporheicDNS}

\end{document}